\documentclass[aps,pra,twocolumn,superscriptaddress, showkeys, nofootinbib, 10pt]{revtex4-2}

\usepackage{amsfonts} 
\usepackage{amsmath} 
\usepackage{amssymb} 
\usepackage{amsthm} 
\usepackage{array}
\usepackage{booktabs} 
\usepackage{graphicx}
\usepackage{dcolumn}
\usepackage{tabularx}
\usepackage{epstopdf}
\usepackage{mathrsfs} 
\usepackage{mathtools}
\usepackage{multirow}
\usepackage{comment}
\usepackage{cancel} 
\usepackage{ragged2e}
\usepackage{physics}
\usepackage{hyperref}
\usepackage{url}
\usepackage{tikz}
\usepackage[caption=false]{subfig} 
\usepackage{diagbox} 
\usepackage{xr} 

\theoremstyle{definition}

\usepackage{silence} 
\hypersetup{colorlinks=true, citecolor=orange, urlcolor=blue, linkcolor=magenta}

\newcommand{\red}[1]{\textcolor{red}{#1}}

\newcommand{\green}[1]{\textcolor{green!75!black}{#1}}

\newlength{\subfigurewidth} 
\newlength{\temp} 
\newcommand{\FixedSize}[2]{\makebox[#1][l]{\ensuremath{#2}}}

\graphicspath{{./images/}} 


\begin{document}

\title{Multi-Scale Node Embeddings for Graph Modeling and Generation}

\author{Riccardo Milocco}
\email[Corresponding author:]{riccardo.milocco@imtlucca.it}
\affiliation{IMT School for Advanced Studies, Piazza San Francesco 19, 55100 Lucca (Italy)}

\author{Fabian Jansen}
\affiliation{ING Bank N.V., Bijlmerdreef 106, 1102 CT Amsterdam (The Netherlands)}

\author{Diego Garlaschelli}
\affiliation{IMT School for Advanced Studies, Piazza San Francesco 19, 55100 Lucca (Italy)}
\affiliation{Lorentz Institute for Theoretical Physics, Leiden University, Niels Bohrweg 2, 2333 CA Leiden (The Netherlands)}

\date{\today}


\begin{abstract}
Lying at the interface between Network Science and Machine Learning, node embedding algorithms take a graph as input and encode its structure onto output vectors that represent nodes in an abstract geometric space, 
enabling various vector-based downstream tasks such as network modelling, 
data compression, 
and community detection. 
Two apparently unrelated limitations affect these algorithms. On one hand, it is not clear what the basic operation defining vector spaces, i.e. the vector sum, corresponds to in terms of the original nodes in the network. On the other hand, while the same input network can be represented at multiple levels of resolution by coarse-graining the constituent nodes into arbitrary block-nodes, the relationship between  node embeddings obtained at different hierarchical levels is not understood.  
Here, building on recent results in network renormalization theory, we address these two limitations at once and define a multiscale node embedding method that, upon arbitrary coarse-grainings, ensures statistical consistency of the embedding vector of a block-node with the sum of the embedding vectors of its constituent nodes. We illustrate the power of this approach on two economic networks that can be naturally represented at multiple resolution levels: namely, international trade between (sets of) countries and input-output flows among (sets of) industries in the Netherlands. 
We confirm the statistical consistency between networks retrieved from coarse-grained node vectors and networks retrieved from sums of fine-grained node vectors, a result that cannot be achieved by alternative methods.
Several key network properties, including a large number of triangles, are successfully replicated already from embeddings of very low dimensionality, allowing for the generation of faithful replicas of the original networks at arbitrary resolution levels.
\end{abstract}

\maketitle

\newpage
\section{Introduction}
Complex networks provide a powerful framework for modeling a variety of socially relevant processes, from economic activities to interactions among brain regions \cite{2005_ScaleFreeSelfSim_Song, 2023_LPRecSysTransData_Yilmaz, 2018_MultiScale_Unfolding_GarciaPerez}. 
Essentially, any (dyadic) interaction can be represented as a network, by properly defining the entities as nodes and their interactions as edges. For example, the transactions between the economic sectors are represented by the Input-Output network (ION) \cite{2015_WION_Cerina}, which interprets the former as edges and the latter as nodes. Similarly, the World Trade Web (WTW) \cite{2010_WTW_Fagiolo}, which tracks the trade between countries, assumes the countries as nodes and the trade flows as edges.

\begin{figure}[tbp]
    \centering
    \includegraphics[width=\linewidth, trim = {3cm 2.3cm 2.8cm 2.1cm}, clip]{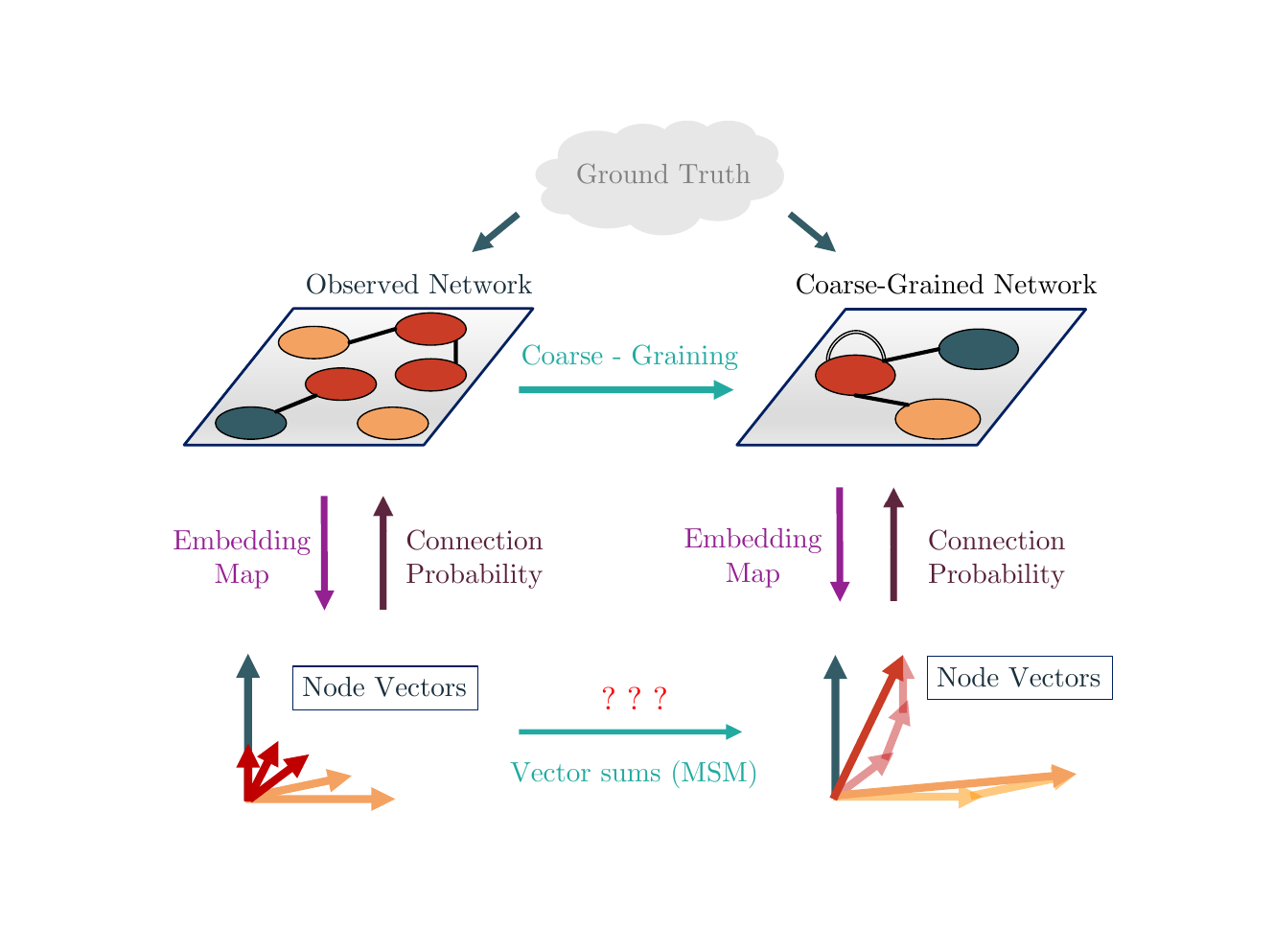} 
    \caption{The ground truth appears at multiple scales, depending on resolution of the dataset. Nevertheless, The macro-scale representation (shown in the plots on the right) can be uniquely obtained by coarse-graining the network at the microscopic scale. \textit{But how node embeddings are connected across different scales?} On the left, one finds the learning procedure of the microscopic node embeddings, whereas on the right, the macroscopic counterparts. In single-scale models, the micro-vectors cannot be directly used to calculate the macro-vectors (indicated by the red question marks). In contrast, the multi-scale model overcomes this limitation, as the macro-vectors are the sum of micro-embeddings (see the parallelogram law on the right).}
    \label{fig:MSNE_ResQuest}
\end{figure}

The downside of this flexibility is that, even when looking at the same generative process, there is no unique definition of the nodes involved in the interaction. By recalling the ION, one may have access to the transactions involving the most detailed sector classification (National Industry), whereas another practitioner to a coarser version of it (Industry) - see \autoref{fig:MSNE_ResQuest}. As expected, the networks at these two resolutions are likely to have different properties. However, they should be described by a model consistently connecting the two scales, since the original Industry level was obtained by aggregating the National Industry one. In general, at a coarser resolution, the network represents the interactions among block-nodes, and it is \textit{uniquely} recovered after specifying the partition $ \Omega$ of the microscopic nodes into blocks. This scheme could be iterated at wish to produce a \textit{multi-scale} representation of the original graph with \textit{nested} partitions: pictorially, this can be thought as a pyramid where the base is the observed network and higher levels are the coarser scale with a lower and lower number of node up to the unique vertex. Lastly, it is worth noticing that the properties of each low-resolution graph change with levels. For instance, the firm graph is less dense than the sector one since there will be fewer nodes to redistribute links to. The same argument applies to other domains (neuroscience, social sciences, $\cdots$), as community structures are widely used to simplifying the heterogeneity of a graph \cite{2008_InfoMap_Rosvall, 2008_Louvain_Blondel}. Therefore, the methodologies exposed in this essay can be applied beyond the examples, we are going to show in the Data and Results sections.

In general, the nodes of a network are known (firms in a country, regions in the brain, people in a social network, $ \dots$), whereas their observed interactions are a single realization of a hidden process. Therefore, by treating the nodes as fixed entities, we aim at modeling their interactions by assigning a probability for every pair of nodes (or edge or link) \cite{2004_StatMecNet_Park}: the higher the probability, the more likely is the edge to exist in a sampled graph. Ultimately, the measurements over the observed graph should coincide with the average over the sampled networks. This exercise is called, in general, \textit{network modeling}, but also in the machine-learning literature \textit{network reconstruction} \cite{2016_SDNE_Wang} or (binary) \textit{edge classification} \cite{2016_EdgeClassification_Aggarwal}.

The machine-learning models tackle this problem by exploiting the \textit{node embeddings} \cite{2016_Grover_n2v,2016_Ou_HOPE,2023_LPRecSysTransData_Yilmaz} which are, in essence, the node parameters on which the probability function depends. Their values are obtained by maximizing the likelihood, or more in general by any other functional optimization. In particular, the general scheme is the one depicted in \autoref{fig:MSNE_ResQuest}. One starts by setting the initial conditions on node embeddings (``embedding map''). Then, the inverse map returns the connection probabilities, giving rise to the likelihood for that set of parameters. In turn, the latter produces the parameter updates towards the optimum. As said, this back-and-forth procedure ends when the likelihood is maximal, or the functional is at an optimum. The vectors are commonly regarded as informative ``features'' to be deployed in different tasks, such as link prediction or community detection\cite{2021_LogReg4SpamNonSpam_Khanday,2021_symLPCA_Chanpuriya,2018_Towards_Interpretation_Dalmia}. 

Two hallmarks of real-world graphs are 1) \textit{low density} (sparsity) and 2) \textit{high triangle density} - many triangles incident to low-degree vertices. Meet them both is challenging, as proven in  \cite{2020_impossibility_of_low_rank_red_Seshandhri}. In particular, every \textit{linear} node embedding methods, such as node2vec \cite{2016_Grover_n2v}, is not capable of reproducing the triangle density of a real network with a low embedding dimension -even with an a posteriori application of a non-linear activation function. Nevertheless, the problem can be solved by \textit{directly} maximizing the likelihood of a \textit{non-linear} logistic function - see LogisticPCA (LPCA) \cite{2021_symLPCA_Chanpuriya}.

As mentioned, a phenomenon may be studied at different scales depending on the resolution level one has access to (see \autoref{fig:MSNE_ResQuest}). Specifically, by combining nodes into communities, one goes from a microscopic to coarse grained scales. Nevertheless, the majority of the models, e.g. LPCA, regard a network only at one scale, providing the optimal embeddings for that level. If nodes were merged into communities, the block vectors have to be re-fitted, implying that these two sets of vectors, e.g. at level $ \ell = \left\{0, 1\right\}$, happened to be completely unrelated as if the models were describing two distinct \textit{processes} - even though we know it was not the case. Moreover, if one wants to re-use the lower-level embeddings, there is no prescription to create an embedding for a community (see \autoref{SI:Inconsistency_LPCA}). For this reason, we are going to call this class of models as ``single-scale'' models (SSM).

Several methods have been proposed in the literature, to solve the multi-scale-classification problem, but they all rely on strong assumptions that limit their use cases \cite{2023_MSNR_Garuccio}. The most promising one \cite{2018_MultiScale_Unfolding_GarciaPerez} assumes that the nodes are embedded in a hyperbolic plane, all the coarser networks are \textit{scale-free} and the block-nodes contain $r$ micro-nodes. The latter restriction doesn't allow to aggregate the nodes with the ``most natural'' partition induced by the studied phenomenon, e.g. geographical distances for the WTW or the sector (industry) for the ION.
To overcome these limitations, the multi-scale model was designed \cite{2023_MSNR_Garuccio} to be generalizable at higher levels, whereas allowing for any arbitrary partition of the microscopic nodes. Specifically, for each block, its \textit{scalar} parameters, or fitnesses, is \textit{uniquely} obtained by summing the fitnesses of its internal nodes (\textit{renormalization rule}). This operation can be seen as the counterpart, in the parameter space, of the aggregation procedure. Note that also the latter operation \textit{uniquely} identifies a community starting from its microscopic members.

In this work, we equip the scalar multi-scale model with node embeddings (MSM), for which the \textit{summation rule} on vectors is shown in \autoref{fig:MSNE_ResQuest}. The benefits are two-folds. On the one hand, it provides an interpretation of the sum of node embeddings, which is rarely addressed in the literature\footnote{The successes of the Natural Language Processing are also due to the effective representation of a phrase obtained by summing the embedding for each word\cite{2013_Word2Vec_Mikolov}. This is hardly replicated in a graph setting, as it is more complex than a language \cite{2017_GCN_Kipf}.} \cite{2018_Towards_Interpretation_Dalmia}. The MSM, instead, assumes that the community vector is the sum of the embeddings of its members. Therefore, the node parameters are now used as vectors, that can be summed as in a proper vector space. The second benefit is that the MSM can be fitted even only at the microscopic level, since the coarser-block embeddings can be obtained under summation. In other words, there is no need to re-fit the model at higher scales. This further implies a lower-computational complexity with respect to a single-scale model that has to be refitted at every level.

The rest of the paper is organized as follows.
In \autoref{sec:Methodology}, we introduce the LogisticPCA and the Multi-Scale Model (MSM) alongside its summation rule.
In \autoref{sec:DatasetDescription}, we describe the ING Input-Output Network and the World Trade Web datasets. Furthermore, we present the \textit{coarse-graining} procedure to obtain the \textit{higher-scale} network.
In \autoref{sec:Results_and_Discussions}, we show the multi-scale results of the two models and discuss the implications of the theoretical results over either network- and machine-learning scores.
Finally, in the Supplemental Material \cite{SuppMat}, we store all the technical details supporting the results in the main text.

\section{Graph Renormalization}
\label{sec:Graph_Renormalization}
In this section, we introduce the mathematical framework for representing a graph and its coarse-graining procedure. The subscript $\ell = 0$ identifies the microscopic graph, while $ \ell > 0$ refers to its aggregated versions. We refer to any quantity defined at level $ \ell$ as an $ \ell$\textit{-quantity}. For instance, the \textit{2-vectors} and \textit{2-nodes} denote the node embeddings and block-nodes at level $ 2$, respectively.

Let us consider a binary, undirected graph with $ N_0$ 
{\raggedright
``microscopic'' nodes $\mathcal{V}_0$ (indexed by $ i_{0} \in [1, N_{0}]$) 
and the connections among them (called edges or links), defined as
$$ \mathcal{E}_0 := \left\{(i_0, j_0) : i_{0} \in \mathcal{V}_0, j_{0} \in [i_0, N_{0}], a^{(0)}_{i_0 j_0} = 1\right\}$$
where $ a^{(0)}_{i_0 j_0} = 1$ indicates the present of an edge between the nodes $ i_0, j_0$ and $ a^{(0)}_{i_0 j_0} = 1$ otherwise.} 
This system is represented by an $ N_{0} \times N_{0}$ symmetric matrix $ \mathbf{A}^{(0)}$, such that $ a^{(0)}_{i_0 j_0} =  a^{(0)}_{j_0 i_0}$, due to the undirected nature of the graph. We do not allow for multiple edges, but self-loops are permitted, and they are stored in the diagonal entries of $ \mathbf{A}^{(0)}$, i.e. $a^{(0)}_{i_0 i_0} = 0, 1$.

\subsection{Coarse-Graining}

To construct a coarse-grained version of the graph, we define a \textit{non-overlapping arbitrary} partition $ \Omega_{0}$ of the microscopic nodes into block-nodes (communities)
\begin{equation*}
    \mathcal{V}_1 := \left\{i_{1} :=  \Omega_{0} (i_{0}) \quad \forall i_{0} \in \mathcal{V}_0 \right\}.
\end{equation*}
Secondly, two blocks are connected if there was at least one edge between any pair of their constituent nodes. Formally,
\begin{equation}
    \label{eq:coarse-graining_01}
    a^{(1)}_{i_1 j_1} = 1 - \prod_{i_{0} \in \Omega_{0}^{-1}(i_{1}), j_{0} \in \Omega_{0}^{-1}(j_{1})} (1 - a^{(0)}_{i_0 j_0})
\end{equation}
where 
$ \Omega_{0}^{-1}(k)$ is the set of microscopic nodes grouped into block-node $ k$. This operation effectively performs a \textit{logical OR} across all edges connecting two communities. That is, only the zeros are preserved, whereas the multiple microscopic edges are seen as one interaction among the two groups. 

Note that $ \Omega_{0}$ is a \textit{surjective} function: multiple nodes $ i_{0}$ belong to exactly one block $ i_{1}$. Self-loops are also allowed at the coarse-grained level ($ i_{1} = j_{1}$), revealing that there was at least one intra-cluster connection -either an interaction between two different nodes or a self-loop at the microscopic level. The process returns a new $ N_{1} \times N_{1}$ adjacency matrix $ \mathbf{A}^{(1)}$ which remains binary and symmetric, as $ \mathbf{A}^{(0)}$. For reference, we denote $ A^{(0)} $ as \textit{0-graph}, $ A^{(1)}$ as \textit{1-graph}, and refer to their nodes as $ 1-nodes$ and $ 0-nodes$, respectively.

\subsubsection{Recursive Coarse-Graining}

This coarsening scheme may be recursively applied. At each level $ \ell + 1 (\geq 1)$, we define a new partition $ \Omega_{\ell}$ that merge $ \ell$-nodes into $ N_{\ell + 1}$ blocks, namely
\begin{equation*}
    \mathcal{V}_{\ell+1} := \left\{i_{\ell + 1} := \Omega_{\ell}(i_{\ell}) \, \forall i_\ell \in \mathcal{V}_{\ell} \right\}.
\end{equation*}
Given a sequence of non-overlapping partitions $ \left\{\Omega_{\ell}\right\}_{\ell \geq 0}$, one can define their composition $ \Omega_{0 \to \ell}$ as
\begin{equation}
    \label{eq:direct_composition}
    \Omega_{0 \to \ell} := \Omega_{\ell} \circ \, \cdots \, \circ \Omega_{0},
\end{equation}
which provides a \textit{direct} mapping from micro $ 0$-nodes to block-nodes $i_{\ell + 1} = \Omega_{0 \to \ell}(i_{0})$.
Therefore, the $ (\ell+1)$-graph $\mathbf{A}^{(\ell + 1)}$ may be recovered either iteratively by means of \autoref{eq:coarse-graining_01}, or \textit{directly} via
\begin{align}
    \label{eq:coarse-graining_ell}
    a^{(\ell + 1)}_{i_{\ell + 1} j_{\ell + 1}} 
    &= 1 - \prod_{i_{\ell} \in \Omega_{\ell}^{-1}(i_{\ell + 1}), j_{\ell} \in \Omega_{\ell}^{-1}(j_{\ell + 1})} (1 - a^{(\ell)}_{i_{\ell} j_{\ell}}) \\
    \label{eq:coarse-graining_0ell}
    &= 1 - \prod_{i_{0} \in \Omega^{-1}_{\ell \to 0}(i_{\ell + 1}), j_{\ell} \in \Omega^{-1}_{\ell \to 0}(j_{\ell + 1})} (1 - a^{(0)}_{i_{0} j_{0}})
\end{align}
where $ \Omega^{-1}_{\ell \to 0} := \Omega^{-1}_{0} \circ \dots \circ \Omega^{-1}_{\ell}$ is the inverse of \autoref{eq:direct_composition}.

The collection of \textit{nested} partitions $ \left\{\Omega_{\ell}\right\}_{\ell \geq 0}$ can be uniquely represent with a dendrogram as shown in \cite{2023_MSNR_Garuccio}. A ``horizontal'' cut through the dendrogram yields the community partitions at the same scale, whereas, by cutting the dendrogram at different heights, it yields the ``multiscale'' clusters. More concretely, the two schemes for the World Trade Web (WTW) consist of merging states nearer than a fixed geographical distance (horizontal clustering) or a state with a continent (multiscale clustering).

In many real-world applications, node partitions are naturally suggested by the problem context, even when an explicit distance metric is missing. An example is the NAICS classification \cite{NAICS_website} of sectors, where each national industry (NI) is identified by a number at $6 $ digits, e.g. Full-Service Restaurants (722511). By removing the last digit, one groups the $6$-digits NI into less-detailed descriptions up to $2 $ digits, e.g. Accommodation and Food Services (72). Specifically, sectors sharing the first $n \in [2, 6]$ digits are grouped into blocks at level $\ell := 6 - n \in [0, 3]$. Interestingly, this induces a Hamming distance between industries, defined as the number of differing digits. Still, this example highlighted that every partition is possible. 
In the absence of a natural partition, one can opt either for community detection algorithms Louvain \cite{2008_Louvain_Blondel,2023_LaplRG_Villegas} or merge at random the 0-nodes using a random distance matrix among them.

\section{Methodology}
\label{sec:Methodology}
In the previous sections, we emphasized that a single \textit{phenomenon} can be analyzed at multiple levels of resolutions. More precisely, since every adjacency matrix $ \mathbf{A}^{(\ell)}$ is binary and undirected, each edge $ (i_\ell,j_\ell)$ can be modeled as a Bernoulli random variable. 

To simplify the notation, we introduce the shorthand $ \xi_{ij} := \xi^{(\ell)}_{i_{\ell} j_\ell}$ to denote a generic quantity $\xi$ at level $\ell \geq 0$, in cases where results hold across all levels. For instance, if $\xi:= \mathbf{A}^{(\ell)}$, then $a_{ij} := a^{(\ell)}_{i_{\ell} j_\ell}$. Hence, for every level $ \ell$, the edge variable $ a_{ij}$ distributes according to
\begin{equation}
    \label{eq:a_ij_Bernoulli_rv}
    a_{ij} =
    \begin{cases}
        1 & \qquad p_{ij} \\
        0 & \qquad 1 - p_{ij}
    \end{cases}
    \qquad \forall i \leq j.
\end{equation}
To capture the hierarchical structure, we used both the LogisticPCA \cite{2021_symLPCA_Chanpuriya} and the vector-based multi-scale model (MSM). We chose LPCA as the representative of the single-scale models, due to its straightforward yet effective derivation. In contrast, to the best of our knowledge, the MSM is the only model that explicitly allows for a \textit{renormalization rule} applicable to arbitrary node partitions.

\subsection{Logistic PCA} 
\label{sec:LPCA}
LPCA\footnote{The authors have historically called LPCA the \textit{directed} logistic model, and no acronym has been assigned for its \textit{undirected} counterpart \cite{2021_symLPCA_Chanpuriya}. In this work, we focus on the symmetric version, and refer to it simply as LPCA for easiness.} 
\cite{2021_symLPCA_Chanpuriya} is a probabilistic model which aims to classify each edge as existing (class $ 1$) or non-existing (class $ 0$). This is analogous to a standard ``spam / non-spam'' classification problem \cite{2021_LogReg4SpamNonSpam_Khanday}, but applied to links, which do not possess explicit features - hence, the need for node embeddings.

The probability of an existing edge between the nodes $i$ and $j$ is given by
\begin{align}
    p_{ij} 
        :&= \sigma(\vec{b}_i, \vec{b}_j, \vec{c}_i, \vec{c}_j) 
        \\[1ex] &= 
        \frac{1}{1 + 
        e^{
        -
        \left(
            \langle \vec{b}_i, \vec{b}_j \rangle - \langle \vec{c}_i, \vec{c}_j \rangle
            \right)
            }}
    \label{eq:LPCA_prob}
\end{align}
where the node embeddings
\begin{equation*}
    \vec{b}_{i} \in \mathbb{R}_{+}^{D_{B} \geq 1}, \vec{c}_{i} \in \mathbb{R}_{+}^{D_{C} \geq 1}
\end{equation*} 
were introduced by the Chanpuriya et colleagues to capture the ``homophily'' and ``heterophily'' natures of each node, respectively (see \autoref{SI:LPCA}).
By defining the non-negative matrices as
$\textbf{B} := 
\begin{bmatrix}
    - \vec{b}_1^T - \\
    \vdots\\
    - \vec{b}_{N}^T -
\end{bmatrix}$ and similarly for $ \mathbf{C}$, the scalar product among them becomes $\mathbf{X} := \mathbf{B}\mathbf{B}^T - \mathbf{C}\mathbf{C}^T$, so that each element is defined by $ x_{ij} := \langle \vec{b}_i, \vec{b}_j \rangle - \langle \vec{c}_i, \vec{c}_j \rangle \in \mathbb{R}$. 
The node embeddings are obtained by maximizing the likelihood \cite{2019_PyTorch_Paszke}
\begin{align}
    \label{eq:BCE_withlogitloss}
    \mathcal{L}\left(\mathbf{X} | \mathbf{A} \right) &:= \mathcal{L}\left(\mathbf{B}, \mathbf{C} | \mathbf{A} \right) = \\
    = \sum_{i < j} \min(x_{ij}, 0) 
    & - \ln( 1 + e^{-\left\vert x_{ij}\right\vert}) - (1 - a_{ij}) x_{ij}
\end{align}
subject to non-negative entries of the vectors $\vec{b}_{ik} \geq 0, \vec{c}_{ik} \geq 0$. It is worth noting that the original likelihood \cite{2021_symLPCA_Chanpuriya} requires summing over all the entries of the adjacency matrix, which is equivalent of summing over the upper triangular elements included the diagonal since $ \mathbf{A}$ is symmetric. However, since the network metrics considered in our study exclude the self-loops, we restricted the optimization to the off-diagonal elements of the adjacency matrix.


\subsection{Multi-Scale Model}
\label{sec:MSM}
A natural approach to modeling the multi-resolution adjacency matrices $\left\{ \mathbf{A}^{(\ell)} : \ell \geq 0 \right\}$ is by means of the multi-scale model (MSM). Specifically, this model is inherently scale-invariant and allows for arbitrary partitions $ \left\{\Omega_{\ell}\right\}_{\ell \geq 0}$ of the microscopic nodes. In this work, we equip the (global) multi-scale model \cite{2023_MSNR_Garuccio} with node embeddings $ \left\{\vec{x}_{i}\right\}_{i \in [1, N]}$, yielding the following connection probability between nodes $ i$ and $ j$:
\settowidth{\temp}{$1- e^{-\frac{1}{2}  \Vert \vec{x}_i \Vert ^2 - w_{i}} \quad $}
\begin{equation}
    \label{eq:MSM_pij}
    p_{ij} :=
    \begin{cases}
        \FixedSize{\temp}{1- e^{-\langle \vec{x}_i, \vec{x}_j \rangle}} i \neq j\\
        \FixedSize{\temp}{1- e^{-\frac{1}{2}  \Vert \vec{x}_i \Vert ^2 - w_{i}}}
    \end{cases}
\end{equation}
where $\Vert \vec{x}_{i} \Vert := \langle \vec{x}_i, \vec{x}_i \rangle$ is the Euclidean norm induced by the scalar product.
Note that the probability is subject to $\vec{x}_{ik} \geq 0, w_{i} \geq - \frac{1}{2}  \Vert \vec{x}_i \Vert ^2$ to ensure it is non-negative. Although one might consider explicitly enforcing $ \langle \vec{x}_i, \vec{x}_j \rangle := \left\Vert \vec{x}_{i}\right\Vert \left\Vert \vec{x}_{j}\right\Vert \theta_{ij} \geq 0$, this condition is naturally satisfied by requiring non-negative components for each vector, i.e. $ \vec{x}_{ik} \geq 0$ (see \autoref{proof:positive_components}). Roughly, $\vec{x}_i$ can be interpreted as the propensity of the node $ i$ to establish a connection with another nodes. More specifically, the magnitude of the embedding can be seen as the node's overall activity, whereas the angle $ \theta_{ij}$ between embeddings indicates similarity between the nodes $ i$ and $ j$. The parameter $ w_{i}$, instead, indicates the propensity of node $ i$ of having a self-interaction. While a precise interpretation of these vectors remains an open question, such discussion falls outside the scope of this work.

The node embeddings are learned by maximizing the following likelihood 
\begin{align}
    \label{eq:MSM_loglikelihood}
    \mathcal{L} \left(\mathbf{X} |\mathbf{A} \right) &:= 
    \sum_{i < j} a_{ij} \ln \left(p_{ij}\right) + \left(1-a_{ij}\right) \ln \left(1-p_{ij}\right) \\
    \nonumber
    &= \sum_{i < j} a_{ij} \ln \left( 1- e^{-  \langle \vec{x}_i, \vec{x}_j \rangle} \right) - \left(1-a_{ij}\right)  \langle \vec{x}_i, \vec{x}_j \rangle
\end{align}
where 
\vspace{1ex}
$
\textbf{X} := 
\begin{bmatrix}
    - \vec{x}_1^T - \\
    \vdots\\
    - \vec{x}_{N_{irred}}^T -
\end{bmatrix} \in \mathbb{R}^{N_{irred} \times D},
$ 
\vspace{1ex}
and $ N_{irred}$ is the number of irreducible \textit{structural inequivalent} nodes (see \autoref{SI:StructE_not_StatE}). We restricted the full likelihood to the off-diagonal likelihood as done for the LPCA. If needed, the MSM allows to fix the self-loops values, after the likelihood maximization, according to the following rule
\settowidth{\temp}{$- \frac{1}{2} \Vert \vec{x}_i \Vert ^2$} 
\begin{equation}
    \label{eq:MSM_w_limits}
    w_{i} = 
    \begin{cases}
        \FixedSize{\temp}{- \frac{1}{2} \Vert \vec{x}_i \Vert ^2} \quad \textnormal{ if }  p_{i i} = 1 \\
        \FixedSize{\temp}{\to \infty} \quad \textnormal{ if } p_{i i} = 0
    \end{cases}
\end{equation}
as explained in see \autoref{SI:sec:self_loops}.

\subsection{Embedding Dimension}
The selection of the \textit{optimal} embedding dimension is still an active Research problem \cite{2019_AIC_Cavanaugh,2023_BIC_Zhang, 2021_Principled_Selection_Gu}. Here, we adopted the ``Minimum Description Length'' principle, approximated by the Bayesian Information Criteria (BIC) \cite{2023_BIC_Zhang}. In the Supplemental Material \cite{SuppMat}, we report the BIC scores for the dimensions $ D = 1,2,8,16$ and, for completeness, also the AIC scores \cite{2019_AIC_Cavanaugh}. According to BIC, the optimal dimension depends on the resolution levels. For the Input-Output Network (ION), $ D^{opt}_{\ell \geq 1} = 1$ whereas $ D^{opt}_{\ell = 0} = 2$. In contrast, for the WTW -a less complex network-, the minimum BIC is obtained for $ D^{opt}_{\ell \geq 0} = 1$ (see Supplemental Material). The fact that for lower levels $ D^{opt}$ increases was somehow expected, since a larger number of nodes introduces more heterogeneity requiring a bigger embedding dimension to capture the observed network. However, since this is the first time the MSM is proposed, we report results for all embedding dimensions $D = 1, 2, 8, 16$ to provide a comprehensive evaluation.

\subsection{Summed Models}

In this section, we present the renormalization procedure used to obtain the scale-invariant versions of the MSM and the LPCA. As previously discussed, the MSM is inherently scale-invariant by design (see \autoref{sec:MSM}). On the other hand, LPCA requires additional modifications to enforce this property (cfr. \autoref{SI:Inconsistency_LPCA}).

\subsubsection{Summed MSM}
To start with, let us consider the MSM. Scale-invariance requires that the renormalized probability $ \mathbf{P}_{sum}^{(\ell)}$ equals the coarse-grained probability $ \mathbf{P}_{cog}^{(\ell)}$, and that the functional form of the model remains unchanged across resolution level (\cite{2023_MSNR_Garuccio}, \autoref{SI:derivation_MSM_probability}). Formally, the two conditions reads
\begin{align}
    \label{eq:self-cons_scale-invariance_conds}
    \begin{cases}
        \mathbf{P}_{sum}^{(\ell)} &\stackrel{!}{=} \mathbf{P}_{cog}^{(\ell)} \\[1ex]
        \mathbf{P}^{(\ell)} &\stackrel{!}{=} \mathbf{P}^{(m)} \quad \forall \ell \geq 0, m \geq 0
    \end{cases}
\end{align}
The fitted vectors at $ \ell = 0$ are denoted as $ \left\{\vec{x}_{i_0}, w_{i_0}\right\}_{i \in [1, N_{0}]}$, while $\left\{\vec{x}_{I}, w_{I}\right\}_{I \in [1, N_{\ell}]}$ are the embeddings at the coarser level $ \ell \geq 0$, where we defined the block-nodes as $ I := \Omega_{0 \to \ell}(i_0)$. Assuming that the block parameters are obtained by summing those at the micro level, namely
\begin{equation}
    \label{eq:MSM_sum_graining_rule}
    \vec{x}_{I} = \sum_{i_0 \in I} \vec{x}_{i_0} \textnormal{ and } w_{I} = \sum_{i_0 \in I} w_{i_0},
\end{equation}
it holds that the scale-invariance conditions (\autoref{eq:self-cons_scale-invariance_conds}) are satisfied by the MSM probability function \autoref{eq:MSM_pij} (see \autoref{SI:derivation_MSM_probability}).
Thus, the MSM at level $ \ell$ is recovered by substituting the summed parameters into \autoref{eq:MSM_pij}, yielding
\settowidth{\temp}{$1- e^{-\frac{1}{2}  \Vert \vec{x}_I \Vert ^2 - w_{I}} \;$} 
\begin{equation}
    \label{eq:MSM_pIJ}
    p_{IJ}:= p(\vec{x}_{I}, \vec{x}_{J}, w_{I}) = 
    \begin{cases}
        \FixedSize{\temp}{1- e^{-\langle \vec{x}_I, \vec{x}_J \rangle}} I \neq J\\
        \FixedSize{\temp}{1- e^{-\frac{1}{2}  \Vert \vec{x}_I \Vert ^2 - w_{I}}}  I = J.
    \end{cases}
\end{equation}

The summation rule also reduces the computational complexity compared to the models that require to be refitted at every scale, e.g. LPCA (see \autoref{SI:Algorithmic_Complexity}).

\subsubsection{Summed LPCA}
\label{sec:LPCA_Renormalization}
In contrast to the MSM, enforcing consistency across different scales in LPCA is less straightforward. 
The key issue is the absence of a \textit{renormalization} rule that allows to solve the system \autoref{eq:self-cons_scale-invariance_conds} using the logistic function (see \autoref{SI:Inconsistency_LPCA}). Additionally, if the vectors\footnote{The \textit{fitted} parameters, at a specific level $ \ell$, are identified by the notation $ \hat{x}_{i}$. Accordingly, $\hat{p}$ refers to the activation function $ \textbf{p}$ with fitted parameters $ \hat{x}_{i}$.} are re-fitted independently at each scale, one looses the dependence of higher level parameters with the lower ones. This effectively treats each level as if it was generated by a different underlying process. This contradicts the hierarchical structure of the network, as the coarse-grain (low-resolution) matrix is \textit{uniquely} derived from the microscopic one. These results indicate that only the $ 0$-vectors to construct a self-consistent, higher-scale LPCA. Moreover, that only the functional invariance (second equation in \autoref{eq:self-cons_scale-invariance_conds}) can be enforced in practice.

Concretely, we adopt the same scheme used in the MSM, as there is no specification of a renormalization rule to be exploited. Specifically, the coarse parameters are
\begin{equation}
    \label{eq:LPCA_summed_bc}
    \vec{b}_{I} := \sum_{i_0 \in I} \vec{b}_{i_0}, \quad \vec{c}_{I} := \sum_{i_0 \in I} \vec{c}_{i_0}
\end{equation}
where $ I := \mathbf{\Omega}_{0 \to \ell}(i_{0}) = (\mathbf{\Omega}_{\ell} \circ \, \cdots \, \circ \mathbf{\Omega}_0) (i_{0})$ denotes the block-nodes at level $ \ell$.
To enforce functional invariance, these summed parameters are substituted in the LPCA activation function, yielding
\begin{equation}
    \label{eq:LPCA_Psummed}
    \sigma(\vec{b}_I, \vec{b}_J, \vec{c}_I, \vec{c}_J) = 
    \frac{1}{1 + 
    e^{
        -\left( \langle \vec{b}_I, \vec{b}_J \rangle - \langle \vec{c}_I, \vec{c}_J \rangle  \right)
    }}
\end{equation}
This approach allows imposing functional invariance while preserving a coherent relationship between embeddings across different scales. Consequently, we can compare the expected values from the summed LPCA and MSM models against the aggregated empirical values.

\section{Data}

\begin{table*}[t]
    \parbox{.45\linewidth}{
    \centering
    \begin{tabular}{c|c|c|c|c}
        \toprule
        Level $ \ell$ & $ N_{\ell}$  & \textit{Reduced} $ N_{\ell}$ & $ L_{\ell}$ & $\rho_{\ell}$ \\
        \midrule
        0 & 972 & 972 & 136887 & 0.29\\
        1 & 647 & 647 & 93659  & 0.45\\
        2 & 303 & 303 & 30861  & 0.67\\
        3 & 87 & 67 & 3364   & 0.89\\
        \bottomrule
        \multicolumn{4}{c}{} \\
        \multicolumn{4}{c}{} \\
    \end{tabular}
    \caption{Table for the ION datasets reporting, for each level $ \ell$, the number of nodes ($ N_{\ell}$), the \textit{reduced} number of nodes, the total number of links and the density.\label{tab:ION_n_nodes_edges}}
    }
    \hfill
    \parbox{.45\linewidth}{
    \centering
    \begin{tabular}{c|c|c|c|c}
        \toprule
        Level $ \ell$ & $ N_{\ell}$  & \textit{Reduced} $ N_{\ell}$ & $ L_{\ell}$ & $\rho_{\ell}$ \\
        \midrule
        0 & 182 & 177 & 9993  & 0.61\\
        1 & 152 & 147 & 7305 & 0.64\\
        2 & 112 & 119 & 4822 & 0.67\\
        3 & 92 & 88  & 2728  & 0.65\\
        4 & 62 & 58  & 1332 & 0.70\\
        5 & 32 & 24 & 400 & 0.80\\
        \bottomrule
    \end{tabular}
    \caption{Table for the WTW datasets reporting, for each level $ \ell$, the number of nodes ($ N_{\ell}$), the \textit{reduced} number of nodes, the total number of links and the density.\label{tab:WTW_n_nodes_edges}}
    }
\end{table*}

\subsection{ING Input-Output Network}
\label{sec:DatasetDescription}
ING Bank N.V. regularly reports the economic transactions of all ING clients for different years. 
We focused on firm-to-firm payments in the year 2022 between ING bank accounts, filtering out the transactions involving \textit{individual} or non-Dutch clients, flows to or from non-ING account and the internal transfers within the same firm, i.e. self-payments. Since ING is the largest bank in The Netherlands \cite{2024_SP_GLobal_Banks_Ranking}, the dataset covers a major portion of the national market. More precisely, we selected the year 2022 to simplify the numerical calculations and to avoid distortions in the data from the aftermath of the COVID-19 pandemic. However, the procedure may be easily extended for other time intervals, such as multi-year spans or quarterly snapshots. 

At the firm-to-firm (f2f) resolution, the dataset consists of $ N_{f2f} \approx 3.4 \cdot 10^{5} \text{ nodes}, L_{f2f} \approx 4 \cdot 10^{6} \text{ links}$, yielding a density $ \rho_{f2f} \approx 3.5 \cdot  10^{-5}$. This classifies the network as \textit{large and sparse}. To analyze the data at a higher level, we aggregated the firms according to their NAICS (North American Industry Classification System) codes, and set an edge among two sectors if there was \textit{at least one} payment among firms belonging to one of them (see \autoref{eq:coarse-graining_ell}). Then, to focus on the production structure of this graph, we excluded sectors such as ``Public Administration'' (92), ``Finance and Insurance'' (52), ``Management of Companies and Enterprises''/``Holdings'' (55) as they are not directly related with a product/service. In particular, ``Public Administration'' fluxes include taxes and fees;  ``Finance and Insurance'', loans, that are not part of the supply chain of a product; and ``Management of Companies and Enterprises''/``Holdings'' include business entities that own shares in multiple firm. Lastly, we mapped, for simplicity, the 6-digits NAICS codes to integers, i.e. $ 111110 \to 1, 111120 \to 2, \cdots$.

Our work is the first application of the MSM to the \textit{multiscale structure} derived from the ION. Therefore, we focused on the \textit{economic relationships} among the sectors, discarding both the directionality and the volume of monetary flow. In this context, each edge is undirected (reciprocated) and binary. For example\footnote{The $i,j$ refers to the observed $ 0$-nodes, namely $ i := i_{0}, j := j_{0}$}, if $ w_{ij}$ refers to the total amount of money sent from $ i \textnormal{ to } j$ at level $ 0$, we \textit{reciprocated} the transaction by setting $ w_{ij} \to w^{'}_{ij} := \frac{w_{ij} + w_{ji}}{2}$ \cite{2010_WTW_Fagiolo}. Furthermore, an edge was regarded as active ($ a_{ij} = a_{ji} = 1$) if $ w^{'}_{ij} > 0$ or absent otherwise. This results in a symmetrical and binary edge, whenever there was \textit{at least} one flow between two sectors. At this stage, the ION consists of $ N_{0} = 972$ sectors and $ L_{0} = 1.4 \cdot  10^{5}$ links, yielding a network density of $ \rho_{0} \approx 0.29 $, roughly 4 order of magnitudes higher than $\rho_{f2f}$.


\subsection{Coarse-Graining The ION}
In the previous section, we constructed the \textit{binary undirected} adjacency matrix $ \mathbf{A}^{(\ell = 0)}$ representing the interactions among the 6-digits sectors (\textit{0-nodes}). Here, we describe the coarse-graining procedure to obtain the \textit{multi-scale} unfolding of $ \mathbf{A}^{(0)}$.

At first, we grouped together the 0-nodes that share the same first $ 6 - \ell$ digits, with $\ell \in [0,4]$. For instance, the sectors $ 111191, 111199$ are merged in the same community $ 11119$ starting from $ \ell = 1$.

Secondly, we set an edge among two $ \ell$-nodes if there existed \textit{at least one} connection between any of their constituent 0-nodes (see \autoref{eq:coarse-graining_0ell}). Our procedure uses a \textit{one-step} approach to easily generate every level directly from the $ \mathbf{A}^{(0)}$, without passing through the intermediate scales. The latter procedure is also allowed by the MSM. 

Note that from $ \ell = 4$ onward, the coarse-grained graphs become \textit{fully-connected}, limiting a meaningful \textit{statistical} modelling up to $ \ell = 3$. For technical details on the evolution of the number of nodes $ N_{\ell},$ links $ L_{\ell}$, and density $ \rho_{\ell} := \frac{2L_{\ell}}{N_{\ell}\left(N_{\ell} - 1\right)}$, refer to \autoref{tab:ION_n_nodes_edges}.

\subsection{World Trade Web}
As a second application, we analyzed the World Trade Web (WTW) tracked by the Gleditsch dataset \cite{2002_Gleditsch}, which reports bilateral trade flows (imports and exports) for every country. We focused on the year 2000 (the most recent one), and excluded the missing states in the BACI-CEPII GeoDist \cite{2011_GeoDist_Mayer}, whose distances are required for the coarse-graining procedure. After this filtering step, the graph has $N_{0} = 185$ 0-nodes. Although we illustrate the method for the year 2000, the methodology applies to any other period of time.

For every pair $( i, j)$, the dataset reports the \textit{exported} volume $ w_{ij}$ and the \textit{imported} $ \omega_{ji}$ one (all in USA dollars). Since the reported export from $ i$ to $ j$ ($ w_{ij}$) generally differ from the import from $ j$ to $ i$ ($ \omega_{ij}$), we redefined $ w_{ij} \leftarrow \frac{w_{ij} + \omega_{ji}}{2}$ \cite{2010_Evo_of_WTW_Weighted_Networks_Fagiolo} as the averaged amount of trade from $ i \textnormal{ to } j$. 

Successively, as for the ION, we \textit{symmetrized} the transaction matrix by mapping every weight $ w_{ij} $ to the average flow between the two directions, namely $ w_{ij} \to w^{'}_{ij} := \frac{1}{2} \left( w_{ij} + w_{ji} \right)$, i.e. the average flow between the two directions. By renaming $ w^{'}_{ij}$ as $ w_{ij}$, the resulting weights are symmetric, i.e. $ w_{ij} = w_{ji}$. Given the high link \textit{reciprocity} of the transaction reported in the Gleditsch dataset \cite{2023_RecEcon_DiVece}, this undirected representation is a sound approximation. Then, we \textit{binarized} the import-export matrix to construct the adjacency matrix $ \mathbf{A} := \Theta(\textbf{W})$ where the $ \Theta(\cdot)$ is the Heaviside function.

\subsection{Coarse-Graining of WTW}
To coarse-grain the level $ 0$, we used the inter-country geographical distances \cite{2011_GeoDist_Mayer} to iteratively merge closer nodes via \textit{single-linkage agglomerative clustering} as in \cite{2023_MSNR_Garuccio}. This procedure returns a dendrogram whose \textit{leaves} are the 0-nodes, whose \textit{branching points} are the block-countries, and whose branch \textit{heights} equal the single-linkage distances between sub-clusters. By cutting this dendrogram at 18 successive levels $\ell\in[0,17]$ we obtain partitions ${\Omega_\ell}{\ell\ge0}$, such that the number of block-countries are approximately $ N_{\ell} := N_0 - 10 \cdot \ell$. For $ \ell \geq 7$ the coarse-grained graph become fully connected. Therefore, a meaningful statistical modelling is possible up to $ \ell = 6$. See \autoref{tab:WTW_n_nodes_edges} for the evolution of the number of nodes, links, and network density.

\section{Results and Discussions}
\label{sec:Results_and_Discussions}
Here, we present the results of the LPCA and MSM models applied to the ION datasets. The WTW results are reported in the Appendix.

\subsection{Scale-Invariance Evidence and Multi-Scale Clustering Coefficient}
\label{sec:ScaleInv_NetMeas}

\begin{figure}[htbp]
    \centering
    \subfloat{
        \includegraphics[width=.6\linewidth]{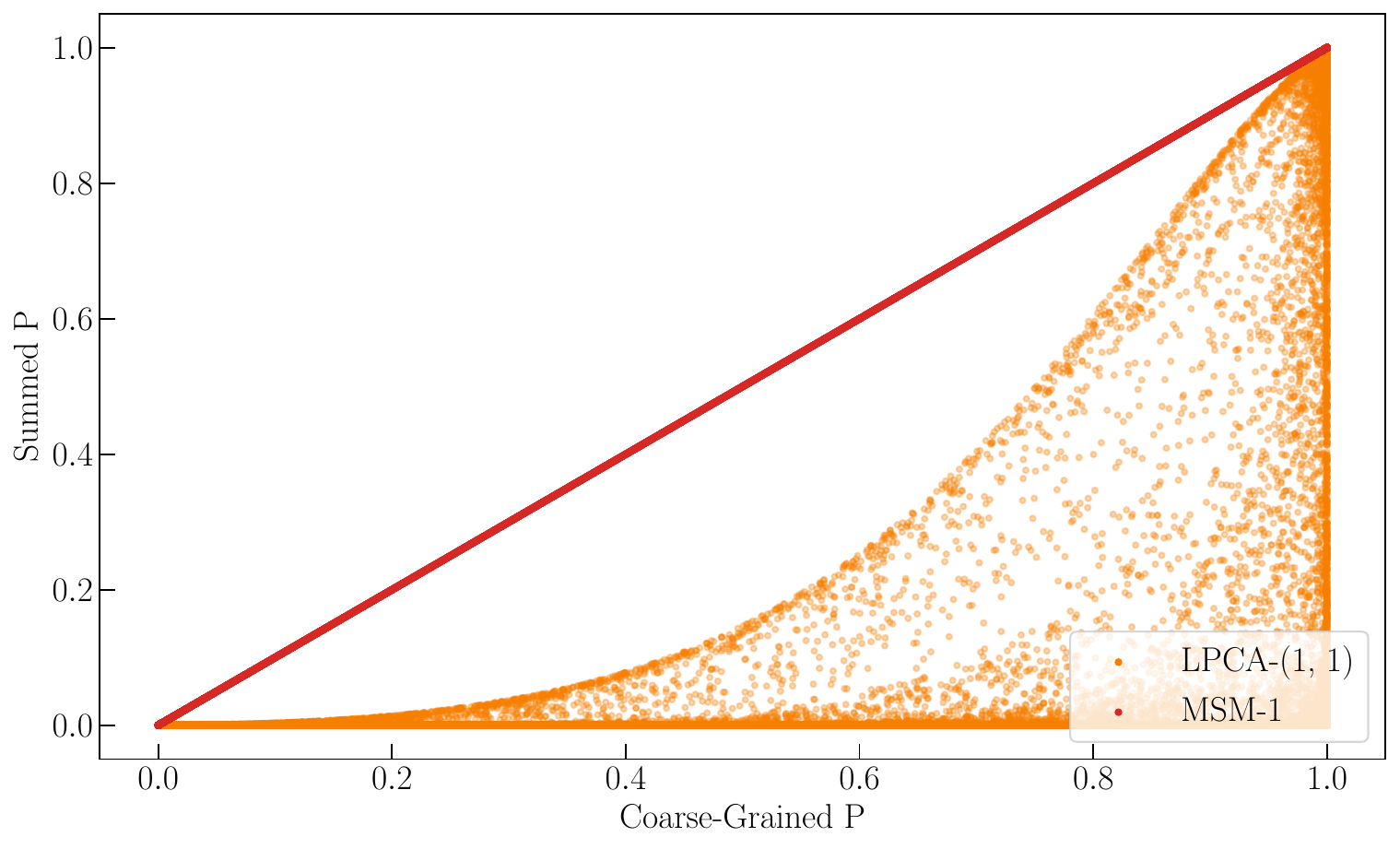}
    }
    \caption{This figure illustrates the numerical evaluation of the two sides of \autoref{eq:MSM_micro_psumVSpcg} at level $\ell = 2$ for the $LPCA$ and $MSM$ models. More precisely, the left-hand side (LHS) is represented on the y-axis while the right-hand side (RHS) on the x-axis.}
    \label{fig:ING_summedP_vs_cgP}
\end{figure}

\newlength{\temph}
\setlength{\temph}{\textheight}
\begin{figure*}[t]
    \centering
    \subfloat[Level 0, 2 - Binary Clustering Coefficient\label{fig:ING_CC_reconstruction}]{
        \includegraphics[width=\linewidth]{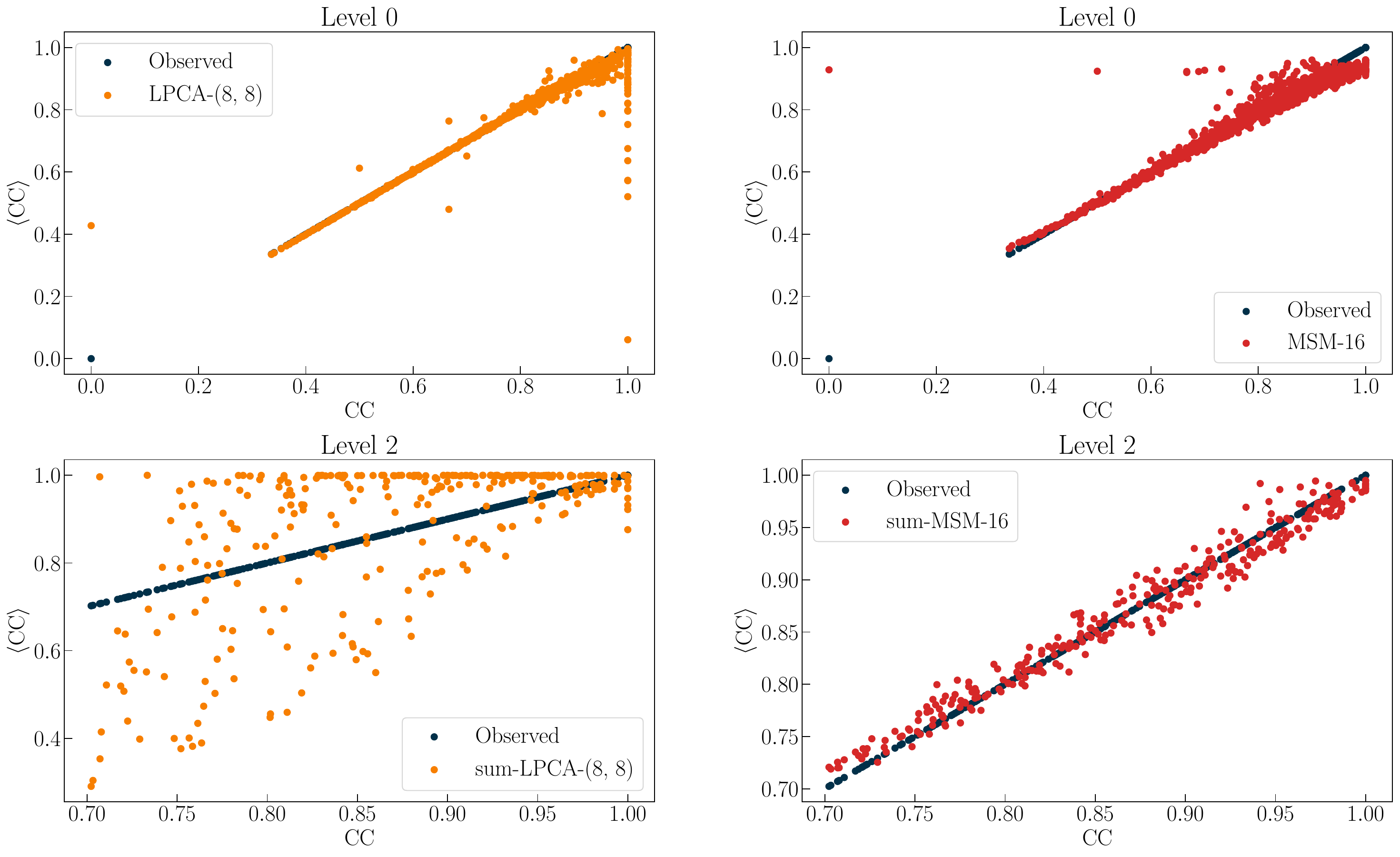}
    }
    \caption{
    Cross comparison of LPCA-(8,8) and MSM-16 in predicting the clustering coefficient (CC) \autoref{SI:sec:NetworkMeasurements}.
    The upper panel reports the expected clustering coefficient at level 0, while the lower panel depicts its corresponding values at level 2. The first column refers to the LPCA-(8,8), whereas the second to the MSM-16.
    } 
    \label{fig:ING_CrossComparison_CC}
\end{figure*}

\begin{figure*}[p]
    \centering
    \subfloat[Level 2 - LPCA-(8,8) - Summed, Fitted, Observed Network Measurements\label{fig:ING_lev2_NetMeas_LPCA_88}]{
        \includegraphics[width=.9\linewidth, trim = {0 0 0 1cm}, clip]{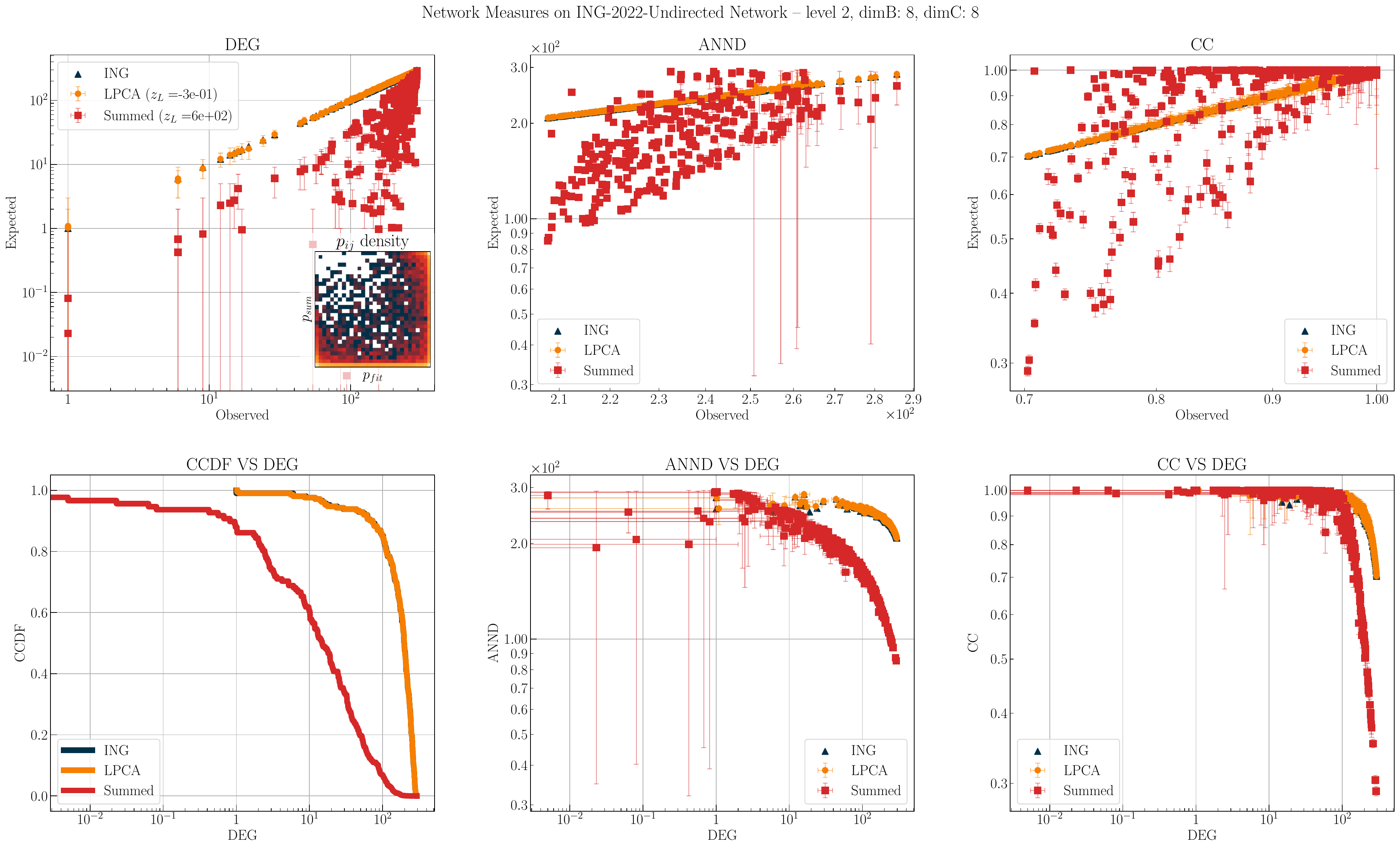}
    }
    \vspace{3ex}
    \subfloat[Level 2 - MSM-16 - Summed, Fitted, Observed Network Measurements\label{fig:ING_lev2_NetMeas_maxlMSM_16}]{
        \includegraphics[width=.9\linewidth, trim = {0 0 0 1cm}, clip]{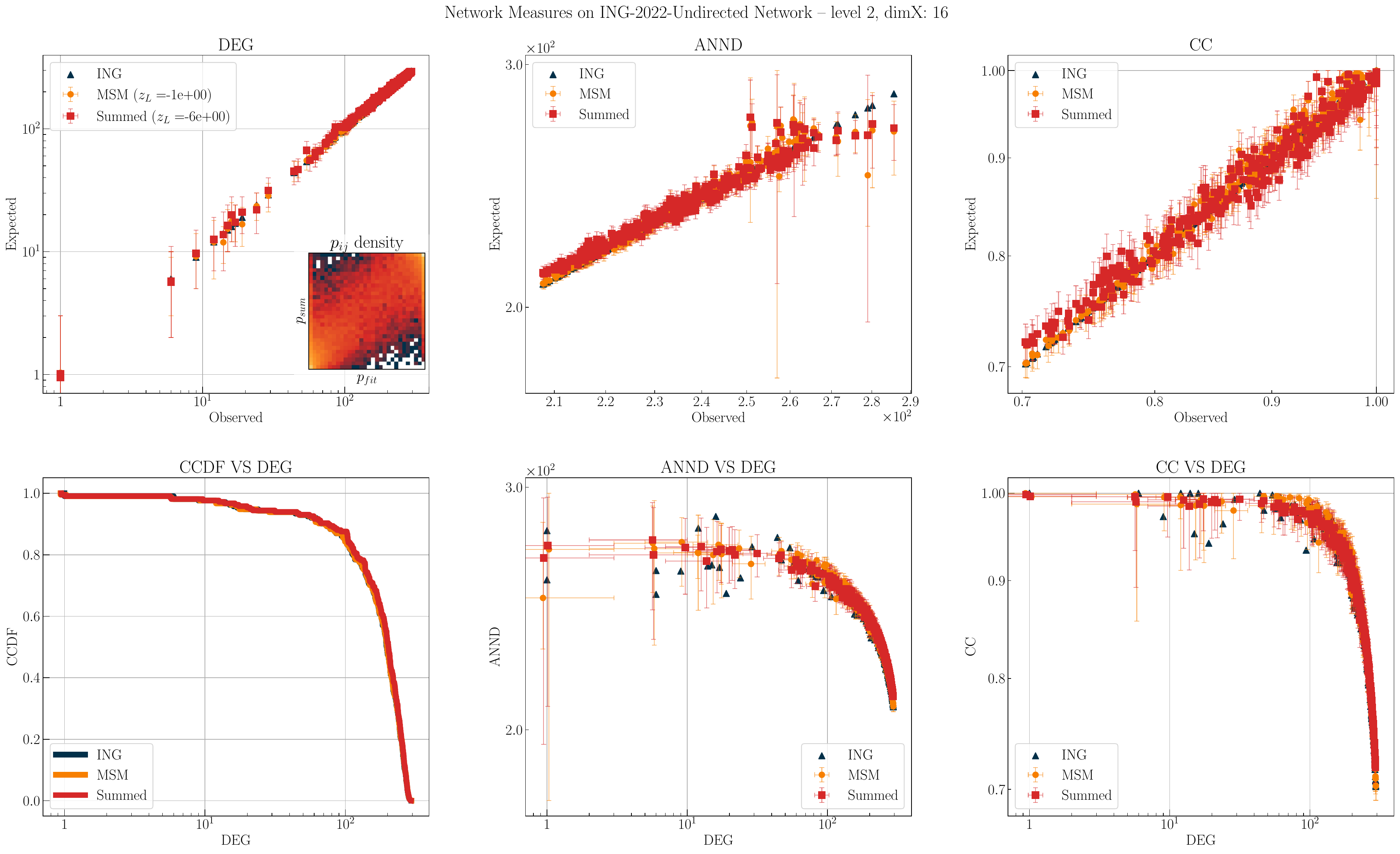}
    }
    \caption{The plots display the predicted network measurements for the ION dataset at level 2 according to LPCA-(8,8) (upper plots) and the MSM-16 (lower plot). 
    In each of the upper panels, the x-axis represents the observed measurements, whereas the y-axis shows the expected ones. From left to right, the plots illustrates the degree (DEG), the average-nearest-neighbor degree (ANND), and the clustering coefficient (CC). The scatter plot comparing $ \mathbf{P}_{sum}^{(2)} $ with $ \hat{\mathbf{P}}^{(2)}$ is reported in the inset. 
    The lower panels show the behavior of the network measurements as the degrees increase. The observed values are colored in blue, while the one calculated using the fitted $ \hat{\mathbf{P}}^{(2)}$ model in red or the summed $ \mathbf{P}_{sum}^{(2)}$ model in orange. Additionally, the z-score of the predicted number of links is indicated in the legend of the upper-left plot.
    }
    \label{fig:ING_NetMeas_DispInt}
\end{figure*}

{
\raggedright \autoref{fig:ING_summedP_vs_cgP} illustrates the relationship among the \textit{summed} probability and its \textit{coarse-grained} counterpart (see \autoref{eq:MSM_micro_psumVSpcg}, \autoref{SI:sec:ComparisonAmongProbabilities}).
}
We selected the lowest embedding dimension $ D_B = D_C = 1$ and $ D = 1$ to highlight the fundamental differences among the models. As shown, only the MSM recovers the identity line for all the pairs, demonstrating its scale-invariant property. In contrast, LPCA consistently underestimates the coarse-grained probability, since the microscopic probabilities, maximizing the likelihood at $\ell = 0$, get lower values than required for scale-invariance.

Subsequently, we increased the embedding dimensions ($ D_B = D_C = 8$ and $ D = 16$) and cross-compared their capability of predicting the multi-level clustering coefficient (see \autoref{SI:sec:NetworkMeasurements}) in \autoref{fig:ING_CC_reconstruction}. Visually, LPCA-(8,8) appears to (slightly) outperform the MSM-16 at level 0, while the MSM excels for $\ell = 2$. Quantitatively, at $ \ell = 0$, LPCA achieves better metrics ($AUC-ROC_{LPCA} \approx 0.92, AUC-PR_{LPCA} \approx 0.89$) than the MSM ($AUC-ROC_{MSM} \approx 0.91, AUC-PR_{MSM} \approx 0.88$). To merge with \cite{2020_LPCA_Chanpuriya}, we exploited the relative Froebenius error between the \textit{fitted} probability $ \hat{\mathbf{P}}^{(0)}$ and the adjacency matrix, defined as \cite{2020_LPCA_Chanpuriya}
\begin{equation}
    \epsilon_2 := \frac{\left\Vert \hat{\mathbf{P}}^{(2)} - \mathbf{A}^{(2)} \right\Vert _{2}}{L_2}.
\end{equation}
where $ \left\Vert \right\Vert _2$ refers to the 2-Euclidean norm. 
Also in this metric, LPCA exploits a better approximation than the MSM ($\epsilon_{LPCA} \approx 48 \% < \epsilon_{MSM} \approx 51 \%$).

Comparable results were also obtained for the other levels and measures, such that the DEG and ANND - for ION at $ \ell = 2$ (see \autoref{fig:ING_NetMeas_DispInt}), but that are not shown here for brevity. Refer to the Supplementary Material for the WTW results.

\subsection{Expected Values of the Network Measurements}

In \autoref{fig:ING_NetMeas_DispInt}, we illustrate key network properties at level $ 2$ (see \autoref{SI:sec:NetworkMeasurements}) computed for the empirical network $ \mathbf{A}^{(2)}$, as well as for its expected values under the summed model $ \mathbf{P}_{sum}^{(2)}$ and the fitted model $ \hat{\mathbf{P}}^{(2)}$ (see \autoref{sec:Methodology} and \ref{sec:LPCA_Renormalization}). For the LPCA, we selected $ D_B = 8, D_C = 8$, and for the MSM, $ D = 16$.

The empirical quantities were computed as in \autoref{SI:sec:NetworkMeasurements}. Focusing on the fitted model $ \hat{\mathbf{P}}^{(\ell = 2)}$, we sampled $N_{\mathcal{A}} = 1000$ realizations to
\\[1ex] compute the ensemble averages $ \left\{\overline{k_{i_2}}, \overline{annd_{i_2}}, \overline{c_{i_2}}\right\}_{i_2 \in [1,N_2]}$ as defined in \autoref{eq:ensemble_average_analytical}. Furthermore, we calculated the dispersion interval for each measurement setting $ c = 95\%$, and represented as error bars in the plots. The same procedure was applied the summed model $ \mathbf{P}_{sum}^{(2)}$ for comparison.

In the upper panel, we compare the observed measurements (x-axis) with their expected values (y-axis) for the summed model (orange points) and the fitted model (red points). The identity line represents the perfect agreement between observed and predicted measures. The inset displays a scatter plot between $ \mathbf{P}_{sum}^{(2)} $ and $ \hat{\mathbf{P}}^{(2)}$, as discussed in \autoref{sec:ScaleInv_NetMeas}. The $ z_{L}$ value is defined as the z-score of links:
\begin{equation}
    \label{eq:z_score}
    z_{L_{\ell}} := \frac{L_{\ell} - \langle L_{\ell} \rangle}{\sigma\left[L_{\ell}\right]}
\end{equation}
where $ \sigma\left[L_{\ell}\right] := \left[\sum_{i < j} p_{ij} \left(1-p_{ij}\right)\right]^{1/2}$ \cite{2011_AnalMax_Squartini}.

As shown in \autoref{fig:ING_summedP_vs_cgP}, the LPCA does not accurately reproduce the empirical measurements. In contrast, the MSM provides a better approximation, with most empirical values falling within the dispersion intervals. Moreover, the inset in the upper-left plot reveals that $ \mathbf{P}_{sum}^{(2)} $ closely approximates $ \hat{\mathbf{P}}^{(2)} $ only for the MSM. This result, which was not theoretically enforced as for \autoref{eq:MSM_micro_psumVSpcg}, accounts for a consistent agreement of the MSM with observed values across scales. 

The lower panel illustrates how network measurements evolve as a function of the degree. From left to right, we show the complementary cumulative distribution function (CCDF) of the degrees, the average nearest neighbors degree (ANND) and the clustering coefficient (CC) \autoref{SI:sec:NetworkMeasurements}. The decreasing trend of the empirical CCDF suggests that the observed network contains more low-degree nodes than the hubs, and it is not \textit{scale-free}, as the CCDF doesn't follow a straight line in log-log scale. Additionally, by looking at the ANND, the ION is \textit{disassortative}, since high-degree nodes tend to trade with loosely-interacting partners \cite{2004_Fitness_WTW_Garlaschelli}.

These plots highlight that the MSM-16 not only captures the CC (see \autoref{fig:ING_CC_reconstruction}), but also the \textit{lower-hops} structural properties. As expected, the LPCA-(8,8) provides a good fit only at the specific scale at which it was fitted.

\subsection{Reconstruction Accuracy and ROC-PR Curves}

\begin{figure*}[tbp]
    \centering
    \subfloat[Level 2 - LPCA-(8,8) - ROC and PR curves for Summed and Fitted models\label{fig:ING_lev2_ROCs_LPCA_88}]{
        \includegraphics[width=\linewidth]{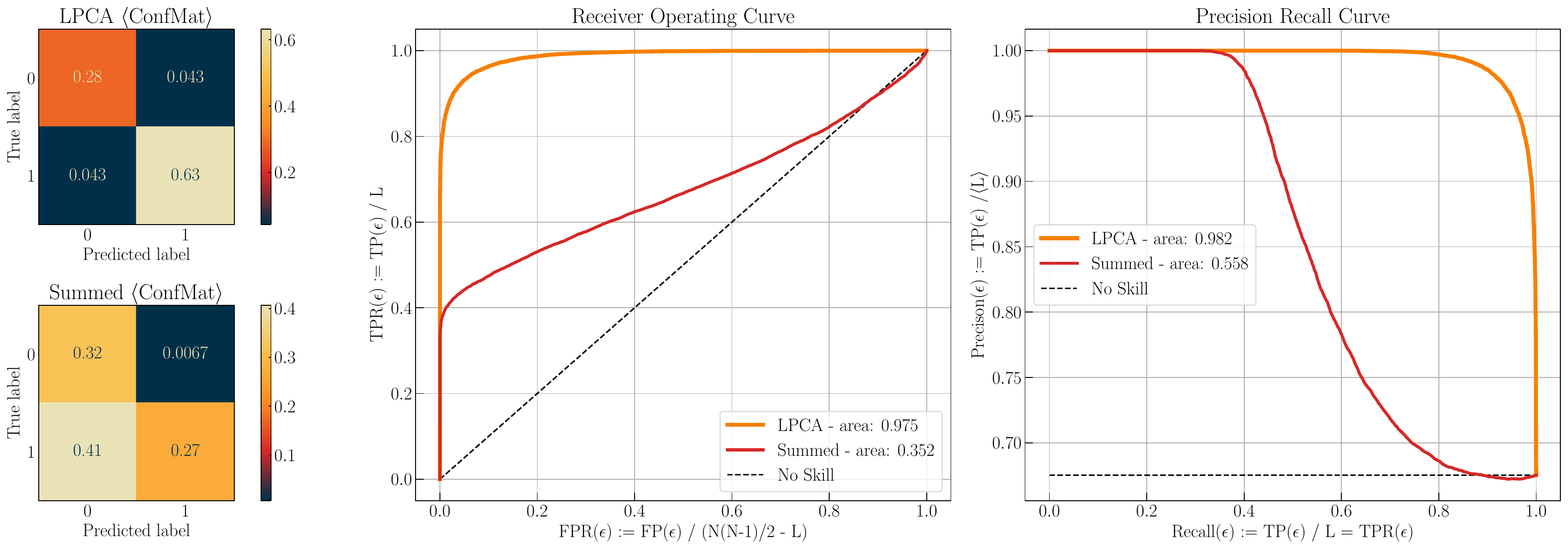}
    }
    \vspace{3ex}
    \subfloat[Level 2 - MSM-16 - ROC and PR curves for Summed and Fitted models\label{fig:ING_lev2_ROCs_maxlMSM_16}]{
        \includegraphics[width=\linewidth]{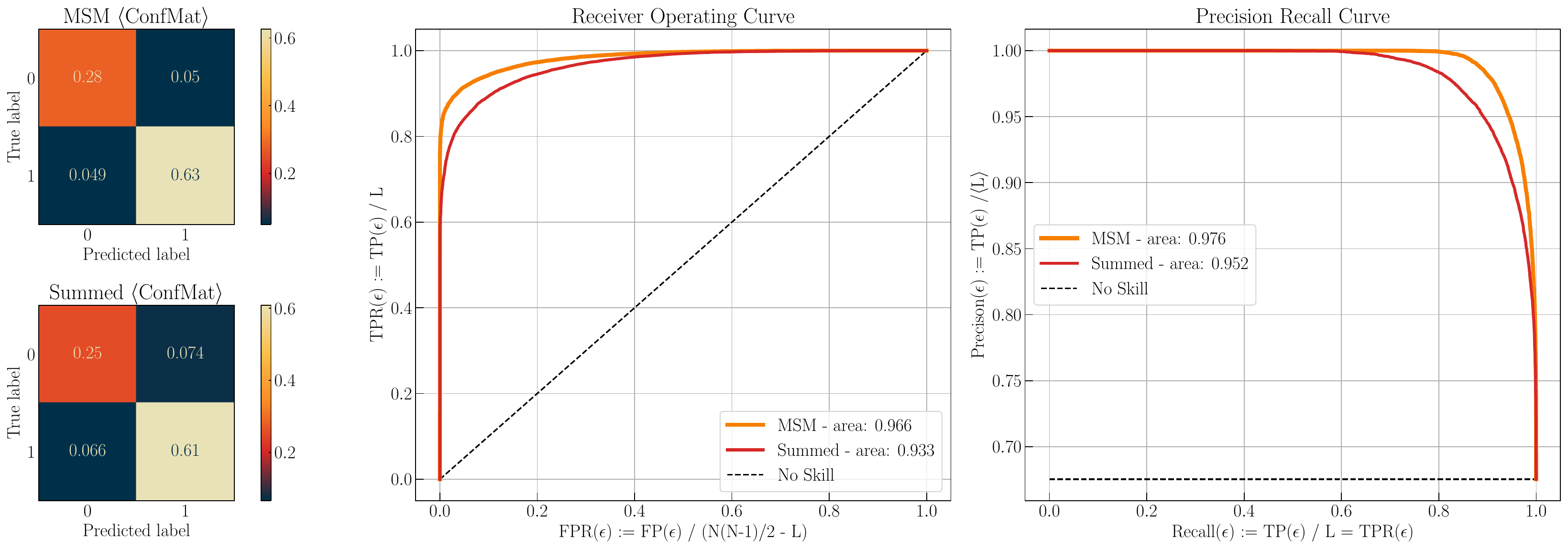}
    }
    \caption{
    The graphs show the Confusion matrices, ROC and PR curves for the ION dataset at level 2 obtained by means of the LPCA-(8,8) (upper plots) and the MSM-16 (lower plot).
    On the left most side, one may display the two confusion matrices for the fitted $ \hat{\mathbf{P}}^{(2)}$ (upper) and the summed $ \mathbf{P}_{sum}^{(2)} $. The middle plot reports the Receiver-Operator Curve, while the right-most plot depicts the Precision Recall curve. As in \autoref{fig:ING_NetMeas_DispInt}, the two curves are associated with the fitted (red) model and the summed (orange) one.
    } 
    \label{fig:ING_ROCs}
\end{figure*}

\begin{figure*}[t]
    \centering
    \subfloat[All level Reconstruction Accuracy\label{fig:ING_rec_acc}]{%
    \includegraphics[height=.28\linewidth]{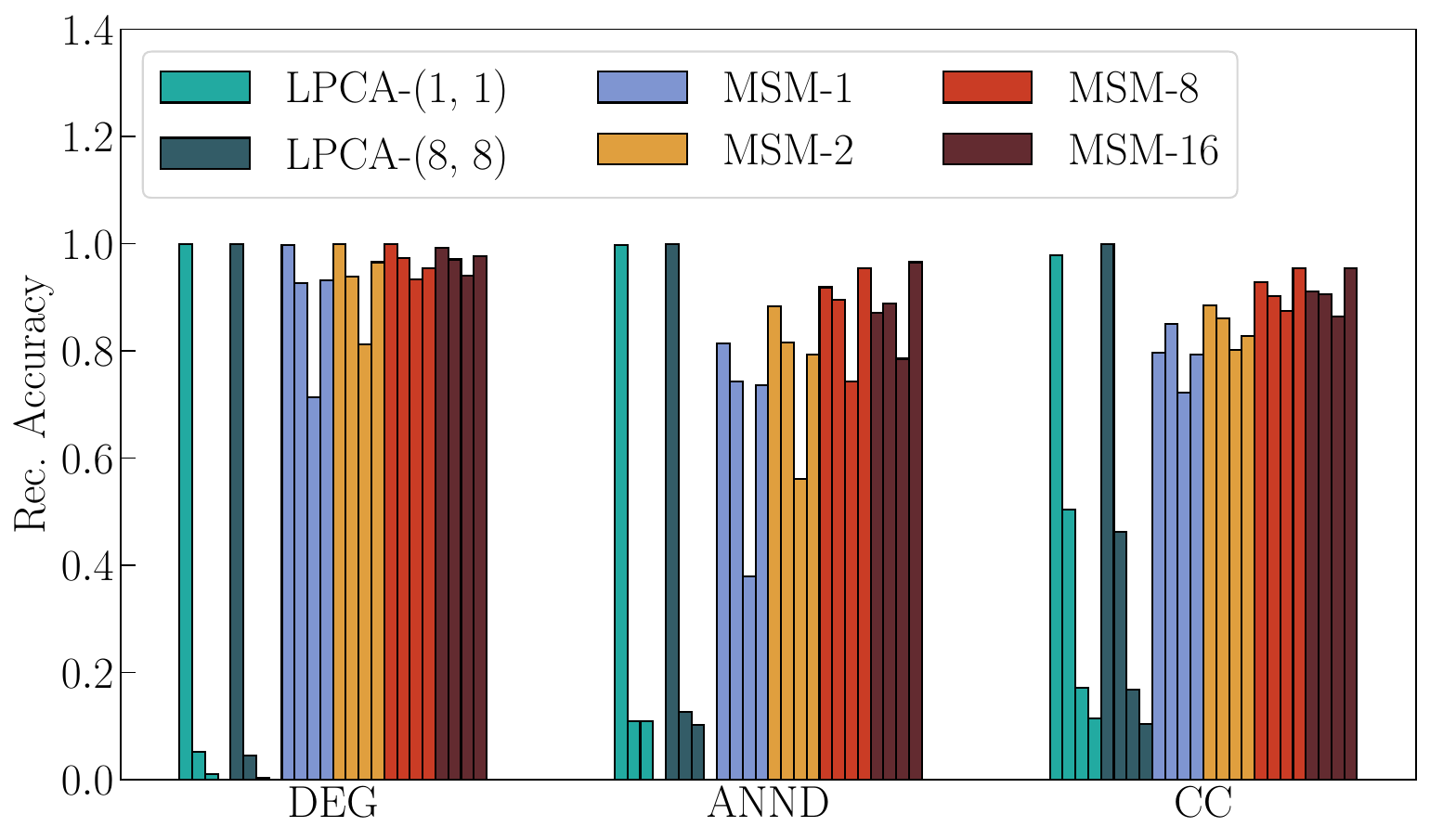}
    }
    \hfill
    \centering
    \subfloat[AUC-ROC and AUC-PR curves through levels \label{fig:ING_auc_roc_prc}]{
        \includegraphics[height=.28\linewidth]{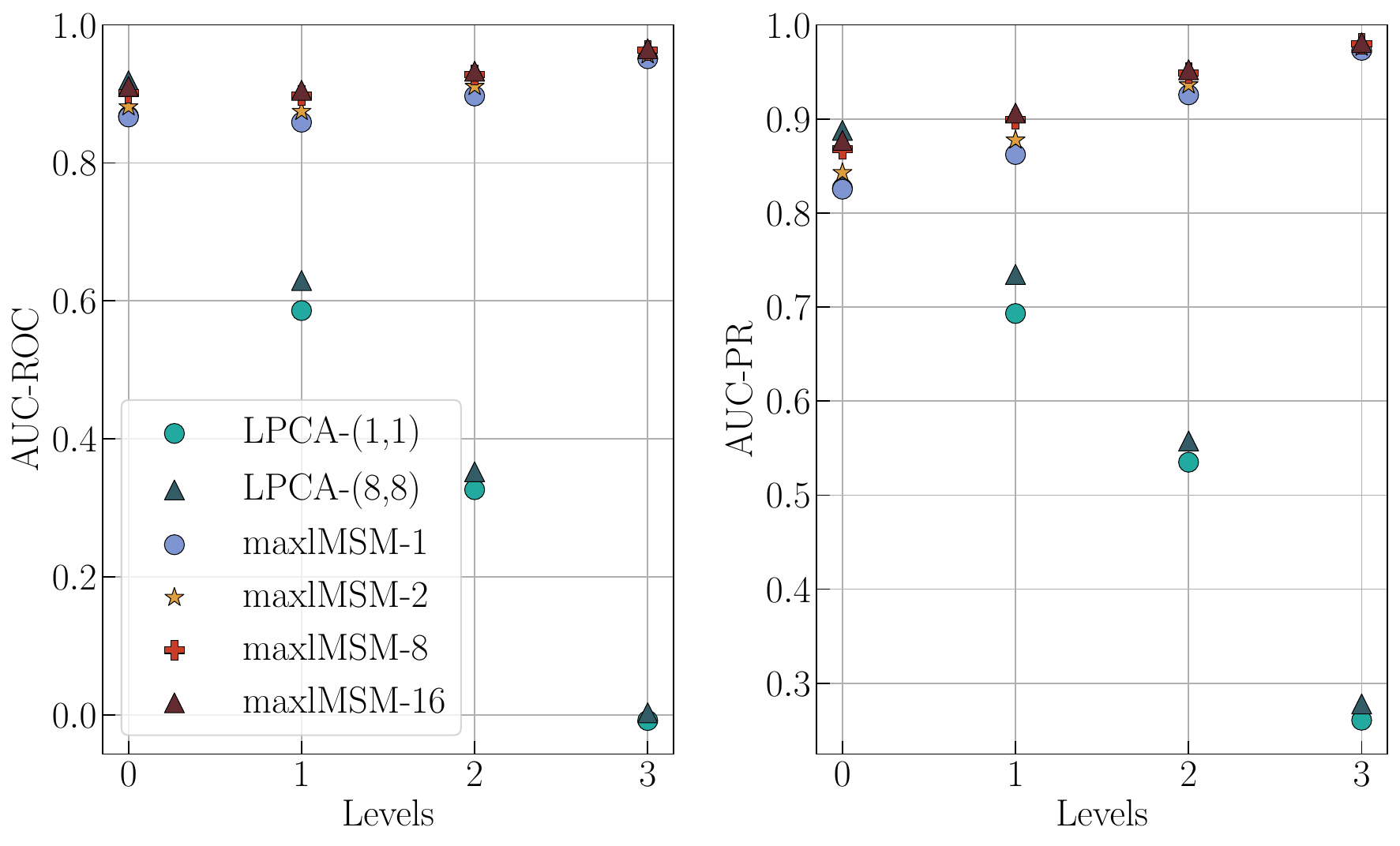}
        }
        \caption{\textit{Left:} Reconstruction Accuracy (y-axis) by model, level and network statistics.
        \textit{Right}: Area Under the ROC and PR curves (y-axis) for the summed models as the scales increase (x-axis) or, equivalently, the numbers of nodes diminish.
        }
        \label{fig:ING_auc_rec_acc}
\end{figure*}
In \autoref{fig:ING_rec_acc}, we show the reconstruction accuracy (see \autoref{eq:RecAcc}) across the available levels ($ \ell = 0, \cdots, 3$) of the ION for the three key topological measures (DEG, ANND, CC). For this exercise, we report only the summed LPCA with dimensions $ D_B = D_C = 1$ and $ D_B = D_C = 8$, and the \textit{summed} MSM with $ D = 1, 2, 8, 16$.
As anticipated, all the models are fitted at $ \ell = 0$, which explains why most trends are peaked at this resolution. For example, MSM-1 has a higher CC at level $ 1$ than at $ 0$. Interestingly, LPCA struggles to produce dispersion intervals that include the observed quantities. Contrarily, the MSM preserve consistency across all levels -except for $\ell = 2$ where the observed topology changed more than expected by the model. 
This fact is more noticeable for $ D = 1$, where the MSM overestimates all the measurements, since the summation of the 0-parameters lead to bigger values than requested parameters at level $ 2$. However, this is not inherently problematic, since the scale-invariance property enforced the \textit{internal} consistency (see \autoref{eq:MSM_micro_psumVSpcg}) over exact recovery of the fitted parameters at every level. Also, the choice of a \textit{pathological} partition would lead to worse results as discussed in \cite{2024_MultiScaleNetRec_Ialongo} and \autoref{sec:Dependence_on_an_arbitrary_partition}.

Increasing the number of parameters ($ D$) generally improves reconstruction accuracy. However, this comes at the price of ``overfitting'', as indicated by the higher BIC scores reported in Supplemental Material \cite{SuppMat}. A potential solution for this mismatch could be to introduce dyadic relationships $ d_{ij}$ between nodes as done in \cite{2023_MSNR_Garuccio}. Still, this remains beyond the scope of the present work.

Similar conclusions can be drawn from \autoref{fig:ING_auc_roc_prc}, which displays the Area-Under the Curve (AUC) of the Receiver Operating Characteristic (ROC) and the Precision-Recall (PR) curves, defined as in \autoref{SI:sec:NetworkMeasurements}. These plots further highlights a form of \textit{phase separation} due to different natures of the two model. While LPCA-(8,8) outperforms all the other candidates at $ \ell = 0$, its performances degrades at coarser levels, as does that of LPCA-(1,1). Conversely, the MSM displays increasing AUC scores at coarser levels. These increasing trends are mainly because, at coarser levels, the empirical number of links decreases $ L_\ell$  (see \autoref{tab:ION_n_nodes_edges}), while the \textit{summed} probabilities $ p_{IJ}$. Due to the probabilities' growth, also the thresholded Predicted Positives ($ \langle L \rangle (\epsilon)$) increase, but at a lower rate than the $ TP(\epsilon)$. As a result, both the $TPR(\epsilon)$ and $PPV(\epsilon)$ increase at coarser levels, leading to higher AUC scores.

Taken together, the results indicates, by selecting a specific functional form, one can prioritize either for single-scale performances with the LPCA or generalize at multiple scales with the MSM.

\subsubsection{Dependence on an arbitrary partition}
\label{sec:Dependence_on_an_arbitrary_partition}
The agreement of the summed model and the observed (coarser) graph depends critically on the chosen partition $ \Omega$. Formally, what has being tested so far is the approximation
\begin{align}
    \label{eq:aIJ_approx_pIJ}
    a_{IJ} &\approx p_{IJ} \stackrel{!}{=} 1 - \prod_{ i_0 \in I, j_0 \in J} \left(1 - p\left(x_{i_0}, x_{j_0}, w_{i_0}\right)\right)
\end{align}
which depends on the partition $ \Omega$ since the block-nodes are defined as $I := \Omega(i), J := \Omega(j)$ (see \autoref{eq:coarse-graining_01} and \autoref{eq:MSM_micro_psumVSpcg} for the notation). As a result, the quality of the above approximation is sensitive of the partition. In fact, one can intentionally construct a partition $\Omega^{\text{diff}}$ that maximizes the discrepancy between the model probability $p_{IJ}$ and the observed adjacency $a_{IJ}$. Formally, this corresponds to set $ \Omega^{diff}$ as
\begin{equation}
    \label{eq:maximal_deviation}
    \Omega^{diff} := \max_\Omega \sum_{I \leq J} \left( 1 - \prod_{ i_0 \in I, j_0 \in J} (1 - p_{i_0j_0}) - a_{IJ} \right)^{2}
\end{equation}
This formulation highlights the fact that poorly chosen partitions can lead to misleading conclusions regarding model fit and predictive accuracy. Hence, the meaningful agreement between the summed model and the observed adjacency matrix, we have appreciated so far, was not expected a priori.

\subsection{Expected Number of Triangles}

\begin{figure*}[t]
    \centering
    \subfloat[Level 0]{
        \includegraphics[width=.49\linewidth]{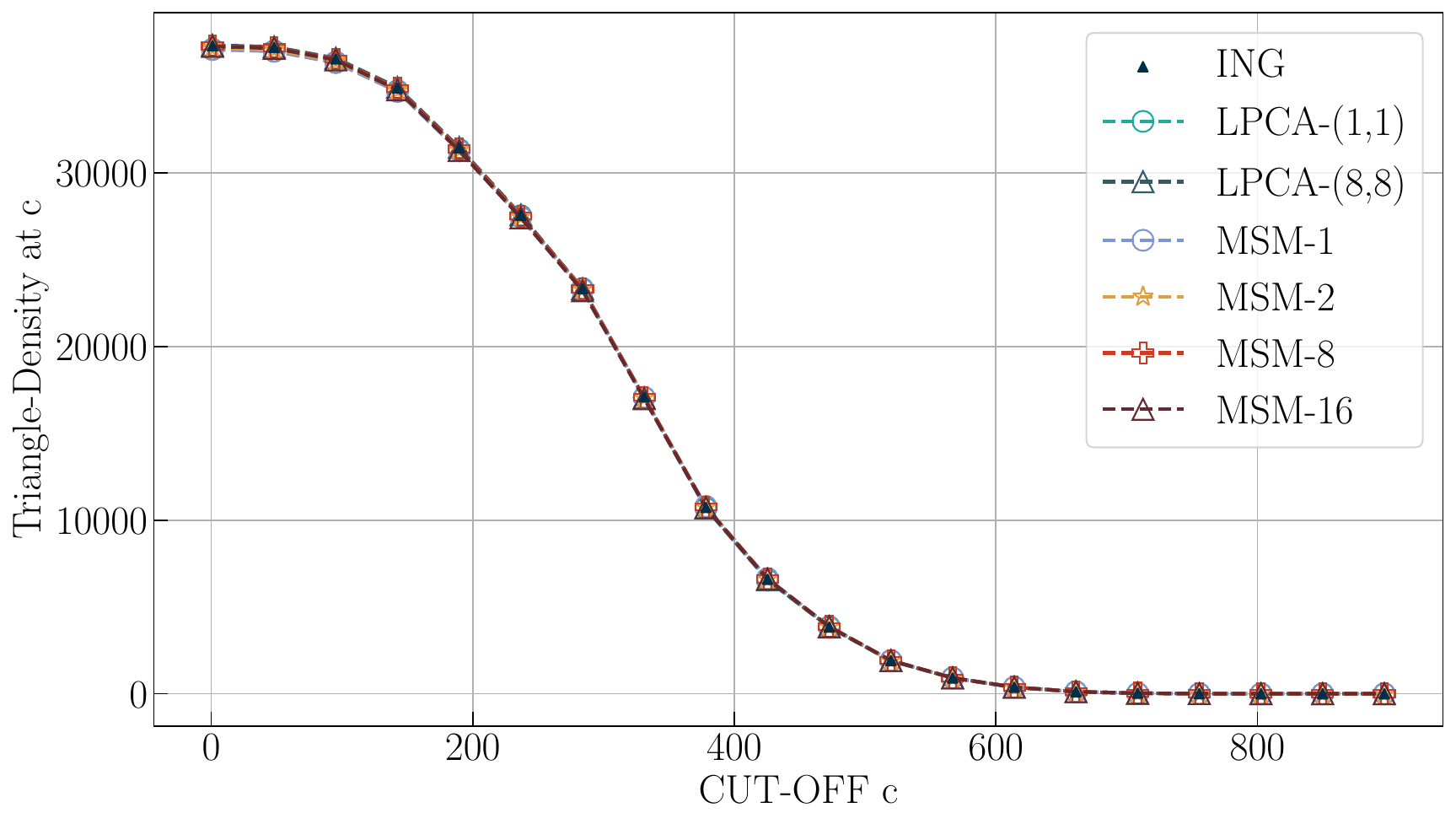}
        \label{fig:ING_RCTriangles_level0}
    }
    \subfloat[Level 2]{
        \includegraphics[width=.49\linewidth]{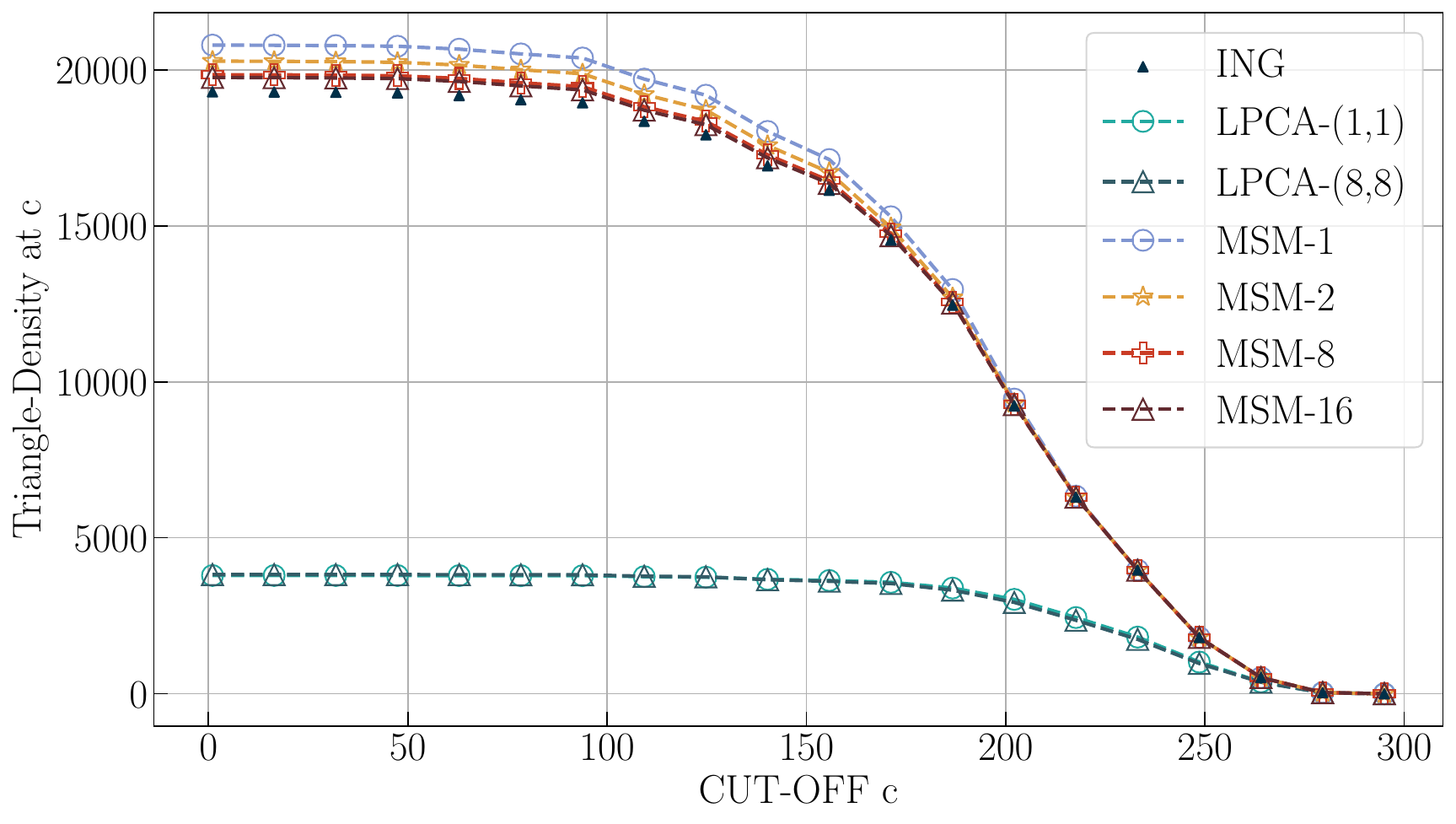}
        \label{fig:ING_RCTriangles_level2}
    }
    \caption{
    \textit{Left:} Triangle Density at level 0. The plot shows the evolution of the $ \tau^{(\ell)}(c)$ with respect to the degree $ c$ in blue solid dots. The other markers identify the expected $\langle \tau^{(\ell)}(c) \rangle$ by the LPCA-(1,1), LPCA-(8,8), MSM-1, MSM-2, MSM-8, MSM-16.
    \textit{Right:} Triangle Density at level 2. 
    }
    \label{fig:ING_RCTriangles_level02}
\end{figure*}

The authors of \cite{2020_LPCA_Chanpuriya} demonstrate that through LPCA, it is possible to reproduce the exact triangle density (TriDens) using a \textit{lower} embedding dimension than the number of nodes (cfr. \cite{2020_impossibility_of_low_rank_red_Seshandhri}). Contrarily, here we illustrate the \textit{expected} TriDens, defined in \autoref{eq:observed_TriDens} and \autoref{eq:expected_TriDens}.

In \autoref{fig:ING_RCTriangles_level02}, the filled blue markers depict the observed TriDens, while the other markers correspond to different models, as described in the legend. \autoref{fig:ING_RCTriangles_level0} reports the case for level $ \ell = 0$ where it is clear that even the smallest embedding dimension ($ D = 1$) provides a good approximation of TriDens. By design, as the cut-off degree $ c$ increases, the difference among the TriDens values diminish, eventually converging at $ c = N - 1$. This is because, in a \textit{disassortative} network, the subgraph composed of hubs is unlikely to contain triangles  (see \autoref{fig:ING_NetMeas_DispInt}). This finding does not contradict previous studies \cite{2020_impossibility_of_low_rank_red_Seshandhri, 2020_LPCA_Chanpuriya}, since our analysis focuses on the \textit{expected} TriDens rather than the \textit{exact} quantities. 

By coarse-graining the network \autoref{fig:ING_RCTriangles_level2} at $ \ell = 2$, the MSM models show a good agreement with the \textit{coarser} TriDens. In particular, as shown in \autoref{fig:ING_NetMeas_DispInt}, the estimates increase with increasing embedding dimension $ D$. In contrast, LPCA tend to underestimate the metrics at coarser level, since it is restricted to the fitted scale.

\section{Conclusions}
\label{Conclusions}
The strength of the graphs-based models lies in their flexibility to capture various type of interactions through suitable definitions of nodes and edges. By grouping microscopic nodes into a single block-node, the resulting network is a \textit{coarse-grained} version of the original graph. Repeating this procedure several times, it produces a multi-scale and hierarchical unfolding of the observed graph. We applied this procedure to both the ION and WTW datasets (see \autoref{sec:DatasetDescription}), demonstrating how a single \textit{generative process} can be represented at different resolutions.

Most of the existing \textit{node-embedding} methods \cite{2023_ZooGuide_Baptista} are designed to find the optimal node representation at a single scale, often underestimating the importance of extrapolating the model across different scales. To illustrate this limitation, we compared LPCA -a \textit{single-scale} method from the machine-learning domain- with our vector-based multi-scale model (MSM) (\autoref{sec:Methodology}). The crucial difference lies in the fact that the MSM is inherently scale-invariant, similarly to the (hidden) generative process it aims to approximate. In this context, a block embedding is computed as the sum of the embeddings of its constituent nodes (\autoref{eq:MSM_sum_graining_rule}), allowing for a principled interpretation of the sum of the embeddings. 

Noticeably, the embeddings are generally thought to live in a vector space that naturally comes with the sum and difference operations. However, despite their relevance, these two operations have received limited attention in the literature.
Here, instead, we were able to provide a solid explanation for the sum of vectors (see \autoref{eq:MSM_sum_graining_rule}).

As exposed in \autoref{sec:Results_and_Discussions}, LPCA may perform at its best only at the scale it is fitted to, and struggles to generalize at higher levels. In contrast, the multi-scale model consistently captures the network structure to coarser resolutions, requiring only the knowledge of how microscopic nodes are organized into communities. 

Taken together, these results highlight the value of adopting a multi-scale perspective, which offers a flexible and robust framework for understanding complex systems by allowing arbitrary adjustment of the scale at which interactions are examined.

\subsection{Future Perspectives}
Although this work focus on the undirected MSM, the framework can be readily adapted to capture other phenomena \cite{2024_RecMSM_Lalli}. Specifically, each node is represented by two embeddings ($ \vec{x}_{i}, \vec{y}_{i}$), which in this context are roughly seen as the capability of ``proposing'' (sender) or ``accepting'' (receiver) an economic relationship, respectively \cite{2020_LPCA_Chanpuriya}. As in the present essay, the block outer (inner) vectors are obtained by summing the $ \vec{x}$ ($\vec{y}$) embeddings of its constituent nodes, ensuring that the sum of vectors remains interpretable.

For the weighted part, further theoretical investigation is required to determine how weights can be integrated into the MSM framework.

\section{DATA AND CODE AVAILABILITY}
The dataset on sector transactions (ION) used in the paper is highly confidential and cannot be
disclosed. However, the world trade is a long-standing public dataset, one may find at \href{http://ksgleditsch.com/exptradegdp.html}{Gleditsch}. Lastly, the code is freely available as the \href{https://github.com/RMilock/multi-scale-node-embeddings/tree/main}{multi-scale-node-embeddings} python package.

\section{ACKNOWLEDGMENTS}
We thank ING Bank N.V. for their support and active collaboration. A special thanks to the whole DataScience team at ING Bank for their advice that helped shape this research. This work is supported by the project “Reconstruction, Resilience and Recovery of Socio-Economic Networks” RECON-NET EP FAIR 005 - PE0000013 “FAIR” - PNRR M4C2 Investment 1.3, financed by the European Union - NextGenerationEU.

%
%
\clearpage
\appendix
\setcounter{page}{1}
\onecolumngrid

{\center
\textbf{Appendices}\\
$\quad$\\
accompanying the paper\\
\emph{``Multi-Scale Node Embeddings for Graph Modeling and Generation''}\\
by R. Milocco, F. Jansen and D. Garlaschelli\\
$\quad$\\
$\quad$\\
}

\section{Non-Negative Logistic PCA}
\label{SI:LPCA}
The non-negative LPCA (LPCA) \cite{2020_LPCA_Chanpuriya} aims to classify every edge $ (i,j)$ as existing (0) or non-existing (1). By treating every entry $ a_{ij}$ of the adjacency matrix $ \textbf{A}$ as a Bernoulli random variable (\autoref{eq:a_ij_Bernoulli_rv}), this comes down to the factorization
\begin{equation}
    \textbf{A} \sim \sigma(\textbf{B} \textbf{B}^T - \textbf{C} \textbf{C}^{T})
    \qquad \textnormal{or} \qquad
    a_{ij} \sim \sigma(\langle \vec{b}_i, \vec{b}_j \rangle - \langle \vec{c}_i, \vec{c}_j \rangle) 
    := \frac{1}{1 + 
        e^{
        -
        \left(
            \langle \vec{b}_i, \vec{b}_j \rangle - \langle \vec{c}_i, \vec{c}_j \rangle
            \right)
            }}
    \quad \forall i,j
\end{equation}
where $ \sigma$ is the logistic function depending on the scalar product of two embedding per nodes assumed to encode the role of each node in the network, namely $ \vec{b}_{i} \in \mathbb{R}_{+}^{D_{B}}, \vec{c}_{i} \in \mathbb{R}_{+}^{D_{C}} \textnormal{ where } D_{B} \geq 0,  D_{C} \geq 0$.
The compact formulation on the LHS was written in terms of the matrices $ \textbf{B}\in \mathbb{R}_{+}^{N \times D_{B}}, \textbf{C} \in \mathbb{R}_{+}^{N \times D_{C}}$ that are created by stacking horizontally the vectors $ \vec{b}_{i} \textnormal{ and } \vec{c}_{i}$ respectively.

Similarly, for the MSM (\autoref{sec:Methodology}), the LPCA vectors are fitted by means by maximizing the log-likelihood estimation'' \cite{2021_DL_StepByStep_Godoy}. In particular, the log-likelihood and its gradient read
\begin{align}
    \label{eq:likelihood_LPCA}
    \mathcal{L}\left(\mathbf{B}, \mathbf{C} | \mathbf{A} \right) 
    := &\sum_{i < j} a_{ij} \ln \left(\sigma_{ij}\right) + \left(1-a_{ij}\right) \ln \left(1-\sigma_{ij}\right) \\
    = &\sum_{i < j} \min(x_{ij}, 0) - \ln( 1 + e^{-\left\vert x_{ij}\right\vert}) - (1 - a_{ij}) x_{ij} \\
    \label{eq:grad_LPCA}
    \partial_{b_{ik}} \mathcal{L}  = &\sum_{j (\neq i)} \left(a_{ij}-\sigma_{ij}\right) b_{jk} \\
    \partial_{c_{ik}} \mathcal{L}  = &-\sum_{j (\neq i)} \left(a_{ij}-\sigma_{ij}\right) c_{jk}
\end{align}
where the last passage is taken from \href{https://www.tensorflow.org/api_docs/python/tf/nn/weighted_cross_entropy_with_logits}{BCE-TensorFlow} for numerical stability. Note that the stationarity conditions 
\begin{align}
    &\partial_{b_{ik}} \mathcal{L} \stackrel{!}{=} 0\\
    &\partial_{c_{ik}} \mathcal{L} \stackrel{!}{=} 0
\end{align}
can't be split in a part dependent only by the adjacency matrix as for the Exponential Random Graph models \cite{2011_AnalMax_Squartini}. Hence, LPCA hasn't a \textit{sufficient} statistics and one has to use the whole adjacency matrix.
The motivation for having two vectors per node boils down to the framework studied in \cite{2020_LPCA_Chanpuriya}. Specifically, the authors analyzed a dating graph reporting the messages exchanged among the male-female users living in two different cities. 
Hence, they have introduced two vectors per node to grasp the heterophily (male $ \leftrightarrow$  female) and homophily (same city $ \circlearrowleft$) ``role'' of each user. In the ION setting, we leave out this interpretation just considering them as parameters to be optimized. 

\subsection{Renormalizing the LPCA}
The LPCA does not have a recipe to renormalize the parameters and produce an ``up-scaled'' version of it. Nevertheless, it is possible to model the multi-level structure $ \left\{ \mathbf{A}_{\ell} : \ell \geq 0\right\}$ either through the \textit{coarse-graining} $ \mathbf{P}_{cog}^{(\ell)}$ or by \textit{fitting} it at every level. Both of them suffers from the lack of scale-invariance, as we show in the following.

Firstly, one may calculate $ \mathbf{P}_{cog}^{(\ell)}$ from the RHS of the \autoref{eq:MSM_micro_psumVSpcg}, namely \begin{equation}
    \label{eq:LPCA_Pcg}
    p^{cog,(\ell)}_{IJ} := 1 - \prod_{i_0 \in I, j_0 \in J} 
    \frac{1}{1 + 
    e^{
        \langle \vec{b}_{i_0}, \vec{b}_{j_0} \rangle - \langle \vec{c}_{i_0}, \vec{c}_{j_0} \rangle
    }}
\end{equation}
where $ I := \mathbf{\Omega}_{0 \to \ell}(i_{0}), J := \mathbf{\Omega}_{0 \to \ell}(j_{0})$ are the block-nodes at level $ \ell$. 

However, it wouldn't be possible to rearrange the $ \mathbf{P}_{cog}^{(\ell)}$ to recover a \textit{logistic} function with renormalized parameters. Specifically, \autoref{eq:LPCA_Pcg} is a product of logistic functions which is not rewritable as a logistic function. Hence, the resulting method would no longer belong to the logistic parametric family. In other words, LPCA is not \textit{self-consistent} under coarse-graining.

Secondly, by refitting LPCA at level $ \ell$, one would find another set of vectors that, in general, are unrelated with the \textit{($\ell$-1)-vectors}. To the point of view of LPCA, the different levels $ \ell$ are realizations of different \textit{generative process}.

As pointed out in \autoref{SI:Algorithmic_Complexity}, in order to compute $ \mathbf{P}_{cog}^{(\ell)} $, the computational complexity is higher than the \textit{summed} $ \mathbf{P}_{sum}^{(\ell)} $. Therefore, for the fairest comparison among the models, we will use the $ \mathbf{P}_{sum}^{(\ell)}$ with LPCA connection probability.
   
\subsection{Inconsistency of the LPCA: a trivial example}
\label{SI:Inconsistency_LPCA}
By referring to \autoref{fig:trivial_cg_network}, the microscopic network of 3 nodes $0,1,2$ merges into the community $ A,B$ containing respectively the nodes $ 0,1$ and the node $ 2$, namely $ A = \Omega(0) = \Omega(1), B = \Omega(2)$.
\begin{figure}[t]
    \begin{tikzpicture}[baseline=(base), node distance={2cm}, thick, main/.style = {fill=blue!20, draw, circle}] 
        \centering
        \node (base) at (0,-.5ex) {};
        \node[main] (0) {$0$}; 
        \node (3) [left of = 0] {\textbf{level 0} };
        \node[main] (1) [below of=0] {$1$};
        \node[main, fill=red!20] (2) [right of=0] {$2$}; 
        \draw[-] (1) to (2);          
    \end{tikzpicture}
    $ \qquad \stackrel{ \textnormal{Coarse-Graining} }{\longrightarrow} \qquad$ 
    \begin{tikzpicture}[baseline=(base), node distance={2cm}, thick, main/.style = {fill=blue!20, draw, circle}] 
        \centering
        \node (base) at (0,-.5ex) {};
        \node[main] (0) {A}; 
        \node[main, fill=red!20] (1) [right of=0] {B}; 
        \node (2) [right of = 1] {\textbf{level 1} };
        \draw[-] (0) to (1);
    \end{tikzpicture}
    \caption{Trivial Network for Inconsistency of LPCA and CM. The blue nodes belong to community $ A$ whereas the red node to the community $ B$.}
    \label{fig:trivial_cg_network}
\end{figure}
In addition, to stress the point we will switch-off the dependence on $ \vec{c}_{i}$. Therefore, the connection probability of the communities $ A,B$ from the \autoref{eq:LPCA_Psummed} as
\begin{equation*}
    \sigma(b_{AB}) = \left(1+e^{-b_{AB}}\right)^{-1}
\end{equation*}
which is different from the coarse-grained (\autoref{eq:LPCA_Pcg})
\begin{align*}
    \sigma^{cog}(b_{AB}) 
    &=  1 - \frac{1}{1 + e^{-b_{0} b_2}} \frac{1}{1 + e^{-b_{1} b_2}} \\
    &= \frac{e^{-b_{0} b_2} + e^{-b_{1} b_2} + e^{-b_{0} b_2}e^{-b_{1} b_2}}{1+e^{-b_{0} b_2} + e^{-b_{1} b_2} + e^{-b_{0} b_2}e^{-b_{1} b_2}}.
\end{align*}
From the previous results, it is clear that from $ \sigma^{cog}(b_{AB})$ one can't recover the $ \sigma(b_{AB})$ by defining $ b_{A} := f(b_{0}, b_{1}), b_B := b_2$ similarly to \autoref{eq:MSM_sum_graining_rule}. Therefore, the model is not renormalizable.



\section{Derivation of the multi-scale probability}
\label{SI:derivation_MSM_probability}
Here, we derive the multi-scale model formulation enhanced with \textit{vectors} (MSM). As in the main text, we will consider a coarse-graining procedure from $0$ to $ \ell \geq 0$ even though the treatment will hold for every pair $m, \ell+1$ with $m \leq \ell$. See \cite{2023_MSNR_Garuccio}, for further details even for the following passages.

Before introducing the model, it is worth to recall the problem settings to generate the observed multi-scale structure.
Concretely, let us consider the \textit{binary} undirected adjacency matrix $\mathbf{A}^{(0)}$ at level $0$ describing the microscopic interactions among the 0-nodes. Subsequently, a \textit{hierarchical and non-overlapping} partition of the microscopic nodes $ \left\{\mathbf{\Omega}_{\ell}\right\}_{\ell \geq 0}$ prescribing the community (block-nodes) membership of the lower-level nodes. Concretely, the block-nodes $I := i_{\ell}$ hosting all the $i_{0}$ nodes is obtained by
$$I := \mathbf{\Omega}_{0 \to \ell - 1}(i_0)$$
where we have defined $\Omega_{0 \to \ell-1} := \Omega_{\ell-1} \circ \dots \circ \Omega_{0}$. Lastly, a \textit{rule} to assign a link among blocks, namely
\begin{equation}
    a_{IJ} = 1 - \prod_{i_0 \in I, j_0 \in J} (1-a_{i_0j_0})
    \label{eq:cg_rule_nextl}
\end{equation}
where 
$a_{IJ} := a^{(\ell)}_{i_{\ell+1}j_{\ell+1}}, a_{i_{0}j_{0}} := a^{(0)}_{i_{0}j_{0}}$ and $J := j_{\ell} = \mathbf{\Omega}_{0 \to \ell-1}(j_{0})$. Therefore, by iterating the procedure it is possible to create the nested set of networks describing the \textit{original} phenomenon at different resolutions.

In order to model this architecture, one needs several assumptions (HP). The \textit{first} one (HP1) requires that the MSM must describe the microscopic matrix $\textbf{A}^{(0)}$. Similarly to \autoref{SI:LPCA}, 
$\textbf{A}^{(0)} \sim P^{(0)}\big(\mathbf{A}^{(0)},\mathcal{X}^{(0)}\big)$ subject to 
$P^{(0)}\big(\cdot,\mathcal{X}^{(0)}\big) = (P^{(0)})^{T}\big(\cdot,\mathcal{X}^{(0)}\big)$ and $\sum_{\textbf{S} \in \left\vert \mathscr{A}^{(0)} \right\vert} P^{(0)} \big(\textbf{S},\mathcal{X}^{(0)}\big) = 1$ where $\mathscr{A}^{(0)}$ is the ensemble of all the binary symmetric graphs with $ N_{0}$ nodes. 

In line with \cite{2023_MSNR_Garuccio}, the \textit{second} hypothesis (HP2) prescribes that
\begin{equation}
    \mathcal{X}^{(0)}_{ij} \stackrel{!}{=}
    \begin{cases}
        \langle \vec{x}_{i}, \vec{x}_{j} \rangle & \textnormal{if } i \neq j \\[2ex]
        \frac{1}{2} \left\Vert \vec{x}_{i}\right\Vert^{2} + w_{i} & \textnormal{if } i = j
    \end{cases}
\end{equation}
is given by a product (to-be-defined) between the $ D-$dimensional vectors $ \left\{\vec{x}_i\right\}_{i \in [1, N_\ell]}$ with additional node-wise parameters $ \left\{w_{i}\right\}_{i \in [1, N_\ell]}$ only active in the self-loop part ($ i_\ell = j_\ell$).
In particular, $ \vec{x}_i$ encodes the capability of node $ i$ to connect to the other nodes; whereas $w_{i}$ its propensity for a \textit{self}-interaction. Since the principles leading to an edge are different to the self-loop ones, we introduced two independent parameters.
The $ \frac{1}{2}$ in the self-loop part is introduced to avoid the double-counting of the self-interaction. 
Indeed, if $ A := \left\{0, 1\right\}$ as in \autoref{fig:trivial_cg_network},
\begin{align*}
    1 - p_{AA} &\stackrel{!}{=} 1 - (1 - p_{00})(1 - p_{11})(1 - p_{01}) 
    \\ &\textnormal{ iff }
    \\
    \frac{1}{2} \left\Vert \vec{x}_{A}\right\Vert^{2} 
    = \frac{1}{2} \left\Vert \vec{x}_{0} \right\Vert^{2} + \frac{1}{2} \left\Vert \vec{x}_{1} \right\Vert^{2} + \frac{2}{2} \langle \vec{x}_{0}, \vec{x}_{1} \rangle 
    &\stackrel{!}{=} 
    \frac{1}{2} \left\Vert \vec{x}_{0}\right\Vert^{2} + \left\Vert \vec{x}_{1} \right\Vert^{2} + \langle \vec{x}_{0}, \vec{x}_{1} \rangle
\end{align*} 
which is, indeed, the case after inserting the factor of $ \frac{1}{2}$ in the self-loop part. Lastly, we discarded a possible dependence of $\mathcal{X}^{(0)}$ on a \textit{dyadic} $ d_{ij}$ factor and the higher order terms. For further details we refer to \cite{2023_MSNR_Garuccio}.

In the following, we will use the notation $ P^{(0)}\big(\mathbf{A}^{(0)},\mathcal{X}^{(0)}\big) := P^{(0)}\big(\mathbf{A}^{(0)}, \mathbf{X}_{w}^{(0)} \big)$ where $\mathbf{X}^{(0)}_w := [\textbf{X}, \vec{w}]$. Technically, $ \mathbf{X}^{(0)} := \left[\vec{x}_{1}, \dots, \vec{x}_{N_0}\right]^{T} \in N_0 \times D$ and $ \vec{w}^{(0)} := \left\{w_{i_0}\right\}_{i \in [1, N_0]}$.

Fitted $ \mathbf{X}^{(0)}_w$ at the ground level, the ensemble generated by the MSM contains multiple configurations that, after coarse-graining, lead to the observed macroscopic $ \mathbf{A}^{(\ell)}$ \cite{2023_MSNR_Garuccio}, i.e. $ \{\mathbf{A}^{(0)}\}\xrightarrow{\mathbf{\Omega_{0 \to \ell-1}}}\mathbf{A}^{(\ell)}$. In turns, this induces the probability of observing $\mathbf{A}^{(\ell)}$ as
\begin{equation}
    P_{\ell}\big(\mathbf{A}^{(\ell)},\mathbf{X}^{(0)}_w\big) :=
    \!\!\!\!\!\!\!\!\!\!\!\!\!\sum_{\{\mathbf{A}^{(0)}\}
    \xrightarrow{\Omega_{0 \to \ell-1}}\mathbf{A}^{(\ell)}}
    \!\!\!\!\!\!\!\!\!\!\!\!\!\!P_{0}\big(\mathbf{A}^{(0)},\mathbf{X}^{(0)}_w\big).
    \label{eq:induced_prob}
\end{equation}
To enforce the \textit{scale-invariance} property, we require (HP3) that the \textit{functional form} of the MSM has to be independent of the chosen scale, i.e. $ P_{\ell}\big(\cdot,\cdot) \stackrel{!}{=} P_{0}\big(\cdot,\cdot) \quad \forall \ell \geq 0$. Furthermore, that the model can generate the possible $\ell$-graphs in two equivalent ways \emph{hierarchically} or \textit{directly} (HP4). The former one refers to \autoref{eq:induced_prob}, and it prescribes to generate the $0$-graph ensemble with probability $P\big(\mathbf{A}^{(0)},\mathbf{X}^{(0)}_w\big)$ and, then, coarse-graining them $\ell$ times via the partitions $\{\mathbf{\Omega}_k\}^{\ell-1}_{k= 0}$. The other way around, the second one requires to \textit{renormalize} the parameters $ \widetilde{\mathbf{X}}_{\widetilde{w}}^{(\ell)}$ and, then, directly model $\mathbf{A}^{(\ell)}$ via $P\big(\mathbf{A}^{(\ell)},\widetilde{\mathbf{X}}_{\widetilde{w}}^{(\ell)}) $.
Imposing both requirements
\begin{align}
    P\big(\mathbf{A}^{(\ell)}, \widetilde{\mathbf{X}}_{\widetilde{w}}^{(\ell)} \big) 
    &\stackrel{!}{=}
    \!\!\!\!\!\!\!\!\!\!\!\!\!
    \sum_{\{\mathbf{A}^{(0)}\}
    \xrightarrow{\mathbf{\Omega}_{0 \to \ell-1}}\mathbf{A}^{(\ell)}}
    \!\!\!\!\!\!\!\!\!\!\!\!\!\!
    P\big(\mathbf{A}^{(0)}, \mathbf{X}_{w}^{(0)} \big)
    \quad \Leftrightarrow \quad
    \mathbf{P}_{sum}^{(\ell)} \stackrel{!}{=} \mathbf{P}_{cog}^{(\ell)} 
\end{align} 
where we have defined the LHS and RHS of the first equation as $\mathbf{P}_{sum}^{(\ell)} \textnormal{ and } \mathbf{P}_{cog}^{(\ell)} $ respectively.
In other words, the form of the $ P(\cdot, \cdot)$ will depend on the scale $ \ell$ only through the renormalized parameters $ \widetilde{\mathbf{X}}_{\widetilde{w}}^{(\ell)} $. Moreover, by assuming that the links are statistically independent, the previous equation yields 
\begin{align}
    \label{eq:MSM_micro_psumVSpcg}
    p_{IJ} &\stackrel{!}{=} p^{cog}_{IJ}
\end{align}
where $p_{IJ} := p_{IJ}(\widetilde{\mathbf{X}}_{\widetilde{w}}^{(\ell)})$
and
\begin{equation}
    \label{eq:MSM_pIJ_cg}
    p^{cog}_{IJ} := 1 - \prod_{i_0 \in I, j_0 \in J} (1-p\left(\vec{x}_{i_0}, \vec{x}_{j_0}, w_{i_0}\right))
\end{equation}
depend, respectively, on the renormalized and the fitted parameters at level $\ell = 0$. Note that we didn't use $ p^{sum}_{IJ}$ because the functional form would be \textit{scale-invariant} and the only dependence on the scale is through the parameters $ IJ$.
The interpretation is similar to the \autoref{eq:cg_rule_nextl}: the probability $ p_{IJ}$ that there is one among the block-nodes $ I, J$ is given by the probability that there is \textit{at least one link} among the microscopic nodes $ i_0 \in I, j_0 \in J$.
Specifically, it requires that the model remains \textit{self-similar} whereas the parameters renormalize under renormalization (\textit{scale-variant}). 

The RHS returns the coarse-grained probability for every model, e.g. the MSM and LPCA. The crucial difference is that for the SSM $ \mathbf{P}_{sum}^{(\ell)} \neq \mathbf{P}_{cog}^{(\ell)}$ (cfr. \autoref{eq:self-cons_scale-invariance_conds}) because they are scale-invariant. For a concrete example, we refer to the sections where the models have been introduced.

By taking the logarithm of both sides of the \autoref{eq:MSM_micro_psumVSpcg}, the only functional form compatible with that constraint \begin{equation*}
    \label{SI:eq:MSM_derivation_p_gI}
    \ln(1-p_{IJ}) = - \langle g( \vec{x}_I), g( \vec{x}_J) \rangle
\end{equation*}
where $ g(x)$ is a positive function such that $ g( \vec{x}_I) := \sum_{i_0 \in I} g( \vec{x}_i)$. Assuming (HP5) that $ g( y) := y$ (identity) for every level, as in the reference derivation, the \textit{summation} rule becomes the one reported in \autoref{eq:MSM_sum_graining_rule}. Furthermore, the product $ \langle \cdot, \cdot \rangle$ has to be bilinear (HP6), as required by the \autoref{eq:MSM_micro_psumVSpcg}, and, since the connection probability is symmetric, one can set $\langle \vec{x}_i, \vec{x}_j \rangle := \vec{x}_{i}^{T} \vec{x}_{j}$ (see \autoref{SI:BilinearityRequirement}). By encoding all these passages in the \autoref{SI:eq:MSM_derivation_p_gI}, one obtains
\begin{equation*}
    \ln(1-p_{IJ}) = - \langle \vec{x}_I, \vec{x}_J \rangle
\end{equation*}
which, eventually, leads to the (off-diagonal) \textit{scale-invariant} probability in \autoref{eq:MSM_pIJ}. For the self-loops part, one may apply the same reasoning.

\subsection{Bilinearity Requirement}
\label{SI:BilinearityRequirement}

\begin{figure}[t]
    \begin{tikzpicture}[baseline=(base), node distance={2cm}, thick, main/.style = {fill=blue!20, draw, circle}] 
        \centering
        \node (base) at (0,-.5ex) {};
        \node[main] (0) {$0$}; 
        \node [left of = 0] {\textbf{level 0} };
        \node[main] (1) [below of=0] {$1$};
        \node[main, fill=red!20] (2) [right of=1] {$2$};
        \node[main, fill=red!20] (3) [right of=0] {$3$};

        \draw[dashed, gray] (0) to (2);
        \draw[dashed, gray] (0) to (3);
        \draw[dashed, gray] (1) to (2);
        \draw[dashed, gray] (1) to (3);
    \end{tikzpicture}
    $ \qquad \stackrel{ \textnormal{Coarse-Graining} }{\longrightarrow} \qquad$ 
    \begin{tikzpicture}[baseline=(base), node distance={2cm}, thick, main/.style = {fill=blue!20, draw, circle}] 
        \centering
        \node (base) at (0,-.5ex) {};
        \node[main] (0) {A}; 
        \node[main, fill=red!20] (1) [right of=0] {B}; 
        \node (2) [right of = 1] {\textbf{level 1} };
        \draw[dashed, gray] (0) to (1);
    \end{tikzpicture}
    \caption{Simple graph where the nodes $ 0,1$ merges into the community $ A$ whereas $ 2,3$ in the block-node $ B$. The gray dashed lines represent the non-existing links.}
    \label{fig:easy_net_4_bilinearity}
\end{figure}

In this subsection, we describe why the $ \langle *,* \rangle$ product must be bilinear and why $ \textbf{M}$ can be taken as the identity matrix given that the probability is symmetric. To start with, in \autoref{fig:easy_net_4_bilinearity}, there have been represented 4 nodes $ 0,1,2,3$  at level $ \ell = 0$ merging, at level $ \ell = 1$, into $A := \left\{0, 1\right\}$ and $ B := \left\{2, 3\right\}$.
Hence, from \autoref{eq:MSM_micro_psumVSpcg}, the non-existence of a link (gray dashed lines) among the communities $ A \textnormal{ and } B$ reads
\begin{equation}
    e^{- \langle \vec{x}_{A}, \vec{x}_{B} \rangle } 
    \stackrel{!}{=} \prod_{i \in A; j \in B} e^{- \langle \vec{x}_i, \vec{x}_j \rangle }
    \quad \textnormal{ iff }
    e^{- \langle \vec{x}_{1} + \vec{x}_{1}, \vec{x}_{2} + \vec{x}_{3} \rangle }
    \stackrel{!}{=} e^{- \langle \vec{x}_{1}, \vec{x}_{2} \rangle + \langle \vec{x}_{1},\vec{x}_{3} \rangle + \langle \vec{x}_{1}, \vec{x}_{2} \rangle + \langle \vec{x}_{1},\vec{x}_{3} \rangle }
\end{equation}
and the rightmost side enforces that the $ \langle *,* \rangle$ has to be a bilinear function. 

As said in the main text, the connection probability is symmetric, namely $ \mathbf{P} \stackrel{!}{=} \mathbf{P}^{T}$. In turns, this leads to $\mathbf{M} = \mathbf{M}^{T}$. In addition, since $ p_{ij} \in [0,1] \, \forall i,j$, $ \textbf{M}$ is also positive semidefined as $ \vec{x}_{i} \textbf{M} \vec{x}_{j} \geq 0 \; \forall i,j$. Therefore, $ \mathbf{M}$ has positive eigenvalues such that 
\begin{equation*}
    \vec{x}_i^{T} \textbf{M} \vec{x}_j = \vec{x}_i^{T} \mathbf{O} \mathbf{O}^{T} \vec{x}_j = \vec{y}_i^{T} \vec{y}_j
\end{equation*}
where $ \vec{y}_i := \mathbf{O}^{T} \vec{x}_i \in \mathbb{R}_+^{D}$ (cfr. Cholesky decomposition). Briefly, choosing an arbitrary (symmetric) $ \mathbf{M}$ matrix will lead to $ \vec{x}_i = \textbf{O} \vec{y}_i$ with $ \vec{y}_i$ are optimized with $ M := \textnormal{Id}_{D \times D}$. Hence, for simplicity, we rely on $ \mathbf{M} := \textnormal{Id}_{D \times D}$.

Lastly, fixing $ \mathbf{M} := \textnormal{Id}_{D \times D}$, allows recovering the product among scalars $ x_{i}x_{j}$ for $ D = 1$. This was a successful way of modelling real world networks, e.g. \cite{2023_MSNR_Garuccio,2011_AnalMax_Squartini}.

\subsection{From Constrains to Bounds}
\label{proof:positive_components}
The MSM probability requires a positive inner product
\begin{equation}
    \label{cond:nonn_scalar_prod_x}
    \langle \vec{x}_i, \vec{x}_j \rangle \geq 0
\end{equation}
for every pair of nodes, in order to guarantee $ p_{ij} \in [0,1] \forall i, j$.  

Here, we will prove that the above constraints is equivalent of setting all vector components to be non-negative, namely $ x_{ik} \geq 0$. Roughly, the spanned region by the embeddings is enclosed in one quadrant of the space, and it is possible to rotate the vectors to lay in the positive quadrant. 

The steps to show this are the following.  First, since we are interested on the sign among the vectors, one may restrict to the set of unit vectors 
$\left\{\vec{e}_{j}\right\}_{i \in [1, N]}$ where $ \vec{e}_{i} := \frac{\vec{x}_{i}}{\left\Vert \vec{x}_{i}\right\Vert} \in \mathbb{R}^{D}$ and assume $ \vec{e}_{0} := [1, 0, \dots]$. 
Taking into consideration also \autoref{cond:nonn_scalar_prod_x}, the considered set is
\begin{equation*}
    \mathbb{S} := \left\{ \vec{e}_i \in \mathbb{R}^{D} \textnormal{ s.t. } 0 
    \leq \langle \vec{e}_i, \vec{e}_j \rangle \leq 1 , \vec{e}_{0} := [1, 0, \dots] \quad \forall  i \in [1, N] \textnormal{ and } j \in [1, N] \right\}
\end{equation*} which by construction has to property that
\begin{equation*}
    \max_{ i \in [1, N_\ell], j \in [1, N_\ell]} \left\vert \arccos(\langle \vec{e}_i, \vec{e}_j \rangle)\right\vert \leq \frac{\pi}{2}
\end{equation*}
If one takes the most ``clockwise'' and ``anticlockwise'' vectors in the set $ \mathbb{S}$ defined as 
\begin{align}
    \vec{e}_{c} &:= \left\{\vec{e}_i : \vec{e}_i \times v \geq 0, \, \forall v \in \mathbb{S}\right\}\\
    \vec{e}_{ac} &:= \left\{\vec{e}_i : \vec{e}_i \times v \leq 0, \, \forall v \in \mathbb{S}\right\},
\end{align}
by construction they form an angle $\theta_{a-ac} \in [-\frac{\pi}{2}, \frac{\pi}{2}]$. Therefore, if $ \vec{e}_{c} = \vec{e}_{0}$ the treatment is finished since the vectors lay in the positive quadrant. On the other hand, if $\vec{e}_{ac} = \vec{e}_{0}$, the vectors lay in the negative quadrant, and they can be rigidly rotated to lie on the positive quadrant, i.e. $ x_{ik} \geq 0$.

\subsection{From full to off-diagonal likelihood}
In the main text, we said the node embeddings should be found by maximizing the \textit{off-diagonal} likelihood, namely \autoref{eq:MSM_loglikelihood}. Here, we show how to simplify the \textit{full likelihood} (running also over the self-loops) to the off-diagonal one as suggested by the stationarity conditions. 

Assuming to know the adjacency matrix, the full likelihood for the MSM reads
\begin{equation}
    \label{eq:MSM_full_likelihood}
    \mathcal{L}(\mathbf{X}_{w}|\textbf{A}) =
    \mathcal{L}( \left\{\vec{x}_{i}, w_{i}\right\}_{i \in [1, N]} |\textbf{A}) =
    \sum_{i \leq j} a_{ij} \ln(p_{ij}) + (1-a_{ij}) \ln(1-p_{ij})
\end{equation}
Its gradient can be calculated by deriving the \autoref{eq:MSM_full_likelihood} with respect to $ x_i$ and $ w_{i}$, namely
\newlength{\itsp}
\setlength{\itsp}{3ex}
\begin{align}
    \begin{cases}
        \partial_{x_{ik}} \mathcal{L}
        &= \sum_{j} \left( \frac{a_{ij}}{p_{ij}} - 1 \right) \frac{ \partial_i p_{ij} }{q_{ij}} \\[\itsp]
        &= \sum_{j} \left(\frac{a_{ij}}{p_{ij}} - 1\right) x_{jk} \\[\itsp]
        \partial_{w_i} \mathcal{L}
        &= \frac{a_{i i}}{p_{i i}} - 1
    \end{cases}
\end{align}
which vanishes when the parameters are at the optimal point $ \vec{x}^{*}, \vec{w}^{*}$, i.e.
\begin{align}
    \label{eq:MSM_full_grad0}
    \begin{cases}
        \partial_{x_{ik}} \mathcal{L} \vert_{\vec{x} = \vec{x}^{*}} = \sum_{j} \left(\frac{a_{ij}}{p^{*}_{ij}} - 1\right)  x^{*}_{jk} &= 0 \\[\itsp]
        \partial_{w_i} \mathcal{L} \vert_{\vec{w} = \vec{w}^{*}} = \frac{a_{ii}}{p^{*}_{ii}} - 1 &= 0
    \end{cases}
\end{align}
where $ p^{*}_{ij} := p(\vec{x}^{*}_{i}\vec{x}^{*}_{j}) \textnormal{ and } p^{*}_{ij} := p(\vec{x}^{*}_{i}, \vec{w}^{*}_{i})$. 
In particular, the second equation suggests that $ \left\{w_{i}^{*}\right\}_{i \in [1, N]}$ are a function of $ \left\{\left\Vert x_{i}\right\Vert^{2}\right\}_{i \in [1, N]}$ at the stationarity point, namely
\settowidth{\temp}{$\to \infty \qquad$}
\begin{equation}
    w_{i} =
    \begin{cases}
        \FixedSize{\temp}{\to \infty} \textnormal{ if } a_{i i} = 1 \\[1ex]
        \FixedSize{\temp}{-\frac{1}{2} \left\Vert x_{i}\right\Vert^{2}} \textnormal{ if } a_{i i} = 0.
    \end{cases}
\end{equation}
In addition, one can reduce the number of parameters by inserting the second equation into the first one, which leads to
\begin{equation}
    \partial_{x_{ik}} \mathcal{L} \vert_{\vec{x} = \vec{x}^{*}} = \sum_{j (\neq i)} \left(\frac{a_{ij}}{p^{*}_{ij}} - 1\right)  x^{*}_{jk} = 0
\end{equation}
and, by integrating over $ x_{ik}$, to
\begin{align}
    \mathcal{L}( \left\{x_{i}\right\}_{i \in [1, N]} |\textbf{A}) 
    &= \sum_{i < j} a_{ij} \ln(p_{ij}) + (1-a_{ij}) \ln(1-p_{ij}) \\
    \label{eq:MSM_likelihood}
    &= \sum_{i < j} a_{ij} \ln(1 - e^{-\langle \vec{x}_i, \vec{x}_j \rangle}) - (1-a_{ij}) \langle \vec{x}_i, \vec{x}_j \rangle.
\end{align}
Therefore, it is enough to find the maximum of the \textit{off-diagonal} likelihood \autoref{eq:MSM_likelihood}, and, then, fix the $ \left\{w_{i}\right\}_{i \in [1, N]}$. In this way, the part of the likelihood summing over the self-loops would be identically maximal, i.e. $ 0$, whereas the off-diagonal part was already maximized by the optimization procedure. 

As a final remark, LPCA \cite{2021_symLPCA_Chanpuriya} models the self-loops via $ \mathbf{B}, \mathbf{C}$, whereas the MSM introduces the $ \left\{w_{i}\right\}_{i \in [1, N]}$ parameters to represent the ``different physics'' of the self-interactions. Hence, for a fair comparison, we fit both LPCA and MSM via the \textit{off-diagonal} likelihoods, and evaluating them on metrics that aren't including the auto-interactions as reported in \autoref{sec:Methodology}, e.g. the DEG.
\subsection{Gradient of the off-diagonal likelihood}
To efficiently calculate the maximum of the MSM likelihood \autoref{eq:MSM_loglikelihood}, one needs the analytical expression of the gradient. In particular, by differentiating with respect to the $ t$-th component of the $ s$-th embedding vector, one gets
\begin{align}
    \partial_{st} \mathcal{L}
    &= \sum_{i < j} \left(\frac{a_{ij}}{p_{ij}} - 1\right) \frac{\partial_{x_{st}} p_{ij}}{q_{ij}}  \\
    &= \sum_{j (\neq i)} \left(\frac{a_{sj}}{p_{sj}} - 1\right) x_{jt}\\
\end{align}
where we have defined
$q_{ij} := 1 - p_{ij} = e^{-\langle \vec{x}_i, \vec{x}_j \rangle}.$ 
By renaming the indexes,
\begin{align}
    \label{eq:MSM_loglikel_grad}
    \partial_{ik} \mathcal{L}
    = \sum_{j (\neq i)} \left( \frac{a_{ij}}{p_{ij}} - 1\right) x_{jk} 
    \equiv \sum_{j (\neq i)} \left(\frac{a_{ij}}{1 - e^{- \langle \vec{x}_i, \vec{x}_j \rangle}} - 1 \right) x_{jk}
\end{align}

Lastly, by leveraging on the gradient and the likelihood, we performed the optimization by means of three optimizers: Adam implemented from \cite{2014_Adam_KIngma}; while Truncated-Conjugate Gradient and L-BFGS-B by means of the SciPy library \cite{2020_SciPy_Virtanen}. The hessian was not calculated since it was used neither for a theoretical analysis nor in the numerical optimization -the SciPy methods can approximate it in a sophisticated way.

\subsection{Structural Equivalence is not Statistical Equivalence for multidimensional-node embeddings}
\label{SI:StructE_not_StatE}
Calculating the gradient \autoref{eq:MSM_loglikel_grad} for nodes $ i \textnormal{ and } i'$ at the maximum of the likelihood, one gets
\begin{equation}
    \partial_{ik} \mathcal{L} \vert_{\vec{x}_{i} = \vec{x}^{*}_{i}} = 0 = \partial_{i'k} \mathcal{L} \vert_{\vec{x}_{i'} = \vec{x}^{*}_{i'}}.
\end{equation}
By further assuming $ i \textnormal{ and } i'$ have the same (external) neighbors they are defined to the (externally) \textit{Structural Equivalent} (StructE). Technically, $ \mathcal{N}(i) := \left\{j : a_{ij} = 1, j \neq i \right\} = \mathcal{N}(i') := \left\{j : a_{i'j} = 1, j \neq i' \right\}$. By removing the start superscript in the optimal parameters, i.e. $ \vec{x}^{*}_{i} \to \vec{x}_{i}, w_{i}^{*} \to w_{i}$, the above equations rearranges into
\begin{equation}
    \phi(x_{ik})
    := \sum_{j \in \mathcal{N}(i)} \frac{x_{jk}}{p_{ij}} + x_{ik}
    = \sum_{j \in \mathcal{N}(i')} \frac{x_{jk}}{p_{i'j}} + x_{i'k} =: \phi(x_{i'k})
\end{equation}
where we have defined
\begin{equation}
    \phi(y_{ik}) = \sum_{j \in \mathcal{N}(i)} \frac{y_{jk}}{1-e^{- \langle \vec{y}_i, \vec{x}_j \rangle}} + y_{ik}.
\end{equation}
For $ D = 1$,
\begin{equation}
    \phi(x_{i}) := \sum_{j \in \mathcal{N}(i)} \frac{x_{j}}{1-e^{- x_i x_j}} + x_{i}.
\end{equation}
In both cases, the function $ \phi(\cdot)$ is not a monotonic function of its argument, so the StructE does not imply \textit{statistical} equivalence (StatE) at the optimal likelihood point. The result won't change including the self-loops (full likelihood), since $ \phi(x_{i}) \to \phi(x_{i}, w_{i})$ which will depend on two independent parameters.

However, if $ i, i'$ have the same (external) neighbors, they have the same outer role in the network, and so the model should have the same $ \vec{x}$ parameters for the two nodes. Formally,
\begin{equation*}
    \mathcal{N}(i) = \mathcal{N}(i')  \Leftrightarrow a_{ij} \equiv a_{i'j} \Rightarrow p_{ij} = p_{i'j}  \Leftrightarrow \vec{x}_i = \vec{x}_{i'} \qquad \forall j \in [1, N_\ell], j (\neq i,i').
\end{equation*}
This result is achieved by considering the \textit{reduced} problem where the same $ \vec{x}_{i}$ are copied for all the structural equivalent nodes and the likelihood is, hence, optimized over a \textit{reduced} number of parameters (see tables in \autoref{sec:DatasetDescription}) of the same cardinality of the number of structural inequivalent nodes.

More specifically, by defining the structurally equivalent nodes as
$$ \mathcal{S}_{i_r} := \left\{j \textnormal{ s.t. } \mathcal{N}(j) = \mathcal{N}(i_r) \right\}$$ 
which index $ i_r$ is the lowest index among the StatE nodes, one may create the set of statistically inequivalent nodes 
$$ \mathcal{S} := \left\{S_{i_r}\right\}_{i_r \in [1, N_{irred}]}$$
where we denoted with $ N_{irred}$ the number of irreducible representatives $i_r$.
More concretely, by referring to \autoref{fig:easy_net_4_bilinearity} and considering the gray dashed lines as existing connection, $ \mathcal{S}_1 := \left\{1,2\right\}$ and $ \mathcal{S}_{2} := \left\{3\right\}, \mathcal{S} := \left\{S_{1}, S_{2}\right\}$ with $N_{irred} = 2$.
Secondly, the vector of the representative parameter $ \vec{x}_{i_r}$ is copied to all the members of the class, namely
\begin{align*}
    \left\{ \vec{x}_{i} := \vec{x}_{i_r} \, \forall i : \mathcal{N}(i) \equiv \mathcal{N}(i_r) \right\} \quad \forall i_r \in [1, N_{irred}].
\end{align*}
Hence, it would be possible to maximize the original likelihood (\autoref{eq:MSM_loglikelihood}), but using fewer parameters than in the original formulation: $ N_{irred} \times D$ instead of $ N \times D$.

Note that, since the selected optimizers (\cite{2020_SciPy_Virtanen}) are not ``stochastic'', one may obtain the StatE between the embeddings by starting from the same initial conditions for StructE nodes (see \autoref{eq:grad_LPCA}, \ref{eq:MSM_loglikel_grad}).

The \textit{complete} structural equivalence requires to have the same connections, that is the same behavior in the self-interaction, formally
In conclusion, to fix the same parameter vector $ \vec{x} $ for all the nodes with the same external neighborhood, one may reduce the problem the inequivalent node parameters and, then, copying the representative values to all the nodes in the same class. 
$$ \left\{a_{ij} \equiv a_{i'j}, \, \forall j \in [1, N]\right\} $$
which implies also to have the same self-loop parameter $ w_{i} = w_{i'}$.

\subsection{Loop parameters estimation}
\label{SI:sec:self_loops}
For this paper, we fitted the MSM vectors $ \left\{\vec{x}_{i}\right\}_{i \in [1, N]}$  by maximizing the \textit{off-diagonal} likelihood (\autoref{eq:MSM_loglikelihood}), which does not account for the self-loops. If for any reason, one needs also to fit the parameters $ \left\{w_{i}\right\}_{i \in [1, N]}$, associated with the self-loops, one can fit them to \autoref{eq:MSM_w_limits}. The reason for this choice is that the self-loop likelihood 
\begin{equation}
    \label{eq:MSM_loglik_selfloops}
    \mathcal{L} \left( \mathbf{X}_{w}| \mathbf{A}^{diagonal}\right) 
    = \sum_{\left\{i \textnormal{ s.t. } a_{i i} = 1\right\}} \ln \left( 1- e^{-\frac{1}{2}  \Vert \vec{x}_{i} \Vert ^2  - w_{i}} \right) - \sum_{\left\{i \textnormal{ s.t. } a_{i i} = 0\right\}} \left( \frac{1}{2}  \Vert \vec{x}_{i} \Vert ^2 + w_{i} \right)
\end{equation}
displays only one term related to the presence ($a_{ii} = 1$) or absence of the self-loop ($a_{ii} = 0$). In other words, there is no ``trade-off'' balancing the $ \left\{w_{i}\right\}_{i \in [1, N]}$ which will eventually match the values in \autoref{eq:MSM_w_limits} after the optimization. In turns, the $\mathcal{L} \left( \mathbf{X}_{w}| \mathbf{A}^{diagonal}\right)$ would be identically zero, preserving the maximum of \textit{full}  likelihood, namely $ \mathcal{L}^{full} = \mathcal{L}^{off-diag} + \mathcal{L}^{diag} = \mathcal{L}^{off-diag}$.
However, since the $ \mathcal{L}^{full}$ doesn't guarantee to have only one maximum, we don't know if the reach maximum would be the global one and not only a well-engineered local one.

\subsection{Removal of deterministic nodes}
Network modelling assumes that the observed $ \mathbf{A}$ is a realization of a random process. However, it may happen that some nodes are \textit{deterministic}, i.e. fully-connected (FC) or disconnected (D) to the other nodes. Therefore, their behavior is \textit{trivially} recovered by setting the hidden variable of the FC nodes to infinity or the D nodes to zeros (see \autoref{eq:MSM_pij}). Their optimized parameters would be $ x_{FC} \to \infty$ and $ x_{D} = 0$, respectively implying that $\left\{p_{FC,j} \equiv 1\right\}_{ j \in [1, N] }$ and $\left\{p_{FC,j} \equiv 0\right\}_{ j \in [1, N] }$. In turn, $ \textnormal{ Var}(a_{FC,j}) = \textnormal{ Var}(a_{D,j}) = 0 \quad \, \forall j$ implying that their sampled edges are fixed, and not contributing to the ensemble fluctuation. Concretely, we have hard coded their variables, as seen before, to account for their roles in the graph. 

A similar reasoning applies to the nodes only connected to FC, the so called \textit{only2fc} nodes ($ \nu$). In particular, their values are undetermined since their likelihood reads (assume $ D = 1$)
\begin{align}
    \mathcal{L}_{\nu} = \sum_{FC} \ln\left(p_{\nu,FC}\right) - x_{\nu}  \sum_{j (\neq FC,\nu)} x_{j} = - x_{\nu} \sum_{j (\neq FC,\nu)} x_{j}
\end{align}
since the $ p_{\nu, FC} \equiv 1 \, \forall FC$. Indeed, setting for convenience $ x_{FC} = \tau^{2} $ with $ \tau \to \infty $, the maximum likelihood solutions require that
\begin{equation*}
    \begin{cases}
        x_{o} x_{FC}  \sim \tau  \\ 
        x_{o} x_{j}  \sim 1/\tau \qquad \forall j \neq FC
    \end{cases}
\end{equation*}
which are trivially solved by setting $ x_{o} \stackrel{!}{=} 1/\tau$ as the $ x_{j} \approx \textnormal{ const}$ with respect to $ x_{o}$.

In conclusion, setting manually these values provides a more stable evaluation of the maximum likelihood since one reduces to source of degeneracy in the model. Additionally, this procedure doesn't spoil the \textit{ renormalization rule}. Including an FC node to a block-node produces an FC community; whereas a vanishing parameter (disconnected or only2fc) gives the freedom to the other terms in the summation (see \autoref{eq:MSM_sum_graining_rule}) to determine the group parameter.

\subsection{Algorithmic Complexity}
\label{SI:Algorithmic_Complexity}
As described in the \autoref{SI:Inconsistency_LPCA}, in order to describe a coarser graph without refitting the parameters, one should to use the RHS of \autoref{eq:MSM_micro_psumVSpcg}\footnote{This spoils its functional form, but this would be the only way to avoid refitting the same model at a higher scale}. Specifically, the algorithmic complexity to obtain one $ p^{cog}_{IJ}$ is $c N_I N_J$ where the evaluation of $ p_{ij}$ is assumed to have a complexity $ c$. Hence, to compute the complexity of $ \mathbf{P}_{cog}^{(\ell)} $, one has to sum over all the pairs, i.e.
\begin{align}
    \mathcal{C}_{SSM} = c \sum_{I > J} N_{I}N_{J} = \frac{c}{2} \sum_{I,J} N_{I}N_{J} = c \frac{N^{2}_{0}}{2}
\end{align}
where $ N_{0}$ are the number of structural inequivalent nodes at $ \ell = 0$. 
On the other hand, assuming the ``sum'' of vectors in \autoref{eq:MSM_sum_graining_rule} of order $ O(1)$, the complexity of $ \mathbf{P}_{sum}^{(\ell)} $ (see \autoref{eq:MSM_pIJ}) reads
\begin{equation}
    \mathcal{C}_{MSM} \approx N_0 (D + 1) + c \frac{N_{\ell}(N_{\ell}-1)}{2}
\end{equation}        
where the summation over the $ w_{i}$ parameters counts as $ N_0$ operations and $ \binom{N_{\ell}}{2}$ are all pairs at the level $ \ell$ for the full $ \mathbf{P}_{sum}^{(\ell)} $.

The improvement is reported by the ratio of the two complexities, namely
\begin{equation}
    \frac{\mathcal{C}_{MSM}}{\mathcal{C}_{SSM}} \approx 
    \frac{2 (D + 1)}{cN_{0}} + \frac{ N_{\ell} \left(N_{\ell} - 1\right) }{N_{0}^{2}}
    \approx \frac{1}{N_{0}} \left[ 1 + \frac{N_{\ell} \left(N_{\ell} - 1\right)}{N_{0}}\right].
\end{equation}        
Focusing on level $\ell = 3, N_{3} = 87, N_0 = 972$,
\begin{equation}
    \frac{\mathcal{C}_{MSM}}{\mathcal{C}_{SSM}} \approx \mathcal{O}(10^{-2})
\end{equation}        
one saves two order of magnitude by proceeding with the summed MSM rather than the coarse-grained LPCA. Hence, by means of \autoref{eq:LPCA_summed_bc}, LPCA recovers the same complexity of summed MSM; thereby enabling for a comparison of equal complexity.

\section{Scores}

\subsection{Comparison Among Probabilities}
\label{SI:sec:ComparisonAmongProbabilities}
For every level $ \ell$, both LPCA and MSM give rise to 3 probability functions\footnote{For the MSM, the symbols on top of $ \mathbf{P}$ refers to different values of the \textit{inner} parameters since its functional form does not vary by construction.}: the \textit{fitted} $ \hat{\mathbf{P}}^{(\ell)} $, \textit{summed} $ \mathbf{P}_{sum}^{(\ell)} $ and the \textit{coarse-grained} $ \mathbf{P}_{cog}^{(\ell)} $. Hence, we produced a cross comparison among them to understand their hallmarks.

Firstly, we will compare for each model how the $ \mathbf{P}_{sum}^{(\ell)} $ relates with $ \mathbf{P}_{cog}^{(\ell)} $, i.e. \autoref{eq:LPCA_Psummed} against \autoref{eq:LPCA_Pcg} and \autoref{eq:MSM_pIJ} against \autoref{eq:MSM_pIJ_cg}.

Secondly,
{
\raggedright
$ \mathbf{P}_{sum}^{(\ell)} $ against $ \hat{\mathbf{P}}^{(\ell)}$, namely \autoref{eq:LPCA_Psummed} against $ \hat{\mathbf{P}}^{(\ell)}_{LPCA} $ and \autoref{eq:MSM_pIJ} against  $ \hat{\mathbf{P}}^{(\ell)}_{MSM} $.
}

The insets, displayed in some figure, are reporting the $ 2D$ histogram of the density of points inside each bin\footnote{We set the number of bins equal to $ 30$ both along the x- and y- axis.}, i.e.
\begin{equation}
    r_{xy} = 2\frac{n_{\textnormal{bin}_{xy}}}{N_2 (N_2 - 1)} \in [0,1]
\end{equation}
where $ n_{\textnormal{bin}_{xy}}$ (``xy'' refers to its center of mass) is the total number of points and $N_2 (N_2 - 1)$ the number of pairs.
Finally, we colored the bins according to $ r_{xy}$ (creating a heatmap) - the bigger the value, the lighter the color.

The missing evaluation of $ \hat{\mathbf{P}}^{(\ell)} $ against $ \mathbf{P}_{cog}^{(\ell)} $ is due to the fact the \textit{coarse-grained} probability spoils the LPCA functional form (see \autoref{sec:LPCA_Renormalization}) whereas $ \mathbf{P}_{cog}^{(\ell)} = \mathbf{P}_{sum}^{(\ell)}$ for the MSM. Therefore, we used it only to check numerically the \autoref{eq:MSM_micro_psumVSpcg} in the first comparison.
    
\subsection{Degree, Average-Nearest Neighbor Degree and Clustering Coefficient}
\label{SI:sec:NetworkMeasurements}
The fundamental topological properties of a network are the \textit{degree}, the \textit{average nearest neighbor degree} (ANND) and the \textit{binary clustering coefficient} (CC) \cite{2004_StatMecNet_Park}. 
Formally, each of such measurements is a function $ Y(\mathbf{A})$ of an $ N \times N$ adjacency matrix representing a graph $ \mathbf{G}$.

Here, we compute them both in the observed network $ \mathbf{A}$ and as expected by the model $ \mathbf{P}(\mathbf{A}|\textbf{X})$. More precisely, the \textit{degree} counts the number of edges that are incident to a node $ i$, i.e.
\begin{equation}
    \label{eq:degree_observed}
    k_{i}(\mathbf{A}) := \sum_{j (\neq i)} a_{ij}
\end{equation}
and its expected value is given by
\begin{equation}
    \langle k_{i} \rangle = \sum_{j (\neq i)} p_{ij}
\end{equation}
where $ \langle \cdot \rangle$ denotes the expected value over the ensemble of graphs sampled from $ \mathbf{P}(\mathbf{A}|\textbf{X})$.
Moving \textit{two-hops} away from $ i$, the \textit{ANND} reports the average degree of the neighbors of the node $ i$, i.e.
\begin{align}
    \label{eq:ANND_observed}
    k^{n n}_{i}(\mathbf{A}) :&= \sum_{j (\neq i)} \frac{a_{ij}k_{j}}{k_{i}} 
    \\ &= \frac{\sum_{j (\neq i), k (\neq j)} a_{ij}a_{jk}}{ \sum_{j (\neq i)} a_{ij}}
\end{align}
whereas its expected value reads
\begin{align}
    \langle k^{n n}_{i} \rangle &:= \langle \frac{\sum_{j (\neq i), k (\neq j)} a_{ij}a_{jk}}{ \sum_{j (\neq i)} a_{ij}} \rangle
    \\ &\approx 1 + \frac{\sum_{j (\neq i), k (\neq j, i)} p_{ij}p_{jk}}{ \sum_{j (\neq i)} p_{ij}}
\end{align}
where in the second passage we took advantage on the first order approximation $ \mathbb{E}[ \frac{X}{Y} ] \approx \frac{\mathbb{E}[X]}{\mathbb{E}[Y]}$  (\textit{delta approximation}) \cite{2011_AnalMax_Squartini}.
Lastly, the \textit{CC} is defined as the ratio among the number of triangles of node $ i$ and its number of wedges, namely 
\begin{align}
    \label{eq:CC_observed}
    c_{i}(\mathbf{A}) 
    &:= \frac{\triangle_i}{\wedge_i} \\
    &= \frac{\sum_{i \neq j \neq k} a_{ij} a_{jk} a_{ki} }{\sum_{j \neq k} a_{ij} a_{ik}}
\end{align}
whereas the expected one is
\begin{align}
    \label{eq:CC_expected}
    \langle c_{i} \rangle &:= \langle \frac{\triangle_i}{\wedge_i} \rangle \\
    \label{eq:CC_expected_approx}
    &\approx \frac{\langle \triangle_i \rangle}{\langle \wedge_i \rangle} \\
    &\approx \frac{\sum_{i \neq j \neq k} p_{ij} p_{jk} p_{ki} }{\sum_{j \neq k} p_{ij} p_{ik}}
\end{align}

\subsection{On the ensemble estimation}
For more complicated measurements, e.g. the variance of the \textit{ANND}, the \textit{delta approximation} won't be valid and one has to estimate them as the average over a \textit{sufficiently large} ensemble $\mathcal{A} := \left\{\mathbf{A}_{s}\right\}_{s \in [1, \mathcal{S}]}$ where $ \mathcal{S}$ is the number of graphs. 
In particular, having optimized the parameters of the model, we can generate unbiased realizations $ \mathcal{A}$ by sampling each $ a_{ij}$ independently with probability $ p_{ij}$ \cite{2011_AnalMax_Squartini,2023_MSNR_Garuccio}.

In the limit of $ \mathcal{S} \to \infty$, the \textit{sampled} average of any measure $ Y_i$ meets its \textit{analytical} estimations $ \langle Y_i \rangle$ \cite{2023_MSNR_Garuccio}, i.e.
\begin{align}
    \label{eq:ensemble_average_analytical}
    \bar{Y_i} 
    :&= \frac{1}{\left\vert \mathcal{A}_{N}\right\vert}
    \sum_{\mathbf{\hat{A}} \in \mathcal{A}}  Y_i(\mathbf{\hat{A}})
    \\[1ex]
    \to 
    \langle Y_i \rangle &= \sum_{\mathbf{B} \in \mathcal{A}_{N}} \mathbf{P}(\mathbf{B}|\textbf{X}) Y_i(\mathbf{B})
\end{align}
where $\mathbf{B} \in \mathcal{A}_{N}$ is a matrix drawn from the set of the undirected binary graphs $ \mathcal{A}_{N}$ of $ N$ nodes. 

Lastly, to estimate the uncertainty of the model over the sampled realizations, we calculated the $ 97.5$-th ($ 2.5$-th) percentile of $ Y_i(\mathcal{A})$ calculated with \textit{linear approximation} (see \cite{2010_NumPy_Harris}). These values are seen as upper and lower bounds of the \textit{dispersion intervals} $ \Delta_c( \langle Y_i \rangle )$ \cite{2023_RecEcon_DiVece} which contains $c = 95 \%$ of the measurements $ Y_i(\mathcal{A}) := \left\{Y_i(\mathbf{A}_{s})\right\}_{s \in [1, \mathcal{S}]}$ over the sampled graphs. Note that in the whole procedure we arbitrarily fix the percentage of ``dispersion'' to $c = 95 \%$ \cite{2023_RecEcon_DiVece}, but other values are also allowed.

\subsection{Triangle Density}
Inspired by \cite{2020_LPCA_Chanpuriya}, we computed the \textit{expected} number of triangles for every model at disposal\footnote{In this essay, our objective was to model \textit{probabilistically} the observed network rather than describing it \textit{exactly}, namely the limit where $ p_{ij} \equiv a_{ij} \, \forall i > j$.}.
Specifically, the expected density of triangles at a certain level $ \ell$ is defined as
\begin{equation}
    \label{eq:observed_TriDens}
    \tau^{(\ell)}(c) := \frac{ \triangle(\mathbf{G}^{(\ell)}_{k_{i_{\ell}} \geq c})}{2N_{\ell}} = \frac{ \sum_{i \neq j \neq k} g_{ij} g_{jk} g_{ki} }{2N_{\ell}}
\end{equation}
where $\triangle( \mathbf{G}^{(\ell)}_{k_{i_{\ell}} \geq c})$ is the number of observed triangles (see \autoref{eq:CC_observed}) calculated on the subgraph $ \mathbf{G}^{(\ell)}_{k_{i_{\ell}} \geq c}$ composed by the nodes $I_c := {\left\{i_\ell : k_{i_\ell} \geq c\right\}}$ with degree lower (or equal) than a threshold $ c$ \cite{2020_impossibility_of_low_rank_red_Seshandhri, 2020_LPCA_Chanpuriya}.
Its expected value reads
\begin{equation}
    \label{eq:expected_TriDens}
    \langle \tau^{(\ell)}(c) \rangle = \frac{ \langle \triangle(\mathbf{G}^{(\ell)}_{k_{i_{\ell}} \geq c}) \rangle}{2N_{\ell}} \approx \frac{ \sum_{i < j < k} \tilde{p}_{ij} \tilde{p}_{jk} \tilde{p}_{ki} }{N_{\ell}}
\end{equation}
where $ i \in I_c, j \in I_c, k \in I_c$ and the probabilities $ \tilde{p}_{ij}$ refers to the summed model $ \mathbf{P}_{sum}^{(\ell)} $.

\subsection{Reconstruction Accuracy}
In order to have a cross-comparison among all the levels and models, we exploited the \textit{reconstruction accuracy} \cite{2023_RecEcon_DiVece}. This measure is defined as the fraction of times an observed statistics $ Y_{i}$ falls within the \textit{dispersion interval} $ \Delta_c( \langle Y_i \rangle )$ (see \autoref{SI:sec:NetworkMeasurements}). More formally, the reconstruction accuracy at level $ \ell$ for the statistics $ Y$ is defined as
\begin{equation}
    \label{eq:RecAcc}
    RA^{\ell}_s := \frac{1}{N_{\ell}} \sum_{i = 0}^{N_{\ell} - 1} \mathbb{I}\left\{Y_i \in \Delta( \langle Y_i \rangle )\right\}
\end{equation}
where $ \mathbb{I}$ is the indicator function\footnote{Further refinements are possible, but we stick with this definition for the sake of simplicity.}. 
Roughly, it counts the frequency at which the sampled ensemble includes the observed statistics. If all the observed statistics, e.g. degrees, were included in the interval, the accuracy would be $1$, whereas the accuracy would be $0$ if none of them were included.

\subsection{Rescaled ROC and PR Curves}
The LPCA and MSM could be seen as \textit{binary classifiers} that predict the presence of a link between two nodes. For this reason, we evaluated them also for the common metrics used in the \textit{Machine Learning} field: the \textit{expected} confusion matrix, the Receiver Operating Characteristic (ROC) and the Precision-Recall (PR) curves \cite{2020_ROC_PR_creation_Cook,2006_ROC_PR_skewed_data_Davis}.
Firstly, the \textit{expected} confusion matrix is a $ 2 \times 2$ matrix that reports the expected value of True Positives (TP), i.e. $ \langle TP \rangle := \sum_{i < j} a_{ij} p_{ij}$, False Positives (FP), i.e. $ \langle FP \rangle := \sum_{i < j} (1-a_{ij}) p_{ij}$, True Negatives (TN), i.e. $ \langle TN \rangle := \sum_{i < j} \left(1-a_{ij}\right) \left((1-p_{ij})\right)$, and False Negatives (FN), i.e. $ \langle FN \rangle := \sum_{i < j} a_{ij} \left(1 - p_{ij}\right)$ \cite{2017_ECAPM_Squartini}. By combining these scores, one recovers the True Positive Rate ($TPR$), the False Positive Rate ($FPR$) and the Positive Predictive Value ($PPV$) \cite{2006_ROC_PR_skewed_data_Davis}, namely
\begin{align}
    TPR :&= \frac{ TP }{Pos}
    = \frac{\sum_{i < j} a_{ij} p_{ij}}{ L }
    \\[1ex] FPR  :&= \frac{ FP }{Neg} = \frac{\sum_{i < j} (1 - a_{ij}) p_{ij}}{ \binom{N}{2} - L }
    \\[1ex] PPV  :&= \frac{TP}{P P} = \frac{\sum_{i < j} a_{ij} p_{ij}}{ \sum_{i < j} p_{ij}}
\end{align}
{
\raggedright
where $ Pos := \sum_{i < j} a_{ij} = L$, $Neg := \sum_{i < j} (1 - a_{ij}) = \binom{N}{2} - L$.
} Fixing a threshold $ \epsilon$, the thresholded adjacency matrix $ \hat{A}(\epsilon)$ is obtained by setting $ \hat{a}(\epsilon) = 1$ if $ p_{ij} \leq \epsilon$ and $ \hat{a}(\epsilon) = 0$ otherwise. As a result, the thresholded scores are defined as
\begin{align}
    TPR(\epsilon) :&= \frac{ TP(\epsilon) }{Pos}
    = \frac{\sum_{i < j} a_{ij} \hat{a}_{ij}}{ L }
    \\[1ex] FPR(\epsilon)  :&= \frac{ FP(\epsilon) }{Neg} = \frac{\sum_{i < j} (1 - a_{ij}) \hat{a}_{ij}}{ \binom{N}{2} - L }
    \\[1ex] PPV(\epsilon)  :&= \frac{TP(\epsilon)}{P P(\epsilon)} = \frac{\sum_{i < j} a_{ij} \hat{a}_{ij}}{ \sum_{i < j} \hat{a}_{ij}}.
\end{align}
Since the choice of a specific threshold is arbitrary, one common practice in Machine-Learning is to compute the $ TPR(\epsilon), FPR(\epsilon), P PV(\epsilon)$ for all the possible thresholds, namely $ \epsilon \in [0,\infty]$ \cite{2020_ROC_PR_creation_Cook}. Furthermore, the ROC and PR curves are obtained by plotting the $TPR(\epsilon)$ against the $FPR(\epsilon)$, and the $TPR(\epsilon)$ against the $PPV(\epsilon)$ (see \autoref{fig:ING_ROCs}), respectively. Interestingly, if $ \epsilon \to \infty$, then $TPR(\epsilon \to \infty) = FPR(\epsilon \to \infty) = 0, P PV(\epsilon \to \infty) := 1$, whereas if $ \epsilon = 0$, $TPR(\epsilon = 0) = FPR(\epsilon = 0) = 1, P PV(\epsilon = 0) = p := \frac{Pos}{Pos + Neg} = \frac{L}{\binom{N}{2}}$.

Since the last point of the $P PV$ is the density, the ``naive classifier'' (NC) - always predicting the majority class - can have a high AUC-PR (On the ROC curve the NC always lies on the diagonal). For this reason, one can rescale the Area Under the Curve (AUC) of each model with respect to the AUC predicted by the NC. Specifically, one defines
\begin{align}
    AUC-ROC_{norm} &= \frac{AUC_{ROC} - 0.5}{0.5}
    \\ AUC-PR_{norm} &= \frac{AUC_{PR} - p}{1 - p}.
\end{align}
Therefore, the perfect classifier still have $ \textnormal{ AUC-ROC } = \textnormal{ AUC-PR } = 1$ but the random one $ \textnormal{ AUC-ROC } = \textnormal{ AUC-PR } = 0$. The \textit{new} AUCs can be negative, as a signal of a worse performance than the random classifier. Furthermore, this novelty does not spoil the ranking of the \textit{summed} models which was the ultimate objective of the AUC scores.

\section{Principled Embedding Dimension via Information Criteria}
\begin{table*}[t]
    \label{SI:tab:ING_norm_AIC_BIC}
    \centering
    \begin{minipage}{0.48\textwidth}
        \centering
        \textbf{(a) AICs by model class for ION}
        \vspace{1em} 
        \begin{tabular}{c|c|c|c|c|c}
            \toprule
            Model & Dim &  Level 0 & Level 1 & Level 2 & Level 3\\
            \midrule
            \multirow[c]{2}{*}{LPCA} & (1, 1) & \red{0.5764} & \red{0.6661} & \red{0.5448} & \green{\textbf{0.3733}} \\
            & (8, 8) & \green{\textbf{0.5072}} & \green{\textbf{0.5779}} & \green{\textbf{0.4025}} & \red{2.6648e+17} \\
            \midrule
            \multirow[c]{12}{*}{MSM} & 1 & \red{0.5784} & \red{0.6658} & \red{0.5408} & 0.4097 \\
            & 2 & 0.5516 & 0.6341 & 0.5129 & 0.3841 \\
            & 3 & 0.5434 & 0.6242 & 0.5012 & \green{\textbf{0.3791}} \\
            & 4 & 0.5378 & 0.6169 & 0.4984 & 0.3831 \\
            & 5 & 0.5336 & 0.6127 & 0.4985 & 0.4046 \\
            & 6 & 0.5311 & 0.6088 & 0.4979 & 0.4232 \\
            & 7 & 0.5304 & 0.6074 & \green{\textbf{0.4968}} & 0.4574 \\
            & 8 & \green{\textbf{0.5287}} & \green{\textbf{0.6052}} & 0.4976 & 0.5118 \\
            & 9 & 0.53 & 0.6073 & 0.498 & 0.5588 \\
            & 10 & 0.5305 & 0.6064 & 0.5002 & 0.6091 \\
            & 11 & 0.5311 & 0.6105 & 0.5117 & 0.6671 \\
            & 16 & 0.5365 & 0.6226 & 0.5328 & \red{0.9697} \\
            \bottomrule
        \end{tabular}
    \end{minipage}
    \hfill
    \begin{minipage}{0.48\textwidth}
        \centering
        \textbf{(a) BICs by model class for ION}
        \vspace{1em} 
        \begin{tabular}{c|c|c|c|c|c}
            \toprule
            Model & Dim &  Level 0 & Level 1 & Level 2 & Level 3\\
            \midrule
            \multirow[c]{2}{*}{LPCA} & (1, 1) & \green{\textbf{0.622}} & \green{\textbf{0.7295}} & \green{\textbf{0.6605}} & \green{\textbf{0.7189}} \\
            & (8, 8) & \red{0.8718} & \red{1.0856} & \red{1.3276} & \red{ 2.6648e+17} \\
            \midrule
            \multirow[c]{12}{*}{MSM} & 1 & 0.6012 & \green{\textbf{0.6975}} & \green{\textbf{0.5987}} & \green{\textbf{0.5825}} \\
            & 2 & \green{\textbf{0.5972}} & 0.6976 & 0.6285 & 0.7296 \\
            & 3 & 0.6117 & 0.7194 & 0.6746 & 0.8974 \\
            & 4 & 0.6289 & 0.7439 & 0.7297 & 1.0741 \\
            & 5 & 0.6476 & 0.7714 & 0.7877 & 1.2684 \\
            & 6 & 0.6678 & 0.7992 & 0.8449 & 1.4597 \\
            & 7 & 0.69 & 0.8295 & 0.9015 & 1.6667 \\
            & 8 & 0.7111 & 0.8591 & 0.9602 & 1.8939 \\
            & 9 & 0.7352 & 0.8929 & 1.0184 & 2.1136 \\
            & 10 & 0.7584 & 0.9237 & 1.0785 & 2.3367 \\
            & 11 & 0.7818 & 0.9595 & 1.1477 & 2.5675 \\
            & 16 & \red{ 0.9011} & \red{ 1.1303} & \red{ 1.4579} & \red{3.7339} \\
            \bottomrule
        \end{tabular}
    \end{minipage}
    \caption{Normalized AIC and BIC scores tables for the models LPCA and MSM at different levels for the ION. The best scores are highlighted in green whereas the worst in red.}
    \label{tab:ING_norm_AIC_BIC}
\end{table*}

\begin{table}[t]
    \label{SI:tab:WTW_norm_AIC_BIC}
    \centering
    \begin{minipage}{\textwidth}
        \centering
        \begin{tabular}{c|c|c|c|c|c|c|c}
            \multicolumn{8}{c}{\textbf{(a) AICs by model class for WTW}} \\
            \toprule
            Model & Dim &  Level 0 & Level 1 & Level 2 & Level 3 & Level 4 & Level 5 \\
            \midrule
            \multirow[c]{2}{*}{LPCA} & (1, 1) & \red{0.6109} & \red{0.5636} & \red{0.572} & \green{\textbf{0.5764}} & \green{\textbf{0.5584}} & \green{\textbf{0.444}} \\
            & (8, 8) & \green{\textbf{0.4536}} & \green{\textbf{0.4309}} & \green{\textbf{0.5289}} & \red{0.7033} & \red{1.0492} & \red{2.0645} \\
            \midrule
            \multirow[c]{12}{*}{MSM} & 1 & \red{0.6146} & 0.5637 & 0.5656 & 0.5649 & 0.5387 & 0.3991 \\
            & 2 & 0.5807 & 0.5332 & 0.5389 & \green{\textbf{0.5285}} & \green{\textbf{0.5016}} & \green{\textbf{0.3665}} \\
            & 3 & 0.5684 & 0.5483 & 0.5528 & 0.5519 & 0.5446 & 0.4753 \\
            & 4 & 0.5567 & 0.5366 & 0.5522 & 0.5642 & 0.5645 & 0.4876 \\
            & 5 & 0.5628 & 0.5295 & 0.5511 & 0.567 & 0.5825 & 0.5279 \\
            & 6 & 0.5674 & 0.5356 & 0.5603 & 0.5946 & 0.6059 & 0.607 \\
            & 7 & 0.5707 & 0.5424 & 0.5679 & 0.6174 & 0.6557 & 0.7085 \\
            & 8 & \green{\textbf{0.5181}} & \green{\textbf{0.4733}} & \green{\textbf{0.4959}} & 0.5363 & 0.5764 & 0.7742 \\
            & 9 & 0.5809 & 0.5649 & 0.6051 & 0.6526 & 0.7167 & 0.8852 \\
            & 10 & 0.5923 & 0.5765 & 0.6198 & 0.6819 & 0.7609 & 0.994 \\
            & 11 & 0.6017 & \red{0.5949} & \red{0.6362} & 0.7114 & 0.7962 & 1.0688 \\
            & 16 & 0.557 & 0.5571 & 0.631 & \red{0.7292} & \red{0.9881} & \red{1.5484} \\
        \bottomrule
        \end{tabular}
        \vspace{5ex}
    \end{minipage}
    \begin{minipage}{\textwidth}
        \begin{tabular}{c|c|c|c|c|c|c|c}
            \multicolumn{8}{c}{\textbf{(b) BICs by model class for WTW}} \\
            \toprule
            Model & Dim &  Level 0 & Level 1 & Level 2 & Level 3 & Level 4 & Level 5 \\
            \midrule
            \multirow[c]{2}{*}{LPCA} & (1, 1) & \green{\textbf{0.7813}} & \green{\textbf{0.7583}} & \green{\textbf{0.8004}} & \green{\textbf{0.8551}} & \green{\textbf{0.922}} & \green{\textbf{0.9868}} \\
            & (8, 8) & \red{1.8166} & \red{1.9881} & \red{2.3555} & \red{2.9326} & \red{3.958} & \red{6.4068} \\
            \multirow[c]{12}{*}{MSM} & 1 & \green{\textbf{0.6974}} & \green{\textbf{0.6578}} & \green{\textbf{0.677}} & \green{\textbf{0.6982}} & \green{\textbf{0.7088}} & \green{\textbf{0.6026}} \\
            & 2 & 0.7463 & 0.7215 & 0.7616 & 0.795 & 0.8417 & 0.7736 \\
            & 3 & 0.8169 & 0.8306 & 0.8869 & 0.9518 & 1.0548 & 1.0859 \\
            & 4 & 0.8881 & 0.9131 & 0.9976 & 1.0972 & 1.2448 & 1.3018 \\
            & 5 & 0.977 & 1.0001 & 1.1079 & 1.2334 & 1.4329 & 1.5456 \\
            & 6 & 1.0645 & 1.1003 & 1.2284 & 1.3942 & 1.6264 & 1.8282 \\
            & 7 & 1.1506 & 1.2013 & 1.3473 & 1.5503 & 1.8462 & 2.1333 \\
            & 8 & 1.1809 & 1.2263 & 1.3867 & 1.6025 & 1.9369 & 2.4025 \\
            & 9 & 1.3265 & 1.4121 & 1.6073 & 1.8521 & 2.2474 & 2.7171 \\
            & 10 & 1.4208 & 1.5177 & 1.7333 & 2.0147 & 2.4616 & 3.0295 \\
            & 11 & 1.513 & 1.6303 & 1.8611 & 2.1774 & 2.6669 & 3.3078 \\
            & 16 & \red{1.8825} & \red{2.0631} & \red{2.4126} & \red{2.8616} & \red{3.7093} & \red{4.8051} \\
        \bottomrule
        \end{tabular}
    \end{minipage}
    \caption{Normalized AIC and BIC scores tables for the models LPCA and MSM at different levels for the WTW. The best scores are highlighted in green whereas the worst in red.}
    \label{tab:WTW_norm_AIC_BIC}
\end{table}

To determine the ``best'' embedding dimension for LPCA and MSM, we used the Akaike Information Criterion (AIC) and the Bayesian Information Criterion (BIC) \cite{2019_AIC_Cavanaugh}, \cite{2023_BIC_Zhang}. These scores are defined as
\begin{align}
    \textnormal{AIC} &:= 2K - 2\ln(\mathcal{L}) \\
    \textnormal{BIC} &:= K\ln(n) - 2\ln(\mathcal{L})
    \label{eq:AIC_BIC}
\end{align}
where $ K$ is the number of parameters, $ n := \binom{N}{2}$ the number of observations and $ \mathcal{L}$ the likelihood of the model. They encode the trade-off between the goodness-of-fit $-\ln(\mathcal{L})$ and the complexity of the model. Therefore, the ``best model'' is the one with the \textit{minimum} AIC or BIC; which one of the two remains a debated choice: the AIC is asymptotically equivalent to the Kullback-Leibler divergence among the \textit{generating} model and a \textit{candidate} one \cite{2019_AIC_Cavanaugh} whereas the BIC to the Description Length (DL) \cite{2023_BIC_Zhang}. As the DL is the only one embodying the ``trade-off'' paradigm, we decided to select the \textit{minimum} BIC criterion. In \autoref{eq:AIC_BIC}, the scores are not comparable across scale, therefore, we \textit{normalized} the AIC and BIC scores as
\begin{align}
    \textnormal{AIC}_{\textnormal{norm}, \ell} 
    &:= \frac{2}{n_{\ell}} K_{\ell} - \frac{2\ln(\mathcal{L})}{n_{\ell}} \\
    & = 4\left[\frac{D}{N_{\ell} - 1} - \frac{\ln(\mathcal{L})}{N_{\ell} \left(N_{\ell} - 1\right)}\right] \\
    \textnormal{BIC}_{\textnormal{norm}, \ell} 
    & := \frac{\ln(n_{\ell})}{n_{\ell}} K_{\ell} - \frac{2 \ln(\mathcal{L})}{n_{\ell}} \\
    & \approx \frac{\ln(N_{\ell}) + \ln(N_{\ell} - 1)}{N_{\ell} - 1} D - 4\frac{ \ln(\mathcal{L})}{N_{\ell}(N_{\ell} - 1)}
\end{align}
where $ K_{\ell} = N_{\ell}D$ and $ n_{\ell} := \binom{N_{\ell}}{2}$. This doesn't affect the ranking, but it provides the AIC and BIC \textit{per pair}.

The results are summarized in \autoref{tab:ING_norm_AIC_BIC} for the ION and \autoref{tab:WTW_norm_AIC_BIC} for the WTW where the best scores (minimum) are highlighted in green whereas the worst (maximum) in red. The comparison is provided among LPCA and MSM as they have a different functional forms especially in the combination of the parameters. We have considered only two levels for LPCA as the benchmark for lowest $ D$ and ``maximum'' $ D$ with respect to our computational facility. On the other hand, we spanned more dimensions for MSM to provide an extensive description of the model performances. Recall that, by increasing the level of coarse-graining, the network tends to be \textit{less complex}: the \autoref{eq:cg_rule_nextl} likely densifies the network implying that the nodes will have more similar roles. Therefore, the \textit{ideal} dimension $ D$ \textit{decreases}. Lastly, the \textit{average} BIC score is calculated to provide a \textit{global} view of the model performances. Thus, the best model is the one with the lowest \textit{average} BIC score across levels. Overall, the best models are \textit{LPCA-(1,1)} and \textit{MSM-1} as BIC penalizes more than AIC the complexity of the model. However, in the following we will display the behavior also of the $ D = 2,8,16$ for MSM and $ D = (8,8)$  to assess the model performances at higher dimensions.

\clearpage
\section{World Trade Web results}

In this section, we report the same results presented in the main text, but for the World Trade Web. The conclusions are the same as for the ING network.

\begin{figure*}[t]
    \centering
    \subfloat[Level 0, 4 - Summed / Renormalized VS Coarse-Grained\label{fig:WTW_ScaleInv_Prop}]{
        \includegraphics[width = .5\linewidth]{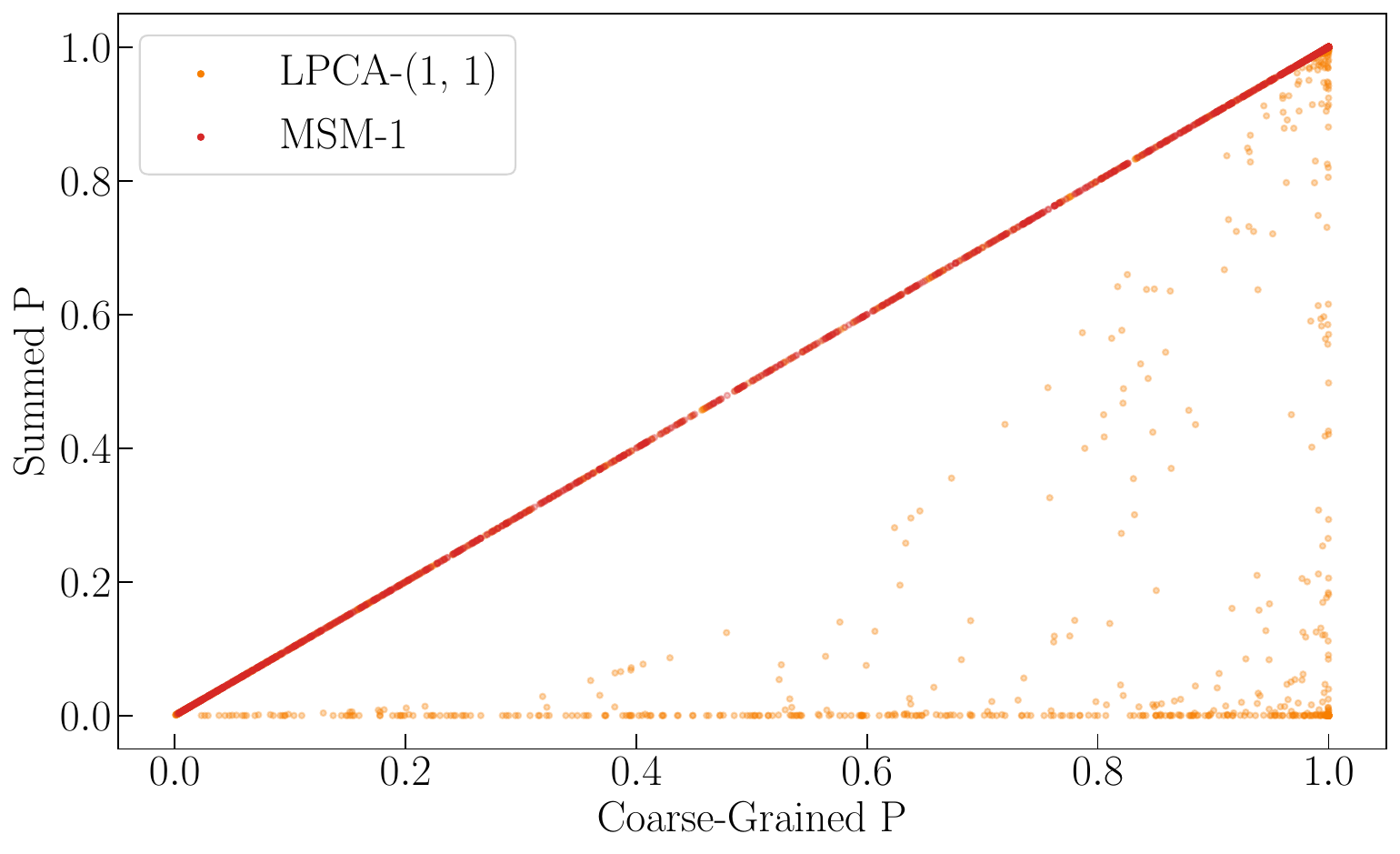}
    }
    \vspace{3ex}
    \subfloat[Level 0, 4 - Binary Clustering Coefficient\label{fig:WTW_bcc_reconstruction}]{
        \includegraphics[width = .8\linewidth]{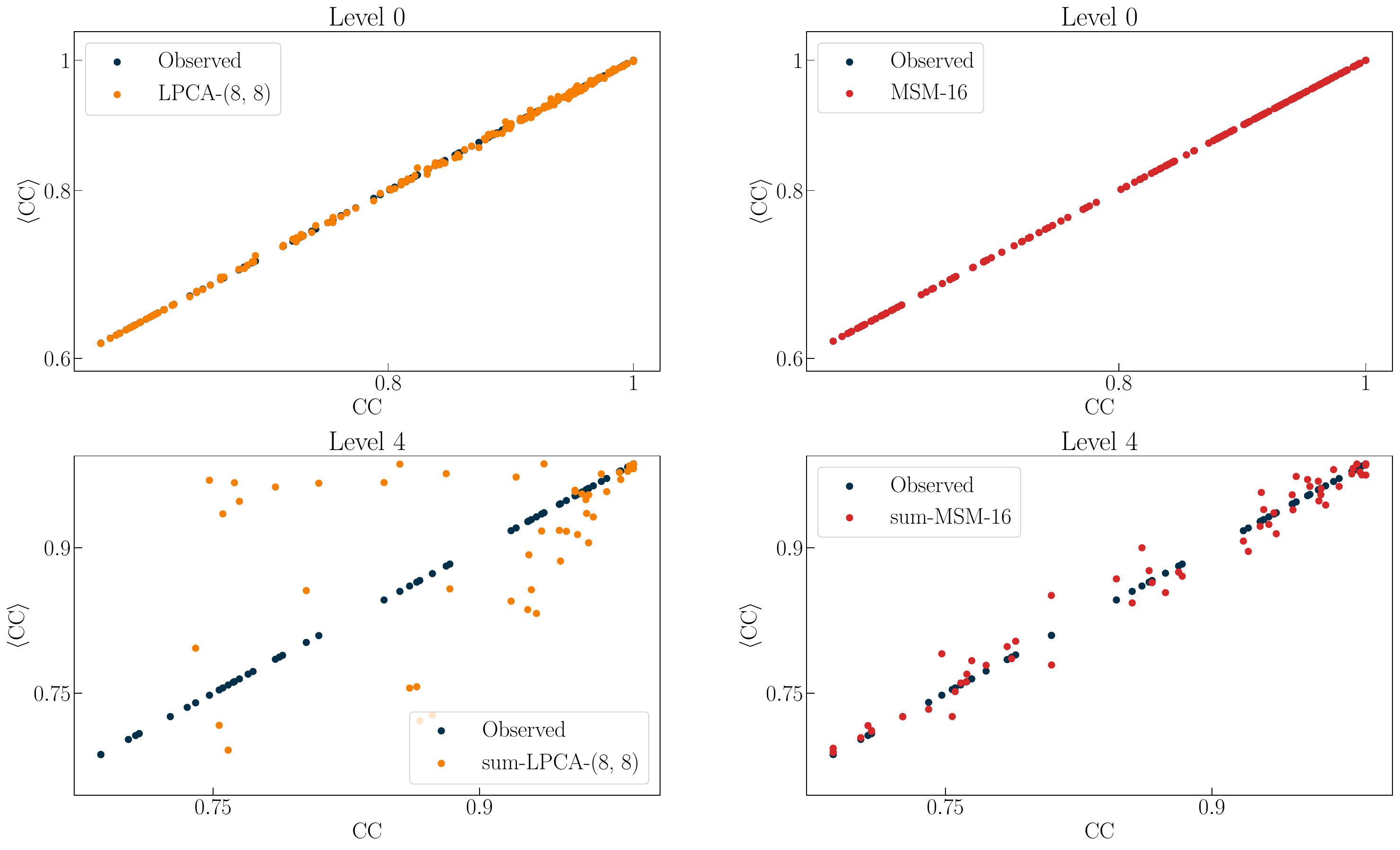}
    }
    \caption{
    \textit{Upper Plot:} This figure illustrates the numerical evaluation of the two sides of the scale-invariant equation at level $\ell = 2$ for the $LPCA$ and $MSM$ models. More precisely, the left-hand side (LHS) is represented on the y-axis while the right-hand side (RHS) on the x-axis.
    \textit{Lower Plot:} Cross comparison of LPCA-(8,8) and MSM-16 in predicting the clustering coefficient (CC) (see the main text).
    The upper panel reports the expected clustering coefficient at level 0, while the lower panel depicts its corresponding values at level 2. The first column refers to the LPCA-(8,8), whereas the second to the MSM-16.
    } 
    \label{fig:WTW_MSProb_CC}
\end{figure*}

\begin{figure*}[t]
    \centering
    \subfloat[Level 4 - LPCA-(8,8) - Summed, Fitted, Observed Network Measurements\label{fig:WTW_lev4_NetMeas_LPCA_88}]{
        \includegraphics[width=.48\linewidth]{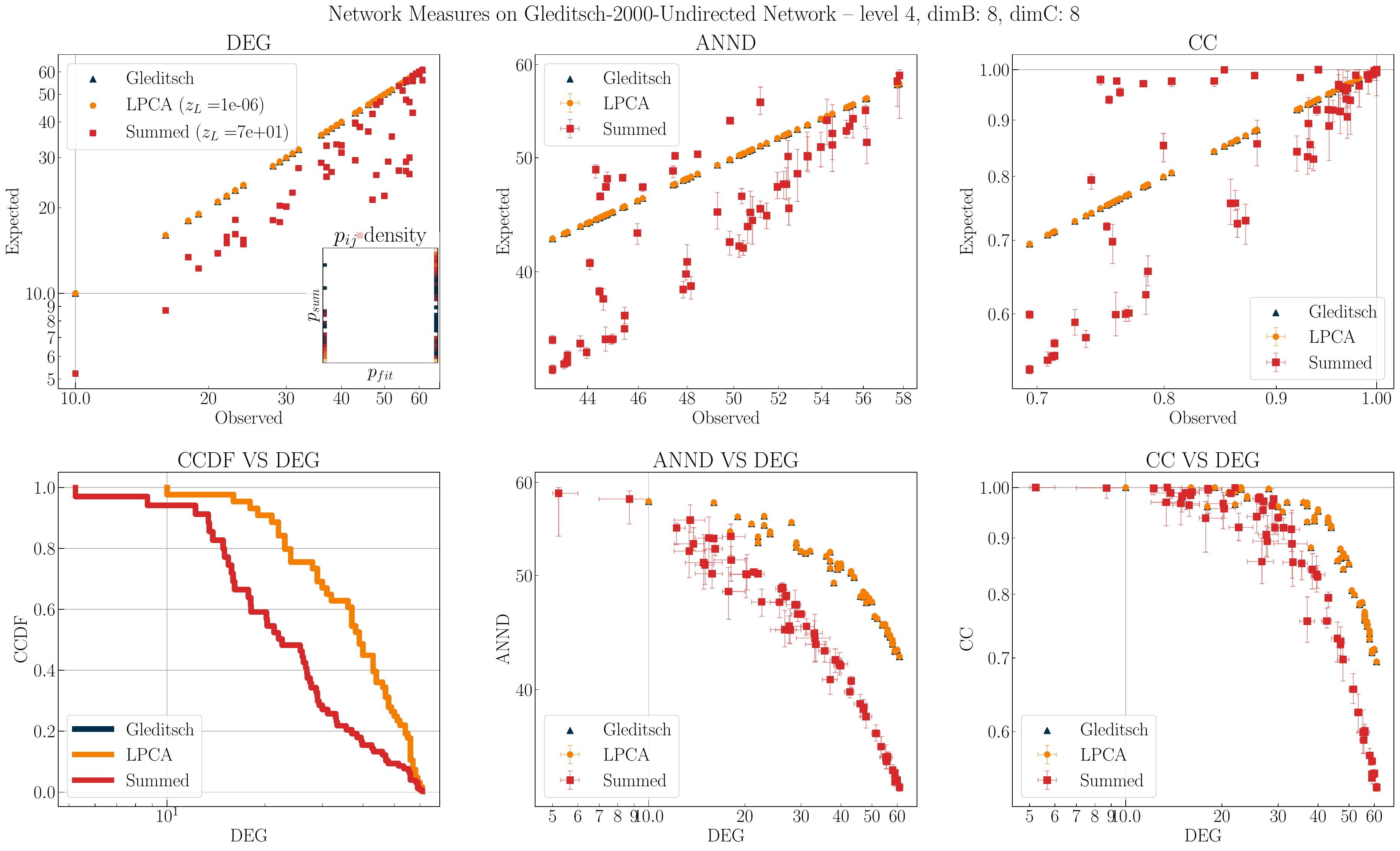}
    }
    \hfill
    \subfloat[Level 4 - MSM-16 - Summed, Fitted, Observed Network Measurements\label{fig:WTW_lev4_NetMeas_maxlMSM_16}]{
        \includegraphics[width=.48\linewidth]{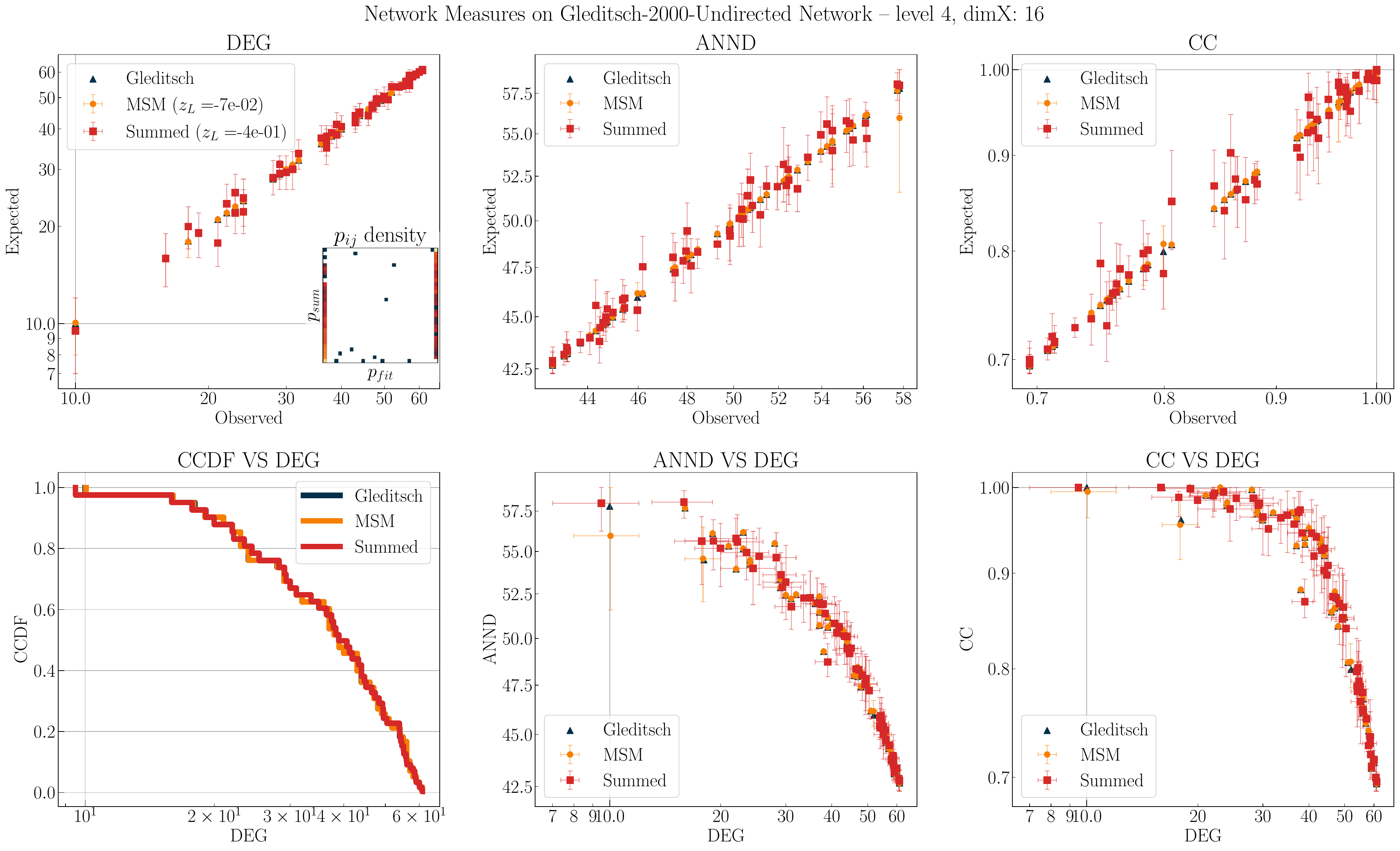}
    }
    \caption{The plots display the predicted network measurements for the WTW dataset at level 4 according to LPCA-(8,8) (left plots) and the MSM-16 (right plot). 
    In each of the upper panels, the x-axis represents the observed measurements, whereas the y-axis shows the expected ones. From left to right, the plots illustrates the degree (DEG), the average-nearest-neighbor degree (ANND), and the clustering coefficient (CC). The scatter plot comparing $ \mathbf{P}_{sum}^{(2)} $ with $ \hat{\mathbf{P}}^{(2)}$ is reported in the inset. 
    The lower panels show the behavior of the network measurements as the degrees increase. The observed values are colored in blue, while the one calculated using the fitted $ \hat{\mathbf{P}}^{(2)}$ model in red or the summed $ \mathbf{P}_{sum}^{(2)}$ model in orange. Additionally, the z-score of the predicted number of links is indicated in the legend of the upper-left plot.
    }
    \label{fig:WTW_NetMeas_DispInt}
\end{figure*}

\begin{figure*}[tbp]
    \centering
    \subfloat[Level 4 - LPCA-(8,8) - Summed, Fitted ROC and PR curves\label{fig:WTW_lev4_ROCs_LPCA_88}]{
        \includegraphics[width=.48\linewidth]{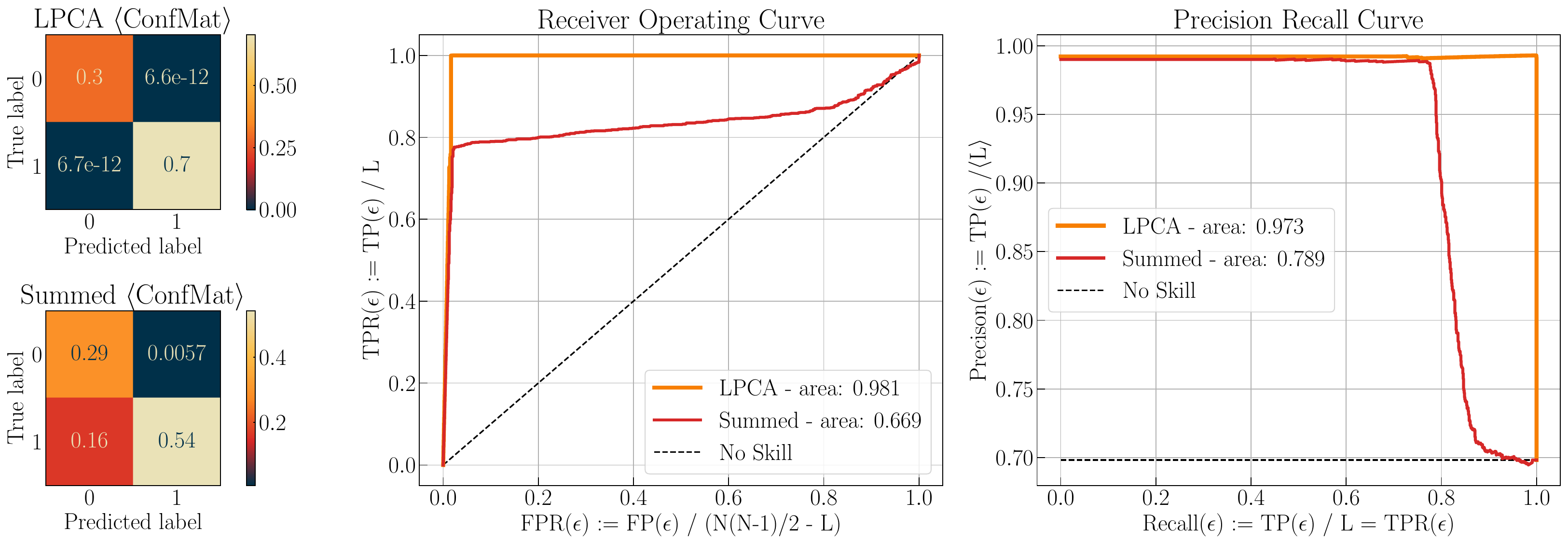}
    }
    \hfill
    \subfloat[Level 4 - MSM-16 - Summed, Fitted ROC and PR curves\label{fig:WTW_lev4_ROCs_maxlMSM_16}]{
        \includegraphics[width=.48\linewidth]{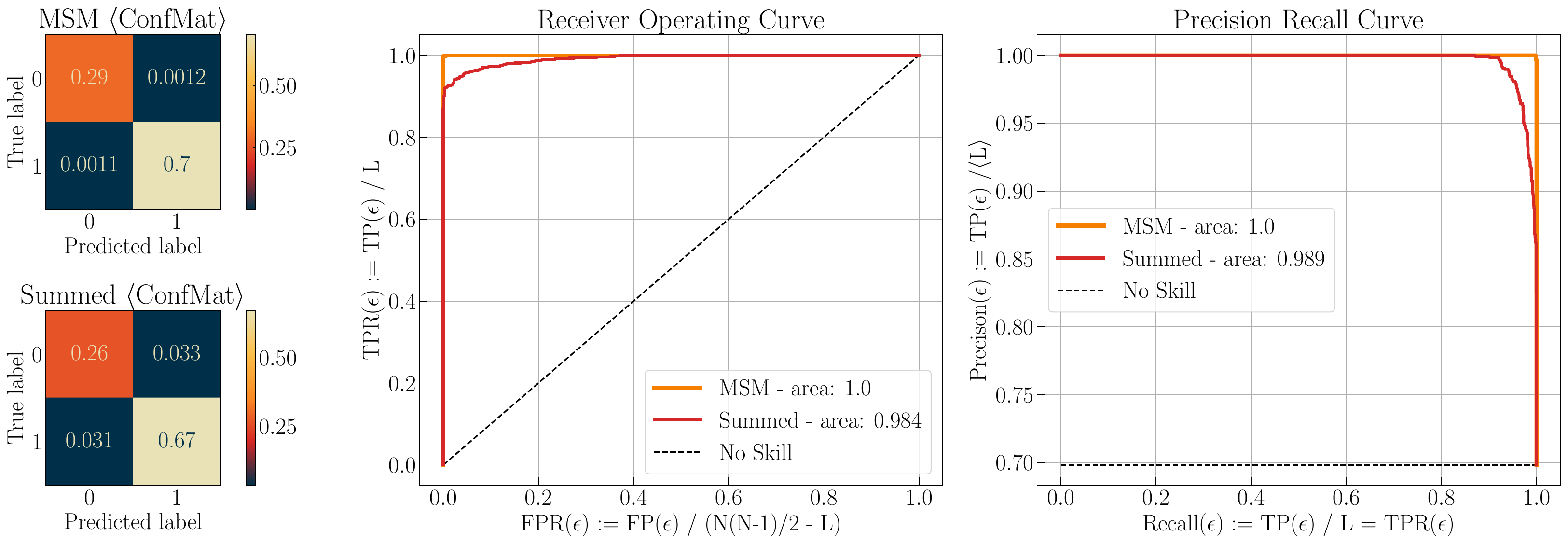}
    }
    \caption{The graphs show the Confusion matrices, ROC and PR curves for the ION dataset at level 2 obtained by means of the LPCA-(8,8) (upper plots) and the MSM-16 (lower plot).
    On the left most side, one may display the two confusion matrices for the fitted $ \hat{\mathbf{P}}^{(2)}$ (upper) and the summed $ \mathbf{P}_{sum}^{(2)} $. The middle plot reports the Receiver-Operator Curve, while the right-most plot depicts the Precision Recall curve. As in the previous plot, the two curves are associated with the fitted (red) model and the summed (orange) one.} 
    \label{fig:WTW_ROCs}
\end{figure*}

\begin{figure*}[t]
    \centering
    \subfloat[All level Reconstruction Accuracy\label{fig:WTW_rec_acc}]{%
    \includegraphics[height=.28\linewidth]{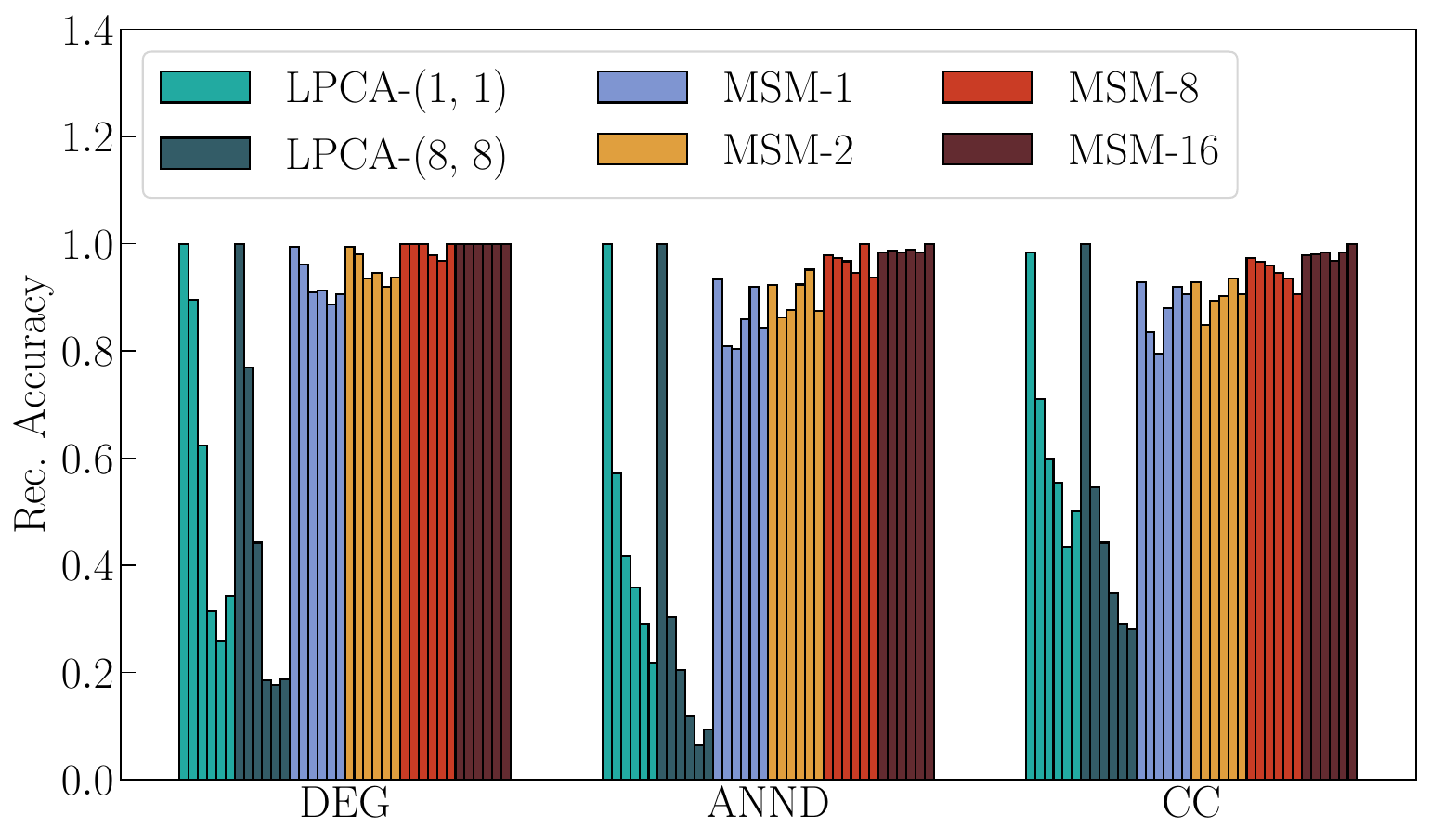}
    }
    \hfill
    \centering
    \subfloat[AUC-ROC and AUC-PR curves through levels \label{fig:WTW_auc_roc_prc}]{
        \includegraphics[height=.28\linewidth]{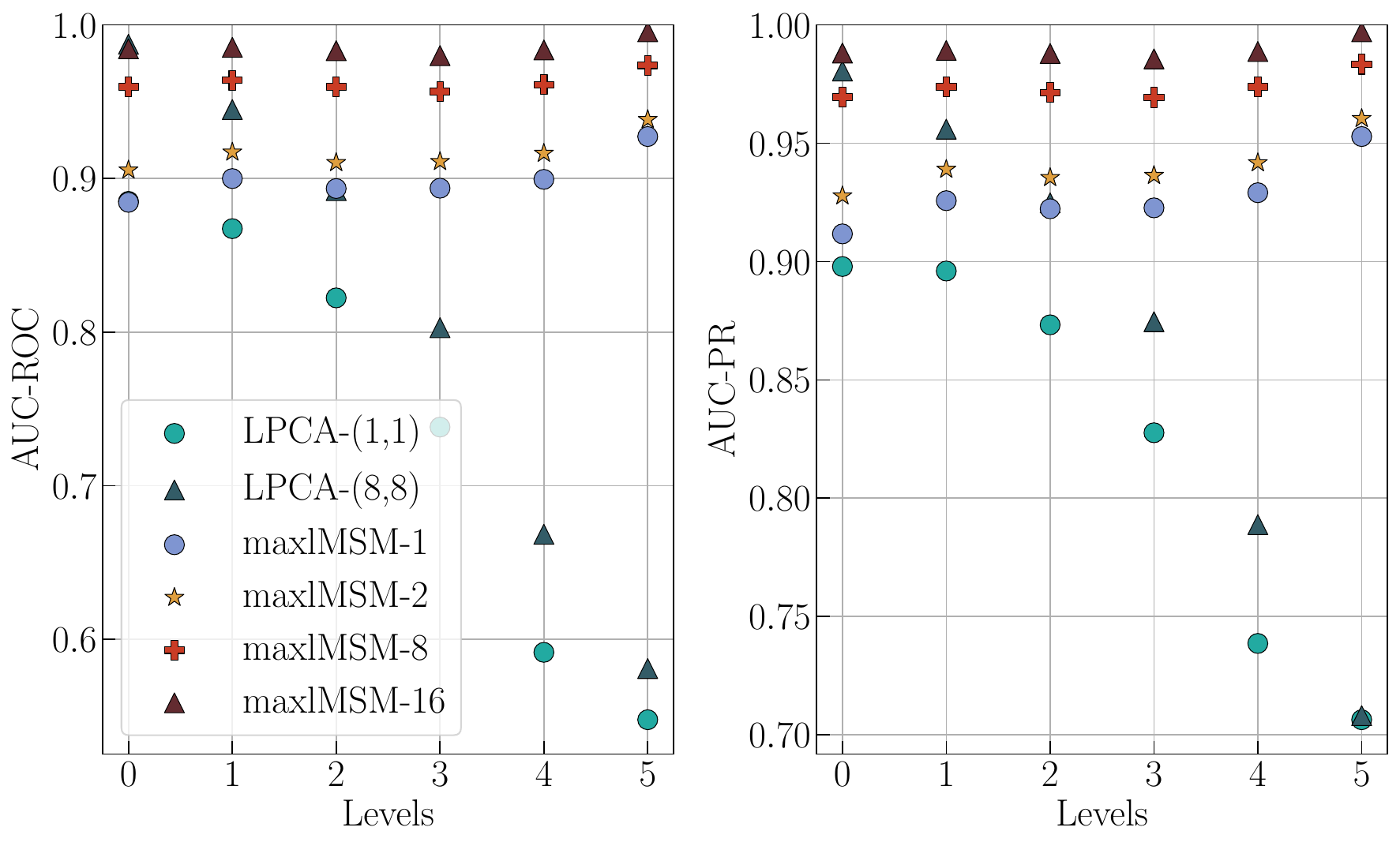}
        }
        \caption{\textit{Left:} Reconstruction Accuracy (y-axis) by model, level and network statistics.
        \textit{Right}: Area Under the ROC and PR curves (y-axis) for the summed models as the scales increase (x-axis) or, equivalently, the numbers of nodes diminish.
        }
        \label{fig:WTW_auc_rec_acc}
\end{figure*}

\begin{figure*}[ht]
    \centering
    \subfloat[Level 0]{
        \includegraphics[width=0.45\linewidth]{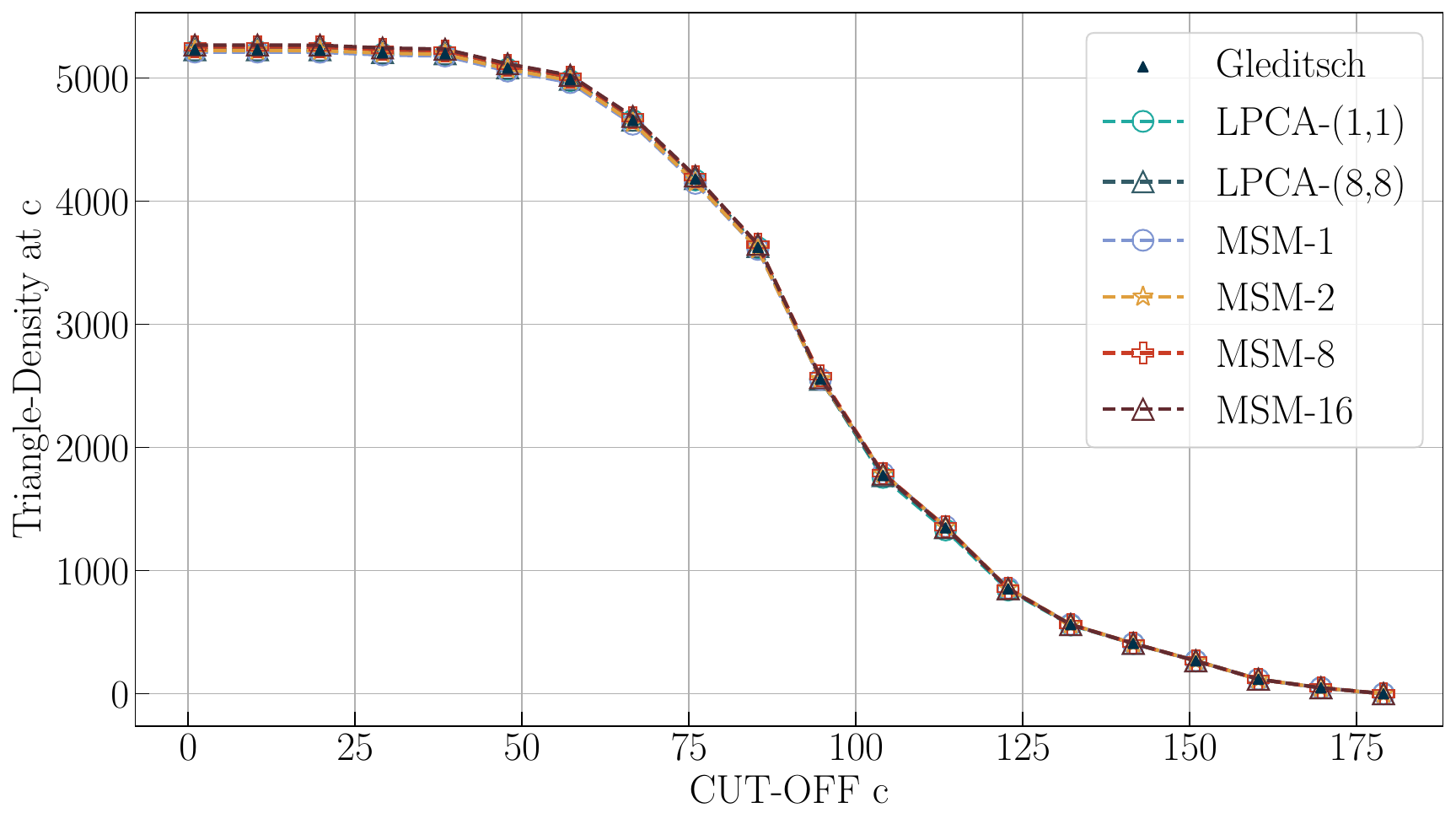}
        \label{fig:WTW_RCTriangles_level0}
    }
    \subfloat[Level 4]{
        \includegraphics[width=0.45\linewidth]{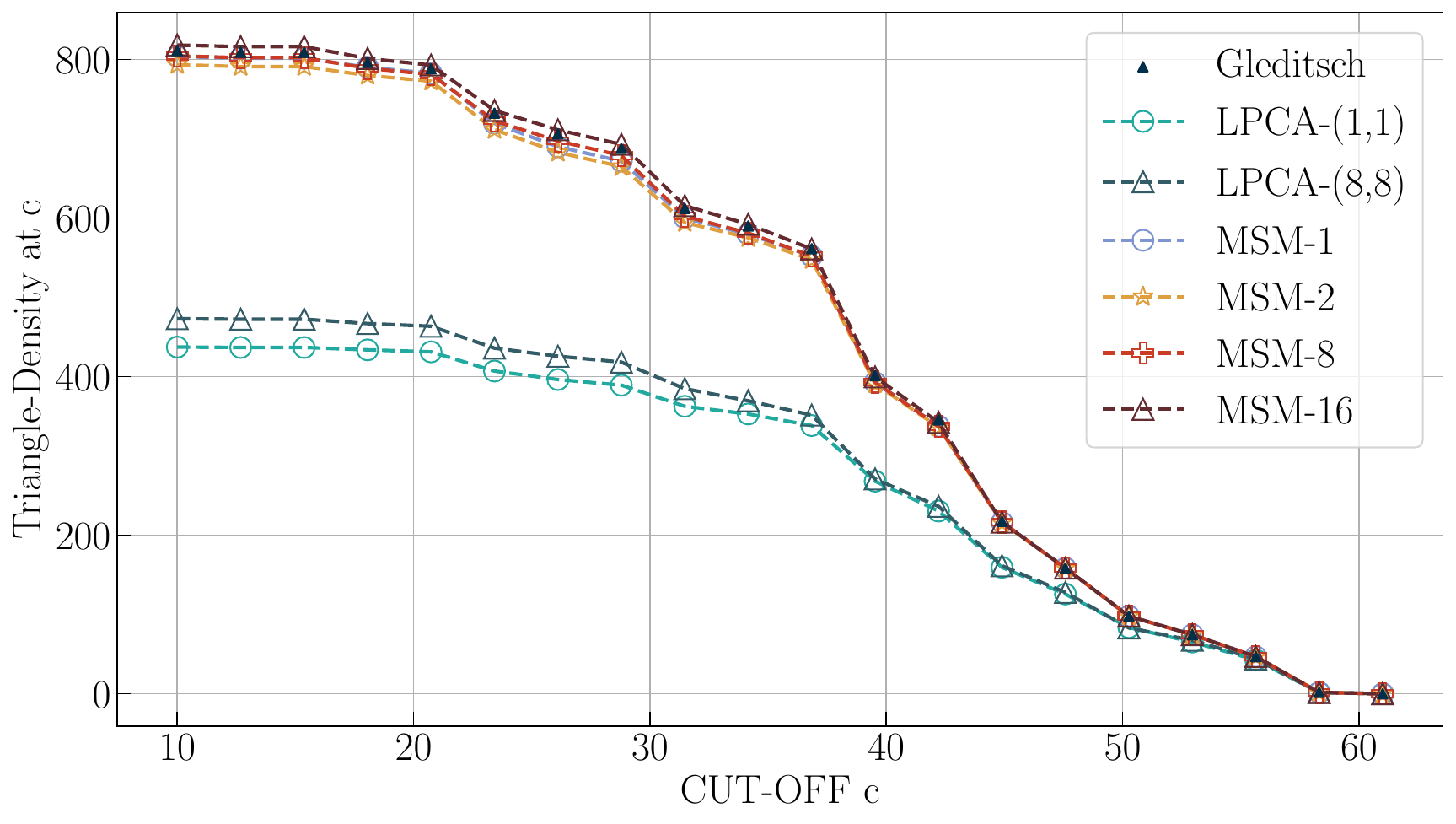}
        \label{fig:WTW_RCTriangles_level4}
    }
    \caption{
    \textit{Left:} Triangle Density at level 0. The plot shows the evolution of the $ \tau^{(\ell)}(c)$ with respect to the degree $ c$ in blue solid dots. The other markers identify the expected $\langle \tau^{(\ell)}(c) \rangle$ by the LPCA-(1,1), LPCA-(8,8), MSM-1, MSM-2, MSM-8, MSM-16.
    \textit{Right:} Triangle Density at level 2. 
    }
    \label{fig:WTW_RCTriangles_level04}
\end{figure*}

\clearpage
\bibliography{bibliography}

\end{document}